%% file: main.tex
\documentclass{lmcs} 
\title{On the Separability Problem of VASS Reachability Languages} 

\input{author_spec.tex}
\input{preamble}
\usepackage{proof}

\begin{document}

\include{paper}

\end{document}

%% file: author_spec.tex
\author{Eren Keskin}
\email{e.keskin@tu-bs.de}

\author{Roland Meyer}
\email{roland.meyer@tu-bs.de}



%% file: preamble.tex
\AtEndPreamble{
    \newtheorem{insight}{Insight}
}

\usepackage{comment}
\usepackage{tabu}
\usepackage{float}
\usepackage{listings}
\usepackage{environ}
\usepackage{tabularx}
\usepackage{hyperref}
\usepackage{cleveref}

\usepackage{scalefnt}
\usepackage{mathtools}
\usepackage{amsfonts}
\usepackage{scalerel}
\usepackage{subcaption}

\usepackage{enumitem}

\usepackage{amsthm}

\usepackage{tikz}
\usetikzlibrary{matrix, shapes,automata,positioning,calc}

\usepackage{styles/defs}
\usepackage{styles/common}
\usepackage{styles/comments}
\usepackage{styles/automata}
\usepackage{styles/vas}

%% file: paper.tex

\begin{abstract}
We show that the regular separability problem of VASS reachability languages is decidable and $\mathbb{F}_{\omega}$-complete. 
At the heart of our decision procedure are doubly-marked graph transition sequences, 
a new proof object that tracks a suitable product of the VASS we wish to separate.
We give a decomposition algorithm for DMGTS that not only achieves perfectness as known from MGTS, but also a new property called faithfulness.
Faithfulness allows us to construct, from a regular separator for the $\ints$-versions of the VASS, a regular separator for the $\nat$-versions.
Behind faithfulness is the insight that, for separability, it is sufficient to track the counters of one VASS modulo a large number that is determined 
by the decomposition. 
\end{abstract}

\maketitle
%
\input{intro_short.tex}
\input{vass.tex}
\input{regsep.tex}
\input{reach.tex}
\input{overview.tex}
\input{dmgts.tex}

\input{algorithm.tex}
\input{separatingaut.tex}
\input{separatingpump.tex}
\input{example.tex}

\input{decomposition.tex}

\input{conclusion.tex}

\bibliography{cited}
\bibliographystyle{alphaurl}

%% file: intro_short.tex
\section{Introduction}\label{Section:Introduction}
Regular separability problems for the languages of infinite-state systems are recently gaining momentum~\cite{Kopczynski16,IntVASS17,CL17,CCLP17,WSTS18,CLP20,CZ20,Baumann23,WSTS23,SizeSep23}. 
These problems take as input two infinite-state systems with languages $\alang_1$ and $\alang_2$, and ask whether $\alang_1\separable\alang_2$ holds, whether there is a regular language~$\areg$ that separates the two in the sense that $\alang_1\subseteq \areg$ and $\areg\cap\alang_2=\emptyset$. 
What makes regular separability problems interesting is that they do not seem to admit a reduction to established problems like emptiness.
Instead, the decision procedure has to analyze the gap between $\alang_1$ and $\alang_2$, and judge whether it is large enough to be described by a regular language.

Despite this challenge, there is a pleasant number of  positive results on regular separability. 
It has been shown that disjoint WSTS languages are always separated by a regular language~\cite{WSTS18,WSTS23}. 
For disjoint VASS coverability languages, matching upper and lower bounds on the size of least separators have been found~\cite{SizeSep23}. 
For Parikh automata~\cite{IntVASS17} and B\"uchi VASS coverability languages~\cite{Baumann23}, regular separability has been shown to be decidable.

Unfortunately, for the main model in this field, namely VASS reachability languages, the search has only brought partial results. 
This includes the decidability of 
the regular separability problem  
for the reachability languages of one-dimensional VASS~\cite{CL17}, 
for $\ints$-VASS reachability languages~\cite{IntVASS17}, 
for the commutative closure of VASS reachability languages~\cite{CCLP17}, 
and for VASS reachability languages from any of the aforementioned classes~\cite{CZ20}. 
The study has also led to important new techniques.
With the transducer trick, one can reduce the regular separability problem to a variant where only one language is taken as input and the second is fixed~\cite{IntVASS17,CZ20}.
For this variant, the basic separator approach tries to determine a limited set of regular languages so that, if separability holds, then a finite combination of these languages will serve as a separator~\cite{CZ20}. 
The techniques turned out widely applicable~\cite{CZ20,SizeSep23,Baumann23}, and the transducer trick will also play a central role in our work.
Related to regular separability is the separability of VASS reachability sets~\cite{CCLP17}. 
A landmark result in this context shows that VASS reachability sets admit Presburger-definable invariants~\cite{Leroux10}, which led to a new algorithm for solving VASS reachability~\cite{Leroux11}. 
To sum up, despite more than a decade of efforts, the decidability of regular separability for VASS reachability languages has remained open.

%
%

We solve the open problem and show that regular separability for VASS reachability languages is decidable and $\fomega$-complete.  
The problem is primitive recursive if the dimension of the input VASS is fixed. 
The class $\fomega$ contains the problems that can be solved with Ackermannian time and space~\cite{Schmitz16,schmitz17}, and the master problem is VASS reachability. 
%
The hardness of VASS reachability has been established only recently~\cite{Leroux21,CO21,SL22}. 
The decidability is a classic result~\cite{Mayr81,Kosaraju82,Lambert92}, with~\cite{ST77} an early attempt, and based on the algorithms proposed in these works, the upper bound has been brought down from $\fof{\omega^3}$~\cite{LerouxSchmitz15} over~$\fof{\omega^2}$~\cite{schmitz17} to $\fomega$~\cite{Leroux19}. 
The algorithms reduce the VASS reachability problem to the reachability problem in $\ints$-VASS, using an iterative decomposition that creates potentially many and potentially large $\ints$-VASS. 
This refinement has been called the KLMST decomposition \cite{LerouxSchmitz15} after its inventors Kosaraju~\cite{Kosaraju82}, Lambert~\cite{Lambert92}, Mayr~\cite{Mayr81}, as well as Sacerdote and Tenney~\cite{ST77}.
We will follow the same strategy, and reduce the regular separability problem of VASS reachability languages to the regular separability problem of $\ints$-VASS reachability languages, with a decomposition that adapts the KLMST decomposition.
%
The latter problem 
has been shown to be decidable in~\cite{IntVASS17}.
For the precise upper bound, we rely on an analysis inspired by~\cite{Leroux19}.   

While this is the overall strategy, it takes new ingredients to make it work that go beyond the toolkit of VASS reachability. 
To explain them, we start from the transducer trick, and fix one of the languages to the Dyck-language $\dycklangn{n}$.
Given our approach of reducing $\N$ separability to $\Z$ separability, we can reasonably expect that our decomposition (i) uses $\Z$-separability to construct $\N$-separators and (ii) deduce $\N$-inseparability from $\Z$-inseparability.

The first direction contains the main obstacle we overcome.
Our intention is to use the KLMST decomposition to refine our separators.
However, on its face, the KLMST decomposition refines VASS (languages) in a way that cannot be captured by regular languages.
Namely, it adds new valuation targets that need to be reached at certain points along the run.
We cannot impose such constraints with regular separators.
The key insight is an invariant, called faithfulness, that is also kept by the KLMST decomposition.
Faithfulness says that relaxing the intermediate reachability constraints to check reachability modulo some base $\mu\geq 1$ does not change the language.
This allows us to implement the intermediate checks in separators. 
Note that the statement does not concern the initial and final configurations, and thus does not imply regularity.
The initial VASS only has targets for initial and final configurations, and thus trivially faithful.
Since the KLMST decomposition preserves faithfulness, so are the results of the decomposition.
Our decomposition inherits this.

In the second direction, we show inseparability.
Similarly to the KLMST decomposition, we can only return a negative answer at the end of the decomposition.
The KLMST decomposition establishes a property called perfectness, where $\Z$-emptiness implies $\N$-emptiness.
Perfectness, via Lambert's iteration lemma \cite{Lambert92}, guarantees certain forms of pumping that make positivity constraints irrelevant.
Our approach establishes a similar notion of perfectness.
Then, for any given DFA $\annfa$, we exploit (the proof of) Lambert's iteration lemma to construct pairs of words that are $\sim_{\annfa}$-equivalent.
Under this definition \cite{Buechi62}, two words are $\sim_{\annfa}$-equivalent, if from any given state, reading either word leads to the same state.
This also implies that they are equivalent in the syntactic monoid of $\langof{\annfa}$.
The equivalence allows for a context-free treatment.
Thanks to this perspective, it suffices to ensure equivalence for each ingredient of the iteration in isolation.

To handle the two directions togeter we work with a new structure, Doubly Marked Graph Transition Sequences (DMGTS).
These objects simultanteously track both the languages we want to separate.
Unlike a product, 
a DMGTS defines two languages $\leftlangof{\admgts}$ and $\rightlangof{\admgts}\subseteq\dycklangn{n}$ (along with their $\Z$-variants).
The goal is to understand the separability of $\leftlangof{\admgts}$ from $\dycklangn{n}$.
In our decomposition, the language $\rightlangof{\admgts}$ is the fragment of $\dycklangn{n}$ that is relevant to deducing $\leftlangof{\admgts}\separable\dycklangn{n}$.
This means that $\leftlangof{\admgts}\separable\rightlangof{\admgts}$ implies $\leftlangof{\admgts}\separable\dycklangn{n}$.
We ensure this by the faithfulness invariant and the structure of the DMGTS.
Namely, a DMGTS is an MGTS \cite{Lambert92} that tracks the counters of both languages $\leftside\uplus\rightside$, equipped with a modulo base $\mu\geq 1$.
The language $\sidelangof{\aside}{\admgts}$ is defined by relaxing the constraints on~$\complementof{\aside}$ counters.
In particular, $\rightlangof{\admgts}$ completely ignores the $\leftside$ constraints, and $\leftlangof{\admgts}$ relaxes the $\rightside$ constraints to modulo-$\mu$ checks.
It is worth noting that DMGTS define four notions of acceptance: languages may focus on one set of counters $\leftside$ or $\rightside$, and be defined over $\N$- or $\Z$-valuations.
To deal with this plurality, we keep our definitions parametrized.
This sometimes leads to dense notation.

Our decomposition breaks the original DMGTS $\admgts$ down into a set of finitely many faithful DMGTS $\admgts_{0}, \ldots, \admgts_{k}$ that preserves one language, $\bigcup_{i\leq k} \leftlangof{\admgts_{i}}=\leftlangof{\admgts}$.
In contrast to the KLMST decomposition, we do not only produce perfect DMGTS, but also DMGTS whose $\leftside$-languages are already separable from $\dycklangn{n}$.
With this, the results of the decomposition satisfy both (i) and (ii) we outlined above. 
Let $\admgts$ be perfect.
If the $\Z$ languages of $\admgts$ are separable, faithfulness allows us to generalize this to the separability of $\leftlangof{\admgts}$ \emph{and $\dycklangn{n}$.} 
Here, the separator is obtained by enriching the $\Z$-separator by the control flow, and modulo-$\mu$ checks used in $\leftlangof{\admgts}$.
For (ii), we rely on perfectness.
Given perfectness, we start from $\Z$-inseparability and deduce $\N$-inseparability as we outlined above.
If $\Z$-inseparability holds, for each DFA $\annfa$, we construct words $w\in\leftlangof{\admgts}$ and $v\in\rightlangof{\admgts}$ that are $\sim_{\annfa}$-equivalent.
This implies that $\leftlangof{\admgts}$ and $\rightlangof{\admgts}\subseteq\dycklangn{n}$ are inseparable, yielding the inseparability $\leftlangof{\admgts}\inseparable \dycklangn{n}$.
To use Lambert's pumping lemma~\cite{Lambert92} for both languages at once, it is crucial that the languages share the same control flow.
%

%

\subsubsection*{Overview}
After an introduction to VASS, reachability languages, and regular separability, we discuss Lambert's decision procedure for VASS reachability in Section~\ref{Section:Reach}. 
It contains a number of concepts that we build on, including MGTS, characteristic equations, and perfectness.
We give an overview of the ideas behind our regular separability algorithm in \Cref{Section:Overview}.
Following this, DMGTS and faithfulness are defined in Section~\ref{Section:DMGTS}. 
The structure of our decision procedure for regular separability is given in Section~\ref{Section:DecisionProcedure}. 
In Section~\ref{Section:SeparabilityResult}, we prove the separability transfer result, relating $\Z$-separability to $\N$-separability.
Before moving on to the technical details of the decomposition, we observe how the it would handle a concrete example in Section~\ref{Section:Example}.
The DMGTS decomposition can be found in Section~\ref{Section:OurDecomposition}. 


%% file: vass.tex
\section{Preliminaries}

We introduce some notation. 
Given a sequence $\sigma\in A^{*}$ over a set~$A$, we use $\sizeof{\sigma}$ for the length and $\coordacc{\sigma}{i}$ for the $i$-th component. 
With the distinguished indices $\firstindex$ and $\lastindex$ we access the first resp. the last component.
Whenever we have a function $f:A\rightarrow\powof{X}$ into a powerset and $B\subseteq A$, we may write $f(B)$ for $\bigcup_{b\in B}f(b)$. 
To reduce unnecessary declarations, if an object $\avas$ contains a component usually named $X$, we also use $\avas.X$ to reference this component.
We use $0, 1 \in\Z^{\counters}$ to refer to the vectors whose entries are all $0$ resp.\ $1$.

\subsection{Vector Addition Systems with States (VASS)} \label{Subsection:VASS}
A vector addition system with states $\avas=(\nodes, \analph, \counters, \edges)$ consists of a finite set of states~$\nodes$, a finite alphabet $\analph$, a finite set of counters~$\counters$, and a finite set of edges $\edges\subseteq\nodes\times\updates\times\nodes$. 
We call $\updates = \analphemp\times\ints^{\counters}$  the set of updates, where $\analphemp=\analph\cup\set{\emptyword}$. 

We will not only work with VASS but also with $\ints$-VASS. 
To define the semantics of both models in one go, let $\countersp\subseteq\counters$ be a subset of the counters. 
A $\countersp$-counter valuation $\aconf\in\nat^{\countersp}\times\ints^{\counters\setminus\countersp}$ gives a non-negative value to the $\countersp$-counters. 
A $\countersp$-configuration is a pair~$(\anode, \aconf)$ consisting of a state $\anode\in\nodes$ and a $\countersp$-counter valuation $\aconf$.  
%
%
%
A $\countersp$-run is a sequence  $\apath=(\anoden{0}, \aconf_{0})\anedge_0(\anoden{1}, \aconf_{1})\ldots(\anoden{l}, \aconf_{l})$ of $\countersp$-configurations and edges where for all $i<l$ we have $\anedge_i=(\anoden{i}, \aletter_i, \avar_i, \anoden{i+1})$ with  $\avar_i=\aconf_{i+1}-\aconf_{i}$.  
We write $\pathsof{\countersp}{\avas}$ for the set of all $\countersp$-runs in $\avas$.  
We also use $\pathsof{\nat}{\avas}$ for this set if $\countersp$ contains all counters, and $\pathsof{\ints}{\avas}$ if $\countersp$ is empty. 
Note that the configurations in a run are already determined by the initial counter valuation and the sequence of edges. 
We may therefore also give a run as $\apath=\aconf.\sigma$ with $\sigma\in \edges^*$. 
We may also emphasize the initial and final configurations and give a run as $\apath=(\anode, \aconf).\sigma.(\anode', \aconf')$.   
We use $\edgelabelof{\apath}\in\analph^*$ for the sequence of letters on the run.    
Two runs are equivalent, $\apath_1\pathequiv\apath_2$, if they only differ in the states they visit. 
We use $\actionbalof{\anedge}\in \ints^{\counters}$ for the counter update done by an edge, and $\actionbalof{\apath}$ for the counter update done by a run. 
A Parikh vector $\psi\in\nat^{\edges}$ associates with each edge an occurrence count, and we define $\actionbalof{\psi}=\sum_{\anedge\in\edges}\coordacc{\psi}{\anedge}\cdot\actionbalof{\anedge}$. 
Note that $\actionbalof{\apath} = \actionbalof{\parikhof{\rho}}$, where $\parikhof{\apath}$ is the Parikh vector induced by~$\rho$. 


We define accepting runs with generalized initial and final configurations. 
Because we employ many related notions of acceptance, we keep our definitions as general as possible.
Let $\natomega=\nat\cup\set{\omega}$ and $\intsomega=\ints\cup\set{\omega}$ extend the natural numbers and the integers by a top element.  
We lift this to counter valuations and call $\aconf\in\nat_{\omega}^{\countersp}\times\intsomega^{\counters\setminus\countersp}$ a generalized $\countersp$-counter valuation. 
We write $\infinitiesof{\aconf}$ for the set of counters $i$ with $\aconf(i)=\omega$.  
We call the pair~$(\anoden, \aconf)$ with state $\anode\in\nodes$ a generalized $\countersp$-configuration if $\aconf$ is a generalized $\countersp$-counter valuation.
To capture multiple notions of acceptance at once,
we define acceptance parametric in a preorder $\sqsubseteq\ \subseteq\intsomega^{d}\times\intsomega^{d}$. 
We lift preorders on $\intsomega$ to generalized counter valuations by a componentwise comparison. 
%
We also lift it to generalized configurations by $(\anoden{1}, \aconf_{1})\sqsubseteq (\anoden{2}, \aconf_{2})$, if $\anoden{1}=\anoden{2}$ and $\aconf_1\sqsubseteq\aconf_2$. 
An important instance is the specialization preorder $\omegaleq\ \subseteq \intsomega\times\intsomega$, which is defined by $k\omegaleq k$ and $k\omegaleq\omega$ for all $k\in\intsomega$.

An initialized VASS $\avasp=(\avas, (\instate, \incounters), (\outstate, \outcounters))$ enriches a VASS $\avas$ with generalized $\nat$-configurations $(\instate, \incounters)$ and~$(\outstate, \outcounters)$ that we call initial and final. 
We speak of an extremal configuration if it is initial or final. 
The runs of $\avasp$ are the runs of $\avas$.
Such a run is~$\sqsubseteq$-accepting, if $\coordacc{\apath}{\firstindex}\sqsubseteq(\instate, \incounters)$ and $\coordacc{\apath}{\lastindex}\sqsubseteq(\outstate, \outcounters)$, the first configuration is smaller than the initial configuration in the given preorder, and the last configuration is smaller than the final configuration. 
For $\mathord{\sqsubseteq}=\mathord{\omegaleq}$, we get the usual notion of reachability acceptance.
The parametrized acceptance definition allows us the extend this notion by modifying $\sqsubseteq$.
We use $\acceptof{\countersp}{\sqsubseteq}{\avasp}$ for the set of all $\countersp$-runs in~$\avasp$ that are $\sqsubseteq$-accepting.

We denote the size of an initialized VASS by~$\sizeof{\avasp}$. 
We measure the size in binary, but this does not matter for the large complexity classes we are concerned with.

With an initialized VASS, we associate the language of all words that label an $\omegaleq$-accepting run: 
\begin{align*}
\speclangof{\countersp}{\avasp}\;\;=\;\;\setcond{\edgelabelof{\apath}}{\apath\in\acceptof{\countersp}{\omegaleq}{\avasp}}\ .
\end{align*}
We use $\speclangof{\nat}{\avasp}$ and $\speclangof{\ints}{\avasp}$ if every counter resp. no counter has to stay non-negative. 
The former are the VASS reachability languages and the latter the $\ints$-VASS reachability languages. 

%
%

%% file: regsep.tex
\section{Regular Separability}\label{Section:RegSep}
We study the regular separability of VASS reachability languages. 
Languages $\alang_1, \alang_2\subseteq\analph^*$ are \emph{separable by a regular language}, denoted by $\alang_1\separable\alang_2$, if there is a regular language $\aseparator\subseteq\analph^*$ that satisfies $\alang_1\subseteq\aseparator$ and $\aseparator\cap\alang_2=\emptyset$. 
The language $\aseparator$ is usually called the separator, and the regular separability problem asks whether a separator exists for given languages.
In the definition of the decision problem, we again make the domain of counter values a parameter: 
\begin{quote}
{\bfseries \large \textsf{$\counterdom$-REGSEP}}\\
{\bfseries Given:} Initialized VASS $\avasp_1$ and $\avasp_2$ over $\analph$.\\
{\bfseries Problem:} Does $\speclangof{\counterdom}{\avasp_1}\separable\speclangof{\counterdom}{\avasp_2}$ hold?
\end{quote}
Our main result is the decidability of regular separability for the reachability languages of ordinary VASS. 
\begin{thm}\label{Theorem:MainResult}
\textsf{$\nat$-REGSEP} is decidable and $\fof{\omega}$-complete. 
We can effectively compute a separator in this time and space bound.
\end{thm}
Recall that $\fof{\omega}$ is the class of problems that can be solved with Ackermannian time and space~\cite{Schmitz16}. 
It is closed under further calls to primitive recursive functions~\cite[Lemma 4.6]{Schmitz16}, and these functions are also used as reductions to define hardness.
Our lower bound for regular separability is by a reduction from the reachability problem in $\nat$-VASS, whose~$\fof{\omega}$-hardness is a recent achievement~\cite{CO21,Leroux21,SL22}. 
It even holds if we promise the input languages to be disjoint and the only separator candidate is $\emptyset$. 
\begin{lem}
    \textsf{$\nat$-REGSEP} is $\mathbb{F}_{\omega}$-hard, even if the input languages are promised to be disjoint and the only separator candidate is $\emptyset$. 
\end{lem}


\begin{proof}
    We show how to reduce the reachability problem for VASS to \textsf{$\N$-REGSEP}. 
    Assume we are given the VASS~$\avasp$ for which we should check reachability. 
    There are VASS $\avasp'$ and $\avasp''$ whose languages are disjoint, $\speclangof{\nat}{\avasp'}\cap\speclangof{\nat}{\avasp''}=\emptyset$, but where regular separability fails, $\speclangof{\nat}{\avasp'}\mathop{\not\hspace{0.00cm}|}\speclangof{\nat}{\avasp''}$. 
    Let 
    $\avasp=(\avas, (\instate, \incounters), (\outstate, \outcounters))$ 
    and $\avasp'=(\avas', (\instate', \incounters'), (\outstate', \outcounters'))$. 
    Assume wlog. that the sets of states resp. counters in $\avasp$ and $\avasp'$ are disjoint. 
    We define the VASS
    \begin{align*}
        \avas_{\mathit{new}}\ =\ \avas_{\epsilon}\uplus\set{\anedge_{\mathit{new}}}\uplus \avas'\ . 
    \end{align*}
    It first executes a version of $\avas$ where all edges are labeled by~$\varepsilon$ and then moves to $\avas'$ using the new edge $\anedge_{\mathit{new}}=(\outstate, \anupdate, \instate')$. 
    The update removes the final counter valuation $\outcounters$ from the counters in $\avas$, and adds the counter valuation $\incounters'$ to the counters in $\avas'$,  $\actionbalof{\anupdate}=(-\outcounters, \zerovec)+(\zerovec, \incounters')$. 
    The labeling is $\lambda(\anupdate)=\varepsilon$. 
    We define $\avasp_{\mathit{new}}=(\avas_{\mathit{new}}, (\instate, (\incounters, \zerovec)), (\outstate', (\zerovec, \outcounters')))$ and claim that 
    \begin{alignat*}{5}
        \acceptof{\nat}{\omegaleq}{\avas}&\neq \emptyset\quad&\text{implies}\quad \speclangof{\nat}{\avasp_{\mathit{new}}} &= \speclangof{\nat}{\avasp'}\\
        \acceptof{\nat}{\omegaleq}{\avas}&= \emptyset\quad&\text{implies}\quad \speclangof{\nat}{\avasp_{\mathit{new}}} &= \emptyset\ .
    \end{alignat*}
        As a consequence, reachability in $\avas$ fails, $\acceptof{\nat}{\omegaleq}{\avas}= \emptyset$,  if and only if $\speclangof{\nat}{\avasp_{\mathit{new}}}$ and $\speclangof{\nat}{\avasp''}$ are separable by $\emptyset$, which is a regular language. 
        The reduction can be computed in logarithmic space. 
\end{proof}




For the upper bound, we use the transducer trick~\cite{IntVASS17,CZ20} 
and reduce \textsf{$\nat$-REGSEP} to a separability problem that only takes one reachability language as input. 
The second language is fixed to the Dyck language $\dycklangn{\dyckdims}$, where $n$ is the number of counters in the second VASS. 
We explain the this language below.

\subsubsection*{Dyck Language and Dyck visibility}
The Dyck language $\dycklangn{n}\subseteq\analph_{n}^{*}$ with $n$ counters is accepted by the initialized VASS $(\dyckvas_{\dyckdims}, (\anode, \zerovec), (\anode, \zerovec))$ with $\dyckvas_{\dyckdims}=(\set{\anode}, \dyckalphn{\dyckdims}, \rightside, \edges)$.  
The set of counters is $\rightside=\set{1, \ldots, \dyckdims}$, the alphabet is $\dyckalphn{\dyckdims}=\setcond{\incdyckn{i}, \decdyckn{i}}{i\in\rightside}$, and the edges move the counters as suggested by the letters, $\edges=\setcond{(\anode, (\incdyckn{i}, \unitvecn{i}), \anode), (\anode, (\decdyckn{i}, -\unitvecn{i}), \anode)}{i\in\rightside}$ with $\unitvecn{i}$ the $i$-th unit vector.
This means we increment counter $i$ when seeing~$\incdyckn{i}$, and decrement $i$ upon $\decdyckn{i}$. 
We call a VASS that sticks with this link between letters and counter updates \emph{Dyck visible}.
%
%
Note that the VASS only has one state, and we also speak of a \emph{VAS}.

The lemmas below follows from standard arguments used in \cite{CZ20}.
The exact details of transducers are not important for our development and can be found in \cite{CZ20}.
For this reason, we abstract away from these details, and treat a transducer $\atransducer$ as the rational transduction \cite[Section 2]{CZ20} $\atransducer\subseteq\analph_{n}^{*}\times\analph^{*}$ it defines.
We can construct a transducer $\atransducer\subseteq \analph_{n}^{*}\times\analph^{*}$ that interprets a sequence of counter commands $\aword\in\analph_{n}^{*}$ as a run in~$\avasp$, and outputs its label.
Assuming that $\avasp$ accepts upon reaching $\zerovec$ from $\zerovec$, we get $\atransducer(\dycklangn{n})=\langof{\avasp}$, since $\dycklangn{n}$ contains the counter history of all accepting runs.
Applying the inverse of $\atransducer$ to both sides of a separability query, allows us to fix one language to $\dycklangn{n}$.
This is stated below.

\begin{lem}[\cite{CZ20}, Proposition 1, and Lemma 12]\label{Lemma:FixToDyck}
    Let $\avas$ and $\avasp$ be initialized VASS over $\analph$, and let $\avasp$ have $\dyckdims$ counters.
    We can compute in elementary time from $\avasp$ a  transducer $\atransducer\subseteq\analph_{n}^{*}\times\analph^{*}$ so that $\langof{\avas}\separable\langof{\avasp}$ if and only if $\atransducer^{-1}(\langof{\avas})\separable\dycklangn{\dyckdims}$.
 \end{lem}
VASS are effectively closed under inverse transductions.
\begin{lem}[\cite{CZ20}, Proposition 1]\label{Lemma:TransducerComputable}
    Let $\avas$ be an initialized VASS over $\analph$ and let $\atransducer:\dyckalphn{\dyckdims}^{*}\times\analph^{*}$.
    We can compute in time elementary in the size of $\avas$ and $\atransducer$ a VASS $\avas'$ so that $\langof{\avas'}=\atransducer^{-1}(\langof{\avas})$.  
\end{lem}          
With the aforementioned closure of $\fof{\omega}$ under primitive recursive functions, Theorem~\ref{Theorem:MainResult} is a consequence of the following result.
\begin{prop}\label{proposition:DyckSeparationDec}
Let $\avas$ be an initialized VASS over $\dyckalphn{\dyckdims}$.
Then $\langof{\avas}\separable\dycklangn{\dyckdims}$ is decidable and we can compute a separator in $\fof{\omega}$. 
\end{prop}

The rest of the paper is devoted to proving Proposition~\ref{proposition:DyckSeparationDec}.
Our algorithm is based on the decision procedure for VASS reachability, whose details we recall in a moment. 
%
A second ingredient is the decidability of regular separability for $\ints$-VASS. 
This is a result we can use in black-box fashion, when it is formulated as follows. 
\begin{thmC}[\cite{IntVASS17}]\label{Theorem:IntsVASSRegSep}
\textsf{$\ints$-REGSEP} is decidable and a regular separator can be computed with elementary resources.
\end{thmC}

%% file: reach.tex
\newcommand{\rankless}{<_{\mathsf{rnk}}}
\newcommand{\rankleq}{\leq_{\mathsf{rnk}}}
\newcommand{\ranking}{\mathsf{rank}}
\newcommand{\rankingof}[1]{\ranking(#1)}
\newcommand{\spanof}[1]{\mathsf{span}#1}
\newcommand{\cyclespaceof}[1]{\mathsf{V}(#1)}

\section{VASS Reachability}\label{Section:Reach}

We recall Lambert's decision procedure for VASS reachability~\cite{Lambert92}, with the recent additions due to Leroux and Schmitz~\cite{Leroux19}.
We base our development on~\cite{Lambert92}, and use some concepts from~\cite{Leroux19} for the Ackermannian upper bound.
The purpose is to introduce concepts that we need later.

A reader familiar with the KLMST decomposition may skip this section.
For this reader, the following points may serve as useful reference for further sections: the MGTS notation introduced in \Cref{Subsection:MGTS}, the parametrized characteristic equations defined in \Cref{Subsection:CharacteristicEq}, the slightly more general statement of Lambert's iteration lemma given in Lemma~\ref{Lemma:ModifiedIteration}.

%
\subsection{Overview}\label{Subsection:KLMSTOverview}
The VASS reachability problem takes as input an initialized VASS~$\avasp$ and asks whether $\acceptof{\nat}{\omegaleq}{\avasp}\neq \emptyset$. 
%
%
The decision procedure is an abstraction-refinement algorithm that computes a sequence 
\begin{align*}
\interacceptof{\ints}{\omegaleq}{S_0}\supseteq\interacceptof{\ints}{\omegaleq}{S_1}\supseteq\ldots \supseteq \interacceptof{\nat}{\omegaleq}{\avasp}\ 
\end{align*}
over a more expressive ($\Z$-)acceptance condition that defines the set of runs $\interacceptof{\nat}{\omegaleq}{\avasp}$ ($\interacceptof{\ints}{\omegaleq}{\avasp}$).
Each over-approximation $S_i$ is a finite set of VASS $\admgts$ and, as agreed, we use~$\interacceptof{\ints}{\omegaleq}{S_i}$ for $\bigcup_{\admgts\in S_i}\interacceptof{\ints}{\omegaleq}{\admgts}$. 
Details on the shape of $\admgts$, and the new acceptance conditions, will follow in a moment.

At each step, the algorithm picks an element $\admgts\in S_i$ and checks reachability. 
If reachability does not hold, $\interacceptof{\ints}{\omegaleq}{\admgts}=\emptyset$,
then the element will not be considered in the future, $S_{i+1}=S_i\setminus\set{\admgts}$. 
If reachability holds, the algorithm checks whether $\admgts$ 
is perfect, a property we discuss below. 
If so, the algorithm concludes that the $\ints$-run it has just found can be turned into an $\nat$-run, and terminates with the verdict \emph{reachable}. 
If $\admgts$ is not perfect, there is a guarantee that $\admgts$ can be refined. 
Following Leroux and Schmitz \cite{LerouxSchmitz15}, the refinement is called KLMST decomposition after the inventors Kosaraju~\cite{Kosaraju82}, Lambert~\cite{Lambert92}, Mayr~\cite{Mayr81}, as well as Sacerdote and Tenney~\cite{ST77}. 
The decomposition computes from $\admgts$ a new and again finite set $S(\admgts)$ of VASS that replace $\admgts$ in the approximation, meaning we have $S_{i+1}=(S_i\setminus\set{\admgts})\cup S(\admgts)$. 
The decomposition guarantees $\interacceptof{\nat}{\omegaleq}{\admgts}=\interacceptof{\nat}{\omegaleq}{S(\admgts)}$ so that we do not lose $\nat$-runs but remain over-approximate. 
If $S_i$ is found to be empty, the algorithm terminates with the verdict \emph{unreachable}. 

What makes the algorithm terminate is a progress guarantee: 
each VASS in the KLMST decomposition $S(\admgts)$ is guaranteed to be strictly smaller than~$\admgts$ in a well-founded order. 
%
Since $S(\admgts)$ is finite, K\"onig's lemma shows termination of the overall algorithm.  

We elaborate on the VASS used in the over-approximation. 
There are two standard techniques to over-approximate reachability in VASS: characteristic equations~\cite{CHEP71,GL73} and coverablity graphs~\cite{KM69}. 
The characteristic equations can track counter values precisely, but they cannot guarantee that intermediate values remain non-negative.  
Coverability graphs are the opposite, they can guarantee that intermediate values remain non-negative, but they cannot track counter values precisely. 
The decision procedure combines the two. 
The VASS are decorated by generalized $\nat$-counter valuations, like coverability graphs. 
If a decoration is~$\omega$, and thus not precise enough for reachability, that counter is treated as a $\ints$-counter and handled by the characteristic equations.

For this combination to solve reachability, we have to be able to transfer the non-negativity guarantee given by coverability graphs to the characteristic equations. 
Behind the non-negativity guarantee given by coverability graphs 
is a translation: if we have a run in the coverability graph, we obtain an $\nat$-run in the underlying VASS by repeating intermediate transition sequences in order to pump counter values arbitrarily high. 
By transferring the guarantee, we mean that also the characteristic equations should admit this pumping: the counters to be pumped as well as the repetition count for the edges on the pumping sequences should be unbounded in the solution space of the characteristic equations.
%
%
If this is the case, the coverability graph is called perfect.

It is easy to check unboundedness in the solution space, and therefore also perfectness: we just have to find a solution to the homogeneous variant of the characteristic equations where the variable is positive. 
More precisely, the characteristic equations will have the shape $A\avar= b\wedge \avar\geq 0$. 
The solutions $\asol$ can have arbitrarily high values $\coordacc{\asol}{i}$ if and only if $A\avarp = 0\wedge \avarp\geq 0$ has a solution $\asol'$ with $\coordacc{\asol'}{i}>0$. 
%
%
To see that homogeneous solutions are necessary, assume there are solutions with arbitrarily high values for a variable. 
The well-quasi order of $\nat^k$ will give us comparable solutions that we can subtract to solve the homogenous equations.

The KLMST decomposition comes in when perfectness fails, but the $\omega$-decorations and pumping sequences do not coincide with the unboundedness in the solution space. 
This yields a bound (on pumping sequences, on the solution space, etc.) that constrains all accepted~$\N$-runs.
The decomposition then breaks the original VASS down into a finite set of VASS, by making this bound explicit in the syntax.
Since imposing the bound does not exclude any accepted $\N$-runs, $\interacceptof{\nat}{\omegaleq}{\admgts}=\interacceptof{\nat}{\omegaleq}{S(\admgts)}$.
The syntax also informs $\Z$-acceptance, so this refines the set of $\Z$-accepted runs and yields $\interacceptof{\Z}{\omegaleq}{\admgts}\supseteq\interacceptof{\Z}{\omegaleq}{S(\admgts)}$.
The new restrictions make the resulting VASS smaller in a well-founded preorder.
We make the ideas formal.

\subsection{MGTS}\label{Subsection:MGTS}
The VASS used for the over-approximation are so-called \emph{marked graph transition sequences (MGTS)} that are defined as follows: 
\begin{align*}
\admgts\;\;::=\;\;\precovering\;\;\mid\;\; \admgts_1.\anupdate.\admgts_2\ .
\end{align*}
An MGTS is an interleaving of precovering graphs $\precovering$ and updates~$\anupdate$.
Precovering graphs are initialized VASS of a form we define in a moment. 
In an MGTS, all precovering graphs share the same alphabet and the same set of counters $\counters$, but the sets of states are pairwise disjoint.
%

A \emph{precovering graph} $\precovering=(\avasp, \bal)$ is an initialized VASS $\avasp$ that is decorated by a consistent assignment~$\bal$.
The VASS should be strongly connected and satisfy $\avasp.\innode=\avasp.\outnode$, called the root of~$\precovering$ and denoted by $\rootstate$. 
Let $\nodes=\nodesof{\avasp}$ be the states, $\edges=\edgesof{\avasp}$ the edges, and $\counters=\countersof{\avasp}$ the counters in $\avasp$. 
An assignment $\bal:\nodes\to \natomega^{\counters}$ of generalized counter valuations to states is \emph{consistent}, if there is a set of counters $\countersp\subseteq\counters$ so that for all states $\anode\in\nodes$ we have $\coordacc{\bal(\anode)}{j}\in\nat$ if and only if $j\in\countersp$, we have $\coordacc{\bal(\rootstate)}{\countersp}=\coordacc{\incounters}{\countersp}=\coordacc{\outcounters}\countersp$, and $\coordacc{\actionbalof{\anupdate}}{\countersp}=\coordacc{\bal(\anodep)}{\countersp}-\coordacc{\bal(\anode)}{\countersp}$ for all $(\anode, \anupdate, \anodep)\in\edges$.  
The consistent assignment is the decoration mentioned in the overview. 
Consistency means all states agree on the set of counters $\countersp$ that are decorated by $\nat$-values.
The remaining counters  are decorated by $\omega$ and we use 
$\infinitiesof{\precovering}=\counters\setminus\countersp$ to refer to them. 
For the counters in $\countersp$, the decoration of the root has to coincide with the initial and the final valuations.
As a consequence, the initial and final valuations may only have less $\omega$-entries. 
The decoration tracks the counter updates performed by the edges.

%
%
%

A counter $i$ may be decorated by $\omega$ in the precovering graph but have a concrete initial value, $i\in \Omega$ with $\Omega = \infinitiesof{\precovering}\setminus\infinitiesof{\incounters}$. 
Then it should be possible to pump this counter in the precovering graph to arbitrarily high values while going from the root back to the root. 
Pumping means the loop starts in a counter valuation~$c_1$ that respects the concrete initial values and ends in a valuation~$c_2$
that is strictly larger in the counters from $\Omega$. 
For the counters in~$\infinitiesof{\incounters}$, there is no requirement and the loop may reduce their values. 
The remaining counters are tracked by the decoration and every loop will leave their valuation unchanged.
We use $c_1<^{\Omega} c_2$ for $c_1(i)<c_2(i)$ for all $i\in \Omega$. 
That pumping should be possible means the following set of \emph{covering sequences} should be non-empty:
%
%
%
\begin{align*}
\coveringseqof{}{\precovering}=\;\{\sigma\in &\ \edges^*\mid\exists c_1, c_2\in \nat^{\counters}. c_1\omegaleq\incounters\wedge c_1<^{\Omega} c_2\\
&\qquad\wedge(\rootstate, c_1).\sigma.(\rootstate, c_2)\in\pathsof{\nat}{\precovering}\}\ .
\end{align*}

We also need to pump down counters $i$ that are decorated $\omega$ in the precovering graph but receive a concrete value in the final configuration, $i\in \Omega$ with $\Omega = \infinitiesof{\precovering}\setminus\infinitiesof{\outcounters}$. 
We reuse the above definition and let
\begin{align*}
\downcoveringseqof{}{\precovering}\;\;=\;\;\reverseof{\coveringseqof{}{\reverseof{\precovering}}}\ .
\end{align*}
The reverse of a run is defined as expected, $\reverseof{(\apath_1.\apath_2)}=\reverseof{\apath_2}.\reverseof{\apath_1}$. 
The reverse of a single edge is $\reverseof{(\anode, a, x, \anodep)}=(\anodep, a, -x, \anode)$, meaning we increment where we have decremented before, and vice versa. 
The reverse runs stem from a reversal of the precovering graph, defined as $\reverseof{\precovering}=((\reverseof{\avas}, (\rootstate, \outcounters), (\rootstate, \incounters)), \bal)$.
Note that the initial and final configurations have changed their roles. 
The reversal of the underlying VASS $\reverseof{\avas}=(\nodes, \analph, \counters, \setcond{\reverseof{\anedge}}{\anedge\in\edges})$ simply reverses the edges. 
Hence, the two reversals in the definition have no effect on the counter updates but just identify the down-pumping runs that end in $\outcounters$ from arbitrarily high values in the counters $\infinitiesof{\precovering}\setminus\infinitiesof{\outmarking}$. 

The emptiness of $\coveringseqof{}{\precovering}$ and $\downcoveringseqof{}{\precovering}$ can be checked using (two different) unboundedness checks~\cite{KM69,Demri13}, which will also provide covering sequences if the sets are non-empty.

In our development, it will be helpful to understand MGTS~$\admgts$ as initalized VASS. 
We simply turn the intermediate updates into edges that connect the final state in one precovering graph with the initial state of the next. 
We write $\nodesof{\admgts}$ for the set of all states in precovering graphs of $\admgts$, and similarly for the alphabet, the counters, and the edges. 
We write $\innodeof{\admgts}$ for $\innodeof{\coordacc{\admgts}{\firstindex}}$, and $\outnodeof{\admgts}$ for $\outnodeof{\coordacc{\admgts}{\lastindex}}$. 
We use $\sizeof{\admgts}$ for the size.
The initial and final configurations of each precovering graph count towards the size.

Let $\mathord{\sqsubseteq}\subseteq \intsomega\times\intsomega$ be a preorder that defines an acceptance condition. 
Seeing MGTS as VASS, we can use $\acceptof{\countersp}{\sqsubseteq}{\admgts}$ to refer to the accepting runs. 
MGTS also have their own notion of \emph{intermediate acceptance}, where the runs not only have to meet the overall initial and final configurations, but the initial and final configurations of every precovering graph.
Since we transition through a sequence of precovering graphs, we also speak of entry and exit rather than initial and final configurations. 
We say that a~$\countersp$-run $\apath\in\pathsof{\countersp}{\admgts}$ is $\sqsubseteq$-intermediate accepting, if for every precovering graph $\precovering$ in~$\admgts$ that is traversed by the infix $\apath_{\precovering}$ of $\apath$, we have $\zerovec\leq\coordacc{\apath_{\precovering}}{\firstindex}\sqsubseteq(\rootstate, \incounters)$ and $\zerovec\leq\coordacc{\apath_{\precovering}}{\lastindex}\sqsubseteq(\rootstate, \outcounters)$. 
Here, $(\rootstate, \incounters)$ and $(\rootstate, \outcounters)$ are the entry and exit configurations of $\precovering$. 
%
We write 
\begin{align*}
\interacceptof{\countersp}{\sqsubseteq}{\admgts}=\setcond{\apath\in\pathsof{\countersp}{\admgts}}{\apath\text{ is $\sqsubseteq$-intermediate accepting}}\ .
\end{align*}
Note that for $\mathord{\sqsubseteq}=\mathord{\omegaleq}$, we have the classical notion of (intermediate-)acceptance for MGTS:
for the infix $\apath_{\precovering}$ that traverses $\precovering$, we have $\zerovec\leq\coordacc{\apath_{\precovering}}{\firstindex}\omegaleq(\rootstate, \incounters)$, and $\zerovec\leq\coordacc{\apath_{\precovering}}{\lastindex}\omegaleq(\rootstate, \outcounters)$.
%
\subsection{Characteristic Equations}\label{Subsection:CharacteristicEq}
The characteristic equations for MGTS are
\begin{align*}
    \chareqof{\precovering} &\;\; =\;\; \basecharof{\precovering}\wedge \initeqof{\precovering}{\omegaleq}\\
    \chareqof{\precovering.\anupdate.\admgts} &\;\; =\;\; \chareqof{\precovering}\wedge 
    \chareqof{\admgts}\\
    &\hspace{0.7cm} \wedge \coordacc{\avar}{\coordacc{\admgts}{\firstindex}, \genter}-\coordacc{\avar}{\precovering, \gexit}=\actionbalof{\anupdate}\ .
\end{align*}
For each precovering graph $\precovering$ and each counter $i$ in the MGTS, we introduce the variables $\coordacc{\avar}{\precovering, \genter, i}$ and $\coordacc{\avar}{\precovering, \gexit, i}$. 
The idea is that the vectors  $\coordacc{\avar}{\precovering, \genter}=(\coordacc{\avar}{\precovering, \genter, 1}, \ldots , \coordacc{\avar}{\precovering, \genter, \sizeof{\counters}})$ and $\coordacc{\avar}{\precovering, \gexit}$  describe the counter valuations upon entering respectively exiting the precovering graph.
The last system of equations says that the counter valuation when entering the first
 precovering graph in $\admgts$ is the valuation when leaving $\precovering$ modified by the counter update $\actionbalof{\anupdate}$. 
We also have a variable $\coordacc{\avar}{\anedge}$ for every edge in a precovering graph, which counts how often the edge is taken in a run. 
For an MGTS $\admgts$, and the discussed set of variables $\varsof{\admgts}$, we say that $\asol\in\Z^{\varsof{\admgts}}$ is a \emph{solution} to $\chareqof{\admgts}$, if the assignment given by $\asol$ satisfies each constraint.  
We call a system that has at least one solution \emph{feasible}.

The system of equations $\basecharof{\precovering}$ captures the runs through the precovering graph $\precovering$. 
It consists of the Kirchhoff equations, the marking equations, and equations that require non-negative values for the edge variables. 
The Kirchhoff equations express the fact that a run must enter and exit every state the same number of times. 
Since precovering graphs are strongly connected, this means the edge vector can be turned into a path provided every single edge is taken at least once. 
Perfectness will make sure this is the case. 
The marking equations say that the counter valuation after the run is the initial counter valuation plus the updates performed by the edges. 
The reader may note that we would only have to track the values for counters in $\infinitiesof{\precovering}$, but this would clutter the presentation. 
Let $\nodes=\precovering.\nodes$, $\edges = \precovering.\edges$, and $\edgesto{\anode}$ and $\edgesfrom{\anode}$ denote the edges leading to and originating from state~$\anode$. We have 
\begin{align*}
    \intertext{$\basecharof{\precovering}$:}
    \sum_{\anedge\in\edgesto{\anode}}\coordacc{\avar}{\anedge}-\sum_{\anedge\in\edgesfrom{\anode}}\coordacc{\avar}{\anedge}&=0\quad \text{for all } \anode\in\nodes\\
    \coordacc{\avar}{\precovering, \genter} +\actionbalof{\coordacc{\avar}{\edges}}-\coordacc{\avar}{\precovering, \gexit}&=\zerovec\\
    \coordacc{\avar}{\edges}&\geq 0\ . 
\end{align*}

The system $\initeqof{\precovering}{\sqsubseteq}$ formulates $\sqsubseteq$-intermediate acceptance, for some preorder $\mathord{\sqsubseteq}\subseteq\intsomega\times\intsomega$. 
Since we define intermediate acceptance relative to a preorder $\sqsubseteq$, this must also be the case for the corresponding equation system.
The system says that the initial counter valuation held by $\coordacc{\avar}{\precovering, \genter}$ should be smaller than the initial counter valuation $\precovering.\incounters$ wrt. $\sqsubseteq$, and the final counter valuation $\coordacc{\avar}{\precovering, \gexit}$ should be smaller than $\precovering.\outcounters$. 
Moreover, both valuations should be non-negative.
as we alluded to in \Cref{Subsection:VASS}, we extend the preorder $\sqsubseteq$ to compare vectors component-wise.
We define 
\begin{align*}
    \intertext{$\initeqof{\precovering}{\sqsubseteq}$:}
    \zerovec\leq\coordacc{\avar}{\precovering, \genter}&\sqsubseteq\precovering.\incounters\\
    \zerovec\leq\coordacc{\avar}{\precovering, \gexit}&\sqsubseteq\precovering.\outcounters\ .
\end{align*}
Note that the characteristic equations $\chareqof{\admgts}$ for the MGTS $\admgts$ capture the classical notion of (intermediate-)acceptance, and thus use $\initeqof{\precovering}{\omegaleq}$.

As explained in the overview given above, to judge whether a variable is unbounded in the solution space of $\chareqof{\admgts}$, we need a homogeneous variant of the characteristic equations. 
Since most equations are homogeneous already, all we have to do is replace by zero the concrete values in the equations for the updates between precovering graphs and in $\initeqof{\precovering}{\sqsubseteq}$.
In the former case, the result of the replacement is $\coordacc{\avar}{\coordacc{\admgts}{\firstindex}, \genter}-\coordacc{\avar}{\precovering, \gexit}=\zerovec$.
In the latter case, we define $\zerovec_{\mathsf{in}}\in\set{0, \omega}^{\counters}$ by $\coordacc{\zerovec_{\mathsf{in}}}{i}=\omega$ if and only if $\coordacc{\incounters}{i}=\omega$ for all $i\in\counters$, and similar for $\zerovec_{\mathsf{out}}$. 
With this, we let
\begin{align*}
    \intertext{$\homeqof{\precovering}{\sqsubseteq}$:}
    \zerovec\leq\coordacc{\avar}{\precovering, \genter}&\sqsubseteq\zerovec_{\mathsf{in}}\\
    \zerovec\leq\coordacc{\avar}{\precovering, \gexit}&\sqsubseteq\zerovec_{\mathsf{out}}\ .
\end{align*}

A support solution of $\chareqof{\admgts}$ is a vector $\asol\in\ints^{\varsof{\admgts}}$ that satisfies the homogeneous  characteristic equations. 
The support $\supportof{\chareqof{\admgts}}$ consists of all (counter and edge) variables $\coordacc{\avar}{c}$ for which there is a support solution $\asol$ with $\coordacc{\asol}{c}\geq 1$. 
We call $\asol$ a maximal support solution, if it gives a positive value to all variables in the support, $\coordacc{\asol}{c}\geq 1$ for all $\coordacc{\avar}{c}\in\supportof{\chareqof{\admgts}}$. 
Since support solutions are stable under addition, we have the following result.

\begin{lem}\label{Lemma:MaximalSupport}\label{Lemma:FullSupport}
There always is a maximal support solution of $\chareqof{\admgts}$.
\end{lem}
\subsection{Perfectness and Reachability}
When it comes to reachability, an $\omegaleq$-intermediate accepting $\nat$-run in an MGTS $\admgts$ immediately yields a solution to the characteristic equations $\chareqof{\admgts}$. 
Lambert's important insight is that also the reverse holds~\cite{Lambert92}: a solution to the characteristic equations yields an $\omegaleq$-intermediate accepting $\nat$-run.  
What is remarkable is that the characteristic equations cannot guarantee non-negativity of the valuations attained within precovering graphs. 
Instead, Lambert achieves non-negativity by pumping covering sequences.
His result needs the hypothesis that covering sequences exist and, moreover, the characteristic equations admit the pumping. 
This is captured by the notion of perfectness. 
The MGTS~$\admgts$ is \emph{perfect}, if for every precovering graph $\precovering$ in $\admgts$,
\begin{itemize}
\item[(i)] $\coveringseqof{}{\precovering}\neq\emptyset\neq \downcoveringseqof{}{\precovering}$, and 
\item[(ii)] $\supportof{\chareqof\admgts}$ justifies the unboundedness in~$\precovering$. 
\end{itemize} 
It remains to define what it means to justify the unboundedness.
We make the definition slightly more general so that we can reuse it later. 
Let $\precovering$ have counters $\counters$, edges~$\edges$, initial valuation $\incounters$, and final valuation $\outcounters$. 
Let $\countersp\subseteq\counters$ be a subset of the counters. 
We say that the set $\supportof{\chareqof\admgts}$ \emph{justifies the unboundedness of $\countersp$ in $\precovering$}, if 
\begin{align*}
\coordacc{\avar}{\precovering, \genter, j}&\in \supportof{\chareqof\admgts}\text{ for all $j\in\countersp$ with $\coordacc{\incounters}{j}=\omega$}\\
\coordacc{\avar}{\precovering, \gexit, j}&\in \supportof{\chareqof\admgts}\text{ for all $j\in\countersp$ with $\coordacc{\outcounters}{j}=\omega$}\\
\coordacc{\avar}{\anedge}&\in \supportof{\chareqof\admgts}\text{ for all $\anedge\in\edges$}\ .
\end{align*}
If $\countersp=\counters$, we say \emph{$\supportof{\chareqof\admgts}$ justifies the unboundedness in~$\precovering$}. 

Perfectness of $\admgts$ is sufficient to construct an $\nat$-run from a solution to the characteristic equations: 
\begin{align*}
\interacceptof{\nat}{\omegaleq}{\admgts}\neq\emptyset\qquad&\text{iff}\qquad\interacceptof{\ints}{\omegaleq}{\admgts}\neq\emptyset\\
&\text{iff}\qquad\chareqof{\admgts}\text{ is feasible}\ . 
\end{align*}
The implication from right to left is Lambert's famous iteration lemma~\cite[Lemma 4.1]{Lambert92}. 
As the key arguments will reappear in our solution to the regular separability problem, we explain them before stating the result. 
For simplicity, assume there is only one precovering graph $\precovering$ whose initial and final counter valuations are concrete, $\incounters, \outcounters\in\nat^{\counters}$.  
Let $\asolsup$ be a maximal support solution that exists by Lemma~\ref{Lemma:FullSupport}. 
Let $\asolfeas$ be a solution to the characteristic equations that exists by feasibility.
Then $\asolsup+\asolfeas=\asol$ solves the characteristic equations and satisfies $\coordacc{\asol}{\anedge}\geq 1$ for every edge.  
As the solution contains every edge, we can turn it into a path $\apath$ with $\parikhof{\apath}=\asol+\asol_{up}$. 
Here, $\asol_{up}$ is for the edges between the precovering graphs.
The path may still fail to be an $\nat$-run, because the $\omega$-decorated counters may become negative. 
By perfectness, however, there is a covering sequence $\upcovering\in\coveringseqof{}{\precovering}$ that produces a positive value on all $\omega$-decorated counters (recall that the initial valuation is concrete). 
The idea is to repeat $\upcovering$ to enable $\apath$. 

Unfortunately, we cannot repeat $\upcovering$ in isolation, otherwise we may end up with a run that no longer solves reachability.
The way out is to work with repetitions of the support solution. 
We also have to involve a sequence $\downcovering\in \downcoveringseqof{}{\precovering}$ for reasons that will become clear in a moment. 
Select $m\in\nat$ so that  
\begin{align}
m\cdot\coordacc{\asolsup}{\edges}-\parikhof{\upcovering}-\parikhof{\downcovering}\geq \onevec\ .\label{Equation:Parikh}
\end{align} 
The condition says that $m$ copies of the support solution contain enough transitions to fit in $\upcovering$, $\downcovering$, and another cycle $\aword$. 
We can form $\aword$ because we still have every edge at least once. 
The idea is to embed $\apath$ into a repetition $\upcovering^k.\apath.\aword^k.\downcovering^k$. 
We first have a sequence that increases the counter values and in the end a sequence that decreases them. 
Since $\upcovering$ and~$\downcovering$ are $\nat$-runs, we know they are executable once we have their initial valuations. 

Unfortunately, we do not even know that $\upcovering.\aword.\downcovering$ forms an $\nat$-run. 
The word $\aword$ may have a negative effect on the $\omega$-decorated counters.
This is where $\downcovering$ comes in. 
We know that $\upcovering.\aword.\downcovering$ has a zero effect on the $\omega$-decorated counters by the shape of the homogeneous characteristic equations. 
Moreover, we know that $\downcovering$ has a strictly negative effect on these counters by the definition of $\downcoveringseqof{}{\precovering}$. 
Then $\upcovering.\aword$ must have a strictly positive effect on the $\omega$-decorated counters.
This means there is $k\in\nat$ so that $\upcovering^k.\aword^k.\downcovering^k$ is an $\nat$-run.
We can choose $k$ large enough for $\upcovering^k.\apath.\aword^k.\downcovering^k$ to form an $\nat$-run.

It is also helpful to consider the case where $\coordacc{\outcounters}{i}=\omega\neq\coordacc{\incounters}{i}$, meaning the precovering graph has to provide arbitrarily large values for counter $i$.  
By perfectness, we know that the variable $\coordacc{\avar}{\precovering, \gexit, i}$ is in the support.
In the above discussion, this means $\upcovering.\aword.\downcovering$ will have a strictly positive effect on this counter. 

To lift the argumentation from precovering graphs 
to composed MGTS, we have to deal with $\omega$-entries in the initial valuation of a precovering graph.
An $\omega$-entry means the covering sequence may have a negative impact on the counter. 
To be able to execute the sequence, we let the precovering graphs which are placed earlier in the MGTS produce a high enough value on the counter as follows. 
By perfectness, the variable for the $\omega$-decorated counter is in the support. 
This means we can scale the support solution $\asolsup$ by a factor $m\in\nat$ that not only achieves Condition~\eqref{Equation:Parikh} from above, but also satisfies the following: 
for all precovering graphs $\precovering$ in~$\admgts$ with $\omega$-decorated counter $i$, $\upcovering\in\coveringseqof{}{\precovering}$ and $\downcovering\in\downcoveringseqof{}{\precovering}$, we have
\begin{align}
\coordacc{m\cdot\asolsup}{\precovering, \genter, i}+\coordacc{\actionbalof{\upcovering}}{i}&\geq 1\label{Equation:Enable}\\
\coordacc{m\cdot \asolsup}{\precovering, \gexit, i}-\coordacc{\actionbalof{\downcovering}}{i}&\geq 1\ .\notag
\end{align}

The following is a slightly strengthened variant of Lambert's iteration lemma that makes explicit the universal quantification over the cycles that can be iterated.
This gives us freedom for our construction in Section~\ref{Section:SeparabilityResult}.
Despite the stronger formulation, the correctness of the lemma still follows from the proof in \cite{Lambert92}.
\begin{lem}[Lambert's Iteration Lemma, Lemma 4.1 in \cite{Lambert92}]\label{Lemma:ModifiedIteration} 
Let $\admgts$ be a perfect MGTS. 
For every $\precovering_i$ in~$\admgts$, let 
 $\upcovering_{i}\in\coveringseqof{}{\precovering_i}$ and $\downcovering_{i}\in\downcoveringseqof{}{\precovering_i}$. 
Let $\asolfeas$ solve $\chareqof{\admgts}$. 
We can compute 
\begin{itemize}
\item a (maximal) support solution $\asolsup$ satisfying~\eqref{Equation:Parikh} and \eqref{Equation:Enable} for every $\precovering_i$, 
\item for every $\precovering_i$, cycles $\apath_{i}$ and $\awordn{i}$ originating in the root, 
\item so that $\coordacc{\asolsup}{\edges_i}=\parikhof{\upcovering_{i}}+\parikhof{\awordn{i}}+\parikhof{\downcovering_{i}}$, 
\item and $\coordacc{\asolfeas}{\edges}= \sum_{\precovering_i\in\admgts}\parikhof{\apath_{i}}$.
\end{itemize}
Moreover, for every $\asolfeas$, $\asolsup$, and $(\upcovering_{i}$, $\apath_{i}$, $\awordn{i}$, $\downcovering_{i})_{\precovering_i\in\admgts}$ that satisfy the above conditions, there is $k_0\in\nat$ so that for all $k\in\nat$ with $k_0\leq k$
    \begin{align*}
    \aconf.\upcovering_{0}^{k}\apath_{0}\awordn{0}^k\downcovering_{0}^{k}\anupdate_{1}\ldots\anupdate_{\lastindex}
    \upcovering_{\lastindex}^{k}\apath_{\lastindex}\awordn{\lastindex}^k\downcovering_{\lastindex}^{k}&\in\interacceptof{\nat}{\omegaleq}{\admgts}\ .
    \end{align*}
\end{lem}
\newcommand{\refine}{<}
\newcommand{\refineruns}{\preceq_{\mathsf{runs}}}
\newcommand{\refineprecision}{\preceq}
\subsection{Decomposition}
%
Recall our discussion at the end of \Cref{Subsection:KLMSTOverview}.
We mentioned that when perfectness fails, then we find a bound that constrains all runs in $\interacceptof{\N}{\omegaleq}{\admgts}$, and implement this bound in the syntax.
Now, we make this precise.
We go through the cases, where perfectness of the MGTS $\admgts$ may be violated.
Let $\admgts$ and let $\precovering$ be the precovering graph in $\admgts$, for which one of the perfectness conditions (i) or (ii) fails.

The first violation to perfectness is $\coveringseqof{}{\precovering}=\emptyset$ or $\downcoveringseqof{}{\precovering}=\emptyset$.
These cases are symmetrical, so only consider $\coveringseqof{}{\precovering}=\emptyset$.
This case implies a bound $B\in\N$, such that all $\N$-runs in $\precovering$ starting from $\precovering.\inmarking$ have at least one counter, where the valuation stays in~$[0, B]$.
Here, KLMST enriches the states $\precovering$ with a bounded counter that tracks the value of some counter $i$.
By this, we mean that for each counter $i\in\counters$,  KLMST constructs a product $\precovering_{i, B}=\precovering\times\annfa_{i, B}$ where $\annfa_{i, B}$ is an NFA that keeps track of counter $i$ up to bound~$B$.
Note that we have not defined such a product operation, so this is only meant to be taken intuitively.
A precise construction would need to take care of some technical details.
It returns the set obtained by replacing $\precovering$ with  ~$\precovering_{i, B}$ in $\admgts$ for all $i\in\counters\setminus\infinitiesof{\precovering.\inmarking}$, and reestablishing the MGTS structure.

Second violation happens, when $\supportof{\chareqof\admgts}$ does not justify the unboundedness in~$\precovering$.
This breaks down into two subcases.
In the first subcase, a counter $i\in\counters$ is marked $\precovering.\inoutmarking[i]=\omega$ but the corresponding variable $\avar[\precovering, \inout, i]$ is bounded in the solution space, where $\inout\in\set{\genter, \gexit}$.
As an implication, this variable may only take finitely many values~$s_{0},\ldots, s_{k}\in\N$.
Then, KLMST returns versions of $\admgts$ where the bounded marking is marked with $s_{\ell}$ instead of $\omega$ for all $\ell\leq k$.
In the final subcase, a variable $\avar[\anedge]$ corresponding to an edge $\anedge\in\precovering.\edges$ may be bounded in the solution space of~$\chareqof{\admgts}$.
This implies a bound $B$ for the number of times a set of edges $\edges'\subseteq\precovering.\edges$ may be taken.
Here, KLMST employs a similar construction to the first violation, and constructs $\precovering_{\edges', B}=\precovering\times\annfa_{\edges', B}$ for the NFA~$\annfa_{\edges', B}$ that counts the number of taken $\anedge\in\edges'$, and only allows $B$ applications.
It inserts the results back into $\admgts$ and returns, as it did for the first violation.
As desired, all of the new restrictions are trivial for the set~$\interacceptof{\N}{\omegaleq}{\admgts}$, but they are meaningful for the set of runs $\pathsof{\Z}{\admgts}$ allowed in the syntax.
These refinement steps are effective as well, and each case can be computed with elementary resources.

To achieve the $\fof{\omega}$ upper bound, we cannot work with the well-founded relation from~\cite{Lambert92}, but rely on recent ideas from~\cite{Leroux19}.
We assign each MGTS with $d$ counters a rank in $\nat^{d+1}$, and define the well-founded relation $\rankless$ to compare the ranks lexicographically.
The rank of an MGTS $\admgts$ is defined inductively.
For a precovering graph, we define $\rankingof{\precovering}$ to be a vector with a single non-zero entry, and this entry holds information about the size of $\precovering$. 
The entry itself is related to a vector space $\cyclespaceof{\precovering}$ that is associated with the precovering graph. 
%
%
This is the space spanned by the cycle effects:
\begin{align*}
\cyclespaceof{\precovering}\ =\ \spanof{\setcond{\actionbalof{\apath}}{\apath=(\anode, \aconf)\ldots(\anode, \aconf')\in\pathsof{\ints}{\precovering}}}\ .
\end{align*}
Intuitively, $\cyclespaceof{\precovering}$ describes an envelop for the cycle effects allowed in the syntax $\pathsof{\Z}{\precovering}$.
The envelop offers a natural notion for measuring its complexity, the dimension.
This underlies the ranking function.
Assume $\cyclespaceof{\precovering}$ is $i$-dimensional. 
We define $\coordacc{\rankingof{\precovering}}{d-i}=\sizeof{\edgesof{\precovering}}+\sizeof{\infinitiesof{\inmarkingof{\precovering}}}+\sizeof{\infinitiesof{\precovering}}+\sizeof{\infinitiesof{\outmarkingof{\precovering}}}$ and $\coordacc{\rankingof{\precovering}}{j}=0$ for $j\neq d-i$.
The inductive case is $\rankingof{\admgts_{1}.\anupdate.\admgts_{2}}=\rankingof{\admgts_{1}}+\rankingof{\admgts_{2}}$. 

The definition allows us to unwind a precovering graph into an MGTS with a number of precovering graphs. 
If the cycle spaces of the new precovering graphs have a smaller dimension, then this makes their non-zero entry in the rank move to the right. 
As a consequence, the well-founded order decreases, even though the new MGTS may have more edges or $\omega$-entries in total.

Our well-founded relation slightly differs from the one in~\cite{Leroux19}, namely, we also include the term $\sizeof{\infinitiesof{\inmarkingof{\precovering}}}+\sizeof{\infinitiesof{\precovering}}+\sizeof{\infinitiesof{\outmarkingof{\precovering}}}$.
The decomposition algorithm in \cite{Leroux19} relies on a clean-up phase to deal with unjustified $\omega$-entries. 
By additing the above term, we can embed the clean-up phase into the decomposition.
Indeed, the term guarantees that the well-founded relation decreases when we remove $\omega$-entries.

%% file: overview.tex
\section{Solving Regular Separability}\label{Section:Overview}

In this section, we discuss the ideas behind our solution.
We recall \Cref{Section:Introduction} and \Cref{Section:RegSep}.
In \Cref{Section:RegSep}, we fix one language to the Dyck Language $\dycklangn{n}$, and make our goal deciding the separability $\langof{\avas}\separable\dycklangn{n}$ for an input VASS $\avas$.
Similarly to KLMST, instead of VASS, we start from an MGTS $\avas$ with concrete input and output markings.
Our approach reduces this separability query to a set of separability queries for $\Z$-VASS.
As discussed in \Cref{Section:Introduction}, this means that we want to 
\begin{itemize}
    \item[(i)] Use $\Z$-separability to construct $\N$-separators,
    \item[(ii)] Deduce $\N$-inseparability from $\Z$-inseparability. 
\end{itemize}

Our algorithm is based on four insights that realize (i) and (ii).
The first three insights allow us to apply the KLMST decomposition to the separability problem.
By this, we mean that we use similar arguments to refine the separability instance at hand.
The insights also lead to our proof object, which is structured to allow for (i).
Our decomposition also establishes a notion of perfectness, which we assume towards (ii).
The fourth insight then allows us to apply the iteration lemma and construct $\N$-inseparability witnesses from $\Z$-inseparability witnesses.
This handles (ii).
We proceed with the first insight.

\begin{insight}\label{Insight:HalfSeparators}
    Over-approximations $\aseparator\subseteq\analph_{n}^{*}$ with $\langof{\avas}\subseteq\aseparator$ simplify the problem from $\langof{\avas}\separable\dycklangn{n}$ to $\langof{\avas}\separable(\dycklangn{n}\cap\aseparator)$. 
    Indeed, any separator $\aseparatorp\subseteq\analph_{n}^{*}$ for the latter implies the separator $\aseparatorp\cap\aseparator$ for the former.
\end{insight}

In line with \Cref{Insight:HalfSeparators}, our algorithm also tracks an NFA $\annfa$ with $\langof{\avas}\subseteq\langof{\annfa}$, and refines it.
This also refines the fragment $\dycklangn{n}\cap\langof{\annfa}$ of $\dycklangn{n}$ that is relevant to $\langof{\avas}\separable\dycklangn{n}$.
A canonical choice for such an overapproximation $\annfa$ is the control flow of the MGTS at hand.
Second insight starts from this choice.
Let $\annfa$ be the NFA that represents the control flow of the MGTS $\avas$.
Consider the Dyck-visible MGTS $\avas_{D}$ with $0$ as input- and output-markings, the control flow $\annfa$, and remaining markings $\omega$.
It is easy to see that this MGTS accepts the relevant fragment $\langof{\avas_{D}}=\dycklangn{n}\cap\langof{\annfa}$ of $\dycklangn{n}$, and has the same control flow as $\avas$.
Thus, we can express our problem as $\langof{\avas}\separable\langof{\avas_{D}}$ where $\avas$ and $\avas_{D}$ share the same control flow.
The second insight gives us hope of maintaining this structure.

\begin{insight}\label{Insight:KLMSTControlFlow}
    In almost all cases, the KLMST decomposition unrolls the control flow.
    In such cases, we may apply a refinement for either of $\avas$ or $\avas_{D}$ to the common over-approximation~$\annfa$, making it visible to both sides.
\end{insight}

Inspired by \Cref{Insight:KLMSTControlFlow}, our goal becomes this.
Apply the refinements to the common finite-state over-approximation for both $\avas$ and $\avas_{D}$.
This would refine the languages, along with the over-approximation, which means a refinement of the separability instance.
With a fitting notion of perfectness, we could get the desired decomposition for the separability problem.
Unfortunately, not all refinements unroll the control flow.
Namely, the KLMST decomposition also introduces new intermediate reachability targets.
In our case, the problem is $\avas_{D}$ receiving such a refinement.
Reachability constraints typically cannot be captured by regular languages, and thus cannot be captured by over-approximations resp. separators.
We overcome this obstacle with our third insight.

Towards this insight, consider a step of the KLMST algorithm, where such an intermediate constraint is introduced.
Even if such constraints would not be expressable in general, KLMST only introduces them if a finite set $K\subseteq\N$ of values are reachable at this marking.
Let $\mu\in\N$ with $\mu>\max K$.
Assuming that the run reaches some $u\in K$, we can recognize whether a precise value $v\in K$ is reached by confirming $u\equiv v \mod\mu$.
This is because $K$ is not large enough to contain two values that are equivalent modulo $\mu$.
The modulo check can be realized by a regular language.
This is our third insight.

\begin{insight}\label{Insight:Faithfulness}
    The intermediate reachability constraints introduced by KLMST decomposition can be realized by counting modulo a large enough $\mu\in\N$.
\end{insight}

We refine this insight with our invariant, faithfulness, which allows us to encode newly introduced intermediate reachability checks using modulo counting.
Intuitively, an MGTS is faithful, if the intermediate reachability constraints have been introduced when only a finite set of values is possible. 
That is, intermediate constraints are faithful to the original reachability constraints.
We formulate faithfulness over our proof object, but the statement below also follows from our development.

\begin{fact}
    Let $\avasp$ be a Dyck-visible MGTS with the only concrete markings $\avasp.\inmarking, \avasp.\outmarking=0$, and let $\avasp$ be a result of the KLMST decomposition for $\avasp$.
    Then, $(\avasp, \mu)$ is faithful for some base $\mu\in\N$: we have $\langof{\avasp}=\dycklangn{n}\cap\langof{\annfa_{\mu}}$, where $\annfa_{\mu}$ is the NFA that follows the control flow of $\avasp$, and checks that markings are reached modulo $\mu$. 
\end{fact}

We combine our first three insights into our proof object, doubly marked graph-transition sequences (DMGTS).
A DMGTS $\admgts=(\openmgts, \mu)$ is an MGTS $\openmgts$ equipped with a natural modulo base $\mu\geq 1$.
This structure captures the MGTS $\avas$ and $\avas_{D}$ above, along with the common over-approximation $\annfa_{\mu}$.
The MGTS $\openmgts$ has the set of counters $\leftside\uplus\rightside$, where $\leftside$ are the counters of the subject language, and $\rightside$ the Dyck-counters.
As the DMGTS $\admgts$ captures the behavior of two MGTS, it has two notions of run-acceptance, intermediate $\leftside$- and $\rightside$-acceptance.
It also defines two languages $\leftlangof{\admgts}$ resp.\ $\rightlangof{\admgts}$ where for $\aside\in\set{\leftside,\rightside}$, the language~$\sidelangof{\aside}{\admgts}$ is the set of words that label intermediate $\aside$-accepting runs in $\admgts$.
The DMGTS encodes the separability query $\leftlangof{\admgts}\separable\dycklangn{n}$ of the subject language $\leftlangof{\admgts}$.
We discuss intermediate $\leftside$-acceptance and thus~$\leftlangof{\admgts}$ in a second.
The over-approximation $\annfa_{\mu}$ is defined as in \Cref{Insight:Faithfulness}.
Namely, it is an NFA requiring that the $\rightside$ markings are reached modulo $\mu$, and follows the control flow of~$\admgts$.
The relevant fragment of $\dycklangn{n}$ is thus $\dycklangn{n}\cap\langof{\annfa_{\mu}}$.
The language $\rightlangof{\admgts}$ is meant to capture this.
A run is intermediate $\rightside$-accepting, if it is an $\N$-run, and reaches the $\rightside$-markings.
Since this allows starting from arbitrarily high values at $\leftside$ counters, positivity in these counters is trivial.
Given faithfulness, which we keep as an invariant, we get $\dycklangn{n}\cap\langof{\annfa_{\mu}}=\rightlangof{\admgts}$.
Note that formally, we define faithfulness only for DMGTS, where the modulo basis $\mu$ is explicit in the syntax.
To complete the picture, we need $\leftlangof{\admgts}\subseteq\langof{\annfa_{\mu}}$.
This is the idea behind the definition of $\leftside$-acceptance.
Namely, a run is intermediate $\leftside$-accepting, if it is an $\N$-run, it reaches the $\leftside$ markings precisely, and $\rightside$ markings modulo $\mu$.
We also define a notion of perfect DMGTS, which we need for (ii).

Our algorithm starts from a DMGTS encoding the separability problem, uses KLMST decomposition on one of $\leftlangof{\admgts}$ or $\rightlangof{\admgts}$ (up to a caveat), and refines accordingly.
As KLMST maintains faithfulness, we naturally capture the refinements in the over-approximation.
Indeed, the DMGTS definition is strong enough to realize (i).
Note that if $\rightlangof{\admgts}$ becomes empty, we can conclude $\leftlangof{\admgts}\separable\dycklangn{n}$ using $\annfa_{\mu}$ as the separator. 
Similarly, even if the $\Z$-relaxations of the languages $\leftlangof{\admgts}$ and $\rightlangof{\admgts}$ become separable, we can enrich $\annfa_{\mu}$ by the $\Z$-separator to conclude $\leftlangof{\admgts}\separable\dycklangn{n}$.
Thus, it only remains to deal with the case where the $\Z$-relaxations are inseparable.

We do not handle the inseparability case, as long as the DMGTS $\admgts$ is not perfect.
If the DMGTS $\admgts$ is perfect, then the $\leftside$- resp. $\rightside$-sides are both perfect in the classical sense.
The implication does not go in the backwards direction, which we discuss shortly.
To get $\N$-inseparability from $\Z$-inseparability, we use the iteration lemma, and construct pairs of words from $\leftlangof{\admgts}$ resp. $\rightlangof{\admgts}$ that require more and more states to distinguish.
We formalize this with the notion of $\sim_{\annfa}$-equivalence \cite{Buechi62}.
We say that two words $\aword, \awordp\in\analph_{n}^{*}$ are $\sim_{\annfa}$-equivalent for a DFA $\annfa$, if for all states in $\annfa$, reading $\aword$ and $\awordp$ leads to the same state.
Note that this implies of $\aword$ and $\awordp$ in the syntactic monoid of $\annfa$.
To state our argument more precisely, given a DFA $\annfa$, we construct $\aword_{\annfa}\in\leftlangof{\admgts}$ and $\awordp_{\annfa}\in\rightlangof{\admgts}$ with $\aword_{\annfa}\sim_{\annfa}\awordp_{\annfa}$.
This shows inseparability, as any separator candidate $\annfa$ accepts resp. rejects both words.

Recall the iteration lemma from the previous section.
A perfect MGTS allows a kind of pumping, where in each precovering graph, an intermediate accepting $\Z$-run $\apath_{i}$ is combined with arbitrary iterations of (a) up- resp. down-pumping sequences $u_{i}$ and $v_{i}$, and (b) a leftover-sequence $w_{i}$ depending on these sequences, to get an intermediate accepting $\N$-run.
To ensure $\sim_{\annfa}$-equivalence of the whole word, we make sure that for each kind of ingredient, the $\leftside$ and $\rightside$ versions are~$\sim_{\annfa}$-equivalent.
The equivalence perspective allows us to soundly treat each ingredient in isolation.
As the shape of the pumping is the same, equivalence of the whole word follows.
We start from $\sim_{\annfa}$-equivalent $\Z$-runs, given by $\Z$-inseparability.
To ensure the equivalence of the pumping derivations, our perfectness prescribes a condition stronger than the perfectness of both sides.
Namely, it requires pumping derivations that pump in both MGTS.
By this, we mean that instead of having a pair of pumping sequences $u^{\aside}_{i}$, $v^{\aside}_{i}$ that pump the $\aside$-counters for each precovering graph~$\precovering_{i}$ in~$\admgts$ and for each $\aside\in\set{\leftside, \rightside}$; we require that there is a pair of pumping sequences $u_{i}$, $v_{i}$ that pump all counters $\leftside\uplus\rightside$ for each precovering graph $\precovering_{i}$ in~$\admgts$.
To achieve the stronger condition, the decomposition makes an exception and considers~$\admgts$ as a product MGTS.
The only unaccounted ingredient is the leftover-sequence~$w_{i}$.
We handle it with our last insight.

\begin{insight}
    In (the proof of) Lambert's iteration lemma, there are very few restrictions for the leftover-sequence.
\end{insight}

Namely, only constraint is on the Parikh image of the edges.
Furthermore, we do not even need to realize a particular Parikh image, we may also realize a multiple of the target.
This allows us enough space to construct equivalent leftover-sequences for both sides.
This realizes (ii), and concludes our argument.

%% file: dmgts.tex
\section{DMGTS}\label{Section:DMGTS}
We introduce \emph{doubly-marked graph transition sequences (DMGTS)} as the data structure behind our decision procedure for regular separability. 
Recall that the goal is to separate the language of a subject VASS 
from the Dyck language.
The idea of DMGTS is to simultanteously track both, the subject VASS and the Dyck VASS, like an MGTS for the intersection would.
The coupling, however, is not as tight as in the case of intersection.
Instead, a DMGTS still defines two languages, one for the subject VASS and one for the Dyck VASS. 
An DMGTS also maintains a common finite-state control, as we discussed in \Cref{Section:Overview}, which serves as the over-approximation for both languages. 
We enrich $\Z$-separators into~$\N$-separators using this over-approximation.
Our decision procedure decomposes an initial DMGTS to achieve a notion of perfectness, separates $\ints$-versions of the two languages, and lifts the resulting separator to the languages of interest.
Interestingly, the decomposition and perfectness do not treat the two components of a DMGTS as symmetric. 
The focus is on the subject VASS whose language a separator has to cover. 
The decomposition may refine the subject VASS in isolation, but it only works on the common finite-state structure when refining the Dyck VASS.
Thus, they agree on the finite-state structure, and the subject VASS maintains approximate information about the Dyck VASS.
With the new notion of faithfulness, this what allows us to lift a $\ints$-separator to an $\nat$-separator.
%
%

%
%
%
%

A DMGTS $\admgts=(\openmgts, \charmod)$ consists of an MGTS $\openmgts$ and a natural number $\charmod\geq 1$. 
The counters in $\openmgts$ form a disjoint union $\leftside\uplus\rightside$ between the counters $\leftside$ in the subject VASS and the counters $\rightside$ in the VASS accepting the Dyck language. 
We use $\aside$ to refer to the counters from either side, $\leftside$ or $\rightside$.  
The idea is this: when we project the runs in $\openmgts$ to $\leftside$, we obtain behavior of the subject VASS, and when we project the runs to $\rightside$, we obtain the effect that this behavior has on the Dyck language. 
To achieve this, we expect that the counters in $\rightside$ are updated \emph{Dyck-visibly}.
Recall that this means a letter $\incdyckn{i}$ leads to an increment of the $i$-th Dyck-counter, and $\decdyckn{i}$ leads to a decrement.
We lift the well-founded relation from MGTS to DMGTS and define $(\openmgts_{1}, \charmod_{1})\rankless(\openmgts_{2}, \charmod_{2})$ by $\openmgts_{1}\rankless\openmgts_{2}$.
The size of the DMGTS is $\sizeof{\admgts}=\sizeof{\openmgts}+\sizeof{\charmod}$, where $\sizeof{\charmod}$ is the length of the binary representation.  
We call~$\admgts$ \emph{zero-reaching}, if $\admgts.\coordacc{\incounters}{\rightside}=\zerovec=\admgts.\coordacc{\outcounters}{\rightside}$. 

During the decomposition, the 
The approximation is formalized through a new form of acceptance, acceptance modulo $\charmod$, and so the number~$\charmod$ in the definition of DMGTS is the precision of this approximation. 
The central new notion is faithfulness: it allows us to conclude ordinary acceptance from acceptance modulo $\charmod$, and will be key to lifting a~$\ints$-separator to an~$\nat$-separator in Section~\ref{Section:SeparabilityResult}.   
%
To formulate acceptance modulo $\charmod$, we define the \emph{modulo~$\charmod$ specialization order} $\omegalequiv{\charmod}\ \subseteq\intsomega\times\intsomega$ by $i\omegalequiv{\charmod} j$, if $j=\omega$ or $i\equiv j$ mod $\charmod$. 
We extend it to counter valuations and to configurations as we have done for the specialization order. 
Given a preorder $\sqsubseteq$ on configurations, we use $\restrictto{\sqsubseteq}{\aside}$ for the restriction of the preorder that only compares the counters in $\aside$, but does not constrain the remaining counters.

Since DMGTS are MGTS, the definitions of runs, acceptance, and intermediate acceptance carry over.
To account for the fact that a DMGTS is meant to represent two languages, one for the Dyck-side and one for the VASS-side, we add the following abbreviations: 
\begin{align*}
    \mathsf{(I)Acc}_{\rightside}(\admgts)\;\;&=\;\;\ \mathsf{(I)Acc}_{\rightside, \restrictto{\omegaleq}{\rightside}}(\admgts)\\ 
    \mathsf{(I)Acc}_{\ints, \rightside}(\admgts)\;\;&=\;\;\ \mathsf{(I)Acc}_{\ints, \restrictto{\omegaleq}{\rightside}}(\admgts)\\ 
    \interacceptofshort{\leftside}{\admgts}\;\;&=\;\;\ \interacceptof{\leftside}{\restrictto{\omegaleq}{\leftside}}{\admgts}\;\cap\; \interacceptof{\rightside}{\restrictto{\omegalequiv{\charmod}}{\rightside}}{\admgts}\ \\
    \interacceptof{\ints}{\leftside}{\admgts}\;\;&=\;\;\ \interacceptof{\ints}{\restrictto{\omegaleq}{\leftside}}{\admgts}\;\cap\; \interacceptof{\ints}{\restrictto{\omegalequiv{\charmod}}{\rightside}}{\admgts}\ .
\end{align*}
We explain these defintions.
Let $\apath$ be a run in $\admgts$.
The run $\apath$ is (intermediate) accepting for the $\rightside$-side, $\apath\in\mathsf{(I)Acc}_{\rightside}(\admgts)$, if the counters in the set $\rightside$ remain non-negative and on these counters the run is (intermediate) accepting in the normal sense, expressed by the relation $\restrictto{\omegaleq}{\rightside}$.
It is intermediate accepting on the $\leftside$-side, $\apath\in\interacceptofshort{\leftside}{\admgts}$, if the same holds for the counters in $\leftside$ and, moreover, the Dyck-side is accepting modulo~$\charmod$. 
Note the common finite-state over-approximation of both acceptances: following the control flow and reaching the $\rightside$-markings modulo $\mu$.
%
%
%
%
We also introduce $\ints$-relaxations of these notions, which we denote by $\mathsf{(I)Acc}_{\ints, \rightside}(\admgts), \interacceptof{\ints}{\leftside}{\admgts}$.  
With acceptance in place, we define the languages
\begin{align*}
    \sidelangof{\aside}{\admgts}\;\;&=\;\;\setcond{\edgelabelof{\apath}}
    {\apath\in\interacceptofshort{\aside}{\admgts}}\;\;\subseteq\;\; \analph_{n}^{*}\\
    \sidelangof{\ints, \aside}{\admgts}\;\;&=\;\;\setcond{\lambda_{\sepsymbol}(\apath)}
    {\apath\in\interacceptof{\ints}{\aside}{\admgts}} \;\; \subseteq\;\; \analph_{n, \sepsymbol}^{*}\ .
\end{align*}
By $\lambda_{\sepsymbol}$, we denote the function that extracts the edge labels except that it replaces the label $\lambda(\anupdate)$ of every update between precovering graphs by $(\lambda(\anupdate), \sepsymbol)$.  
This will allow us to uniquely identify the current precovering graph in a run, which comes handy in \Cref{Subsection:TheInseparability}.
As a consequence, the resulting language has the alphabet $\analph_{n, \sepsymbol}=\analph_{n}\times\set{\varepsilon, \sepsymbol}$.
We treat $(a, \varepsilon)\in\analph_{n, \sepsymbol}$ as $a$ in our development.
The following holds.
\begin{lem}\label{Lemma:EasyInclusion}
If $\admgts$ is zero-reaching, we have $\rightlangof{\admgts}\subseteq\dycklangn{\dyckdims}$. 
One can construct a $\ints$-VASS that accepts the language $\sidelangof{\ints, \aside}{\admgts}$.
\end{lem}
The former is a direct consequence of the definitions.
The latter holds, since $\sidelangof{\ints, \aside}{\admgts}$ requires reaching $\N$ resp.\ a precise $u\in\N$ at only a bounded number of locations along the run.
We can test these conditions using additional counters. 

We also define characteristic equations for each side that mimic the notions of acceptance we have just defined:
\begin{align*}
    \rightcharof{\precovering, \charmod} &\;\; =\;\; \basecharof{\precovering}\wedge \initeqof{\precovering}{\restrictto{\omegaleq}{\rightside}}\\
    \leftcharof{\precovering, \charmod} &\;\; =\;\; \basecharof{\precovering}\wedge \initeqof{\precovering}{\restrictto{\omegaleq}{\leftside}}\\
    &\hspace{2.2cm}
\wedge \initeqof{\precovering}{\restrictto{\omegalequiv{\charmod}}{\rightside}}
    \\
    \sidecharof{\aside}{\precovering.\anupdate.\admgts, \charmod} &\;\; =\;\; \sidecharof{\aside}{\precovering, \charmod}\wedge 
    \sidecharof{\aside}{\admgts, \charmod}\\
    &\hspace{0.2cm} \wedge \coordacc{\avar}{\coordacc{\admgts}{\firstindex}, \genter}-\coordacc{\avar}{\precovering, \gexit}=\actionbalof{\anupdate}\ . 
\end{align*}
We also define the support $\supportof{\sidecharof{\aside}{\admgts}}$. 
As in the case of reachability, we first define a homogeneous variant of the above system by replacing concrete values in $\initeqof{\precovering}{\sqsubseteq}$ and in the equations for the updates between precovering graphs with zero.
Then we collect the variables that receive a positive value in a solution to the homogeneous characteristic equations. 

The central new notion for regular separability is faithfulness.
%
%
A DMGTS~$\admgts$ is \emph{faithful}, if it is zero-reaching and
\begin{align}
\acceptof{\ints}{\rightside}{\admgts}\ \cap\ \interacceptof{\ints}{\restrictto{\omegalequiv{\charmod}}{\rightside}}{\admgts} \quad\subseteq\quad \interacceptof{\ints}{\rightside}{\admgts}\ .\label{Equation:Faithfulness}
\end{align}
This inclusion says the following.
In a faithful DMGTS $\admgts$, $\Z$-runs $\apath$ that (1) reach the initial and final targets for $\rightside$ precisely ($\apath\in\acceptof{\ints}{\rightside}{\admgts}$) and (2) reach the intermediate targets for $\rightside$ modulo $\mu$ ($\apath\in\interacceptof{\ints}{\restrictto{\omegalequiv{\charmod}}{\rightside}}{\admgts}$), reach these intermediate target precisely as well ($\apath\in \interacceptof{\ints}{\rightside}{\admgts}$).
In other words, the intermediate targets on the $\rightside$ counters can be relaxed to modulo checks, without adding new accepting runs.
Note that membership to~$\interacceptof{\ints}{\restrictto{\omegalequiv{\charmod}}{\rightside}}{\admgts}$ still expects non-negative values on entries and exits of precovering graphs.
%
%
%
The reverse inclusion is indeed readily checked, but will not be needed.

To foreshadow the use of faithfulness for proving separability in Section~\ref{Section:SeparabilityResult}, 
note that the left-hand side of Inclusion~\eqref{Equation:Faithfulness} mimics an intersection between the Dyck language, given by~$\acceptof{\ints}{\rightside}{\admgts}$, and a regular language, given by $\interacceptof{\ints}{\restrictto{\omegalequiv{\charmod}}{\rightside}}{\admgts}$. 
%
The inclusion ensures that the language $\rightlangof{\admgts}$ is not too small, that is $\rightlangof{\admgts}=\dycklangn{n}\cap\aseparator$ for some regular $\aseparator\subseteq\analph_{n}^{*}$, derived from $\interacceptof{\ints}{\restrictto{\omegalequiv{\charmod}}{\rightside}}{\admgts}$. 
Without this guarantee, $\leftlangof{\admgts}\separable\rightlangof{\admgts}$ (resp.~$\Z$-separability) may not imply $\leftlangof{\admgts}\separable\dycklangn{n}$.
To make this argument, it would actually be sufficient to have a version of faithfulness that keeps the Dyck counters non-negative, meaning we could have replaced $\ints$ by $\rightside$ in Inclusion~\eqref{Equation:Faithfulness}. 
%
%
%
The resulting notion of faithfulness, however, would be weaker in that the $\ints$-version of Inclusion~\eqref{Equation:Faithfulness} implies the $\rightside$-version, but not the other way around. 
As the overhead for proving the $\ints$-version is negligible, and as we believe it will come in handy in future developments, we opted for the stronger result.

A DMGTS $\admgts$ is \emph{perfect}, if it is faithful and for every precovering graph $\precovering$ in $\admgts$, for every $\aside\in\set{\leftside, \rightside}$, 
\begin{itemize}
    \item $\coveringseqof{}{\precovering}\neq\emptyset\neq \downcoveringseqof{}{\precovering}$ and
    \item $\supportof{\sidecharof{\aside}{\admgts}}$ justifies the unboundedness of $\aside$ in $\precovering$.
\end{itemize} 
Note that the edge variables are in the support of both, the VASS-side and the Dyck-side.

%% file: algorithm.tex
\section{Deciding Regular Separability}\label{Section:DecisionProcedure}
Our decision procedure for regular separability decomposes the DMGTS $\admgts$ of interest until the regular separability $\leftlangof{\admgts}\separable\dycklangn{\dyckdims}$ reduces to the regular separability of the $\ints$-VASS approximations $\leftsolutions(\admgts)\separable\rightsolutions(\admgts)$.
The latter problem can be solved with the algorithm from~\cite{IntVASS17}, as stated in Theorem~\ref{Theorem:IntsVASSRegSep} above.
Behind our decision procedure are two key results.
The first says that we can decompose faithful DMTGS into two finite sets of DMTGS. 
\begin{lem}\label{Lemma:DecoupledDecomposition}
    We can decompose a faithful DMTGS $\admgts$ in  $\fof{\sizeof{\leftside\cup\rightside}+4}$ into two finite sets  $\opensystems$ and $\decidedsystems$ of DMTGS, where all $\openmgts\in\opensystems$ are perfect, all $\decidedmgts\in\decidedsystems$ satisfy $\leftlangof{\decidedmgts}\separable\dycklangn{n}$, and 
    \begin{align*}
        \leftlangof{\admgts}\quad=\quad\leftlangof{\opensystems\cup\decidedsystems}\ .
    \end{align*}
\end{lem}
The second is a transfer result, saying that separability can be checked on the $\ints$-approximations as soon as we have perfectness. 
\begin{lem}\label{Lemma:PerfectSeparation}
    If $\admgts$ is faithful, then $\leftsolutions(\admgts)\separable\rightsolutions(\admgts)$ implies 
    $\leftlangof{\admgts}\separable\dycklangn{\dyckdims}$.
    If $\admgts$ is perfect, also the reverse holds. 
\end{lem}
%
%
%
%
Since perfect DMGTS are faithful, the following is a consequence.  
\begin{cor}\label{Corollary:PerfectSeparationDec}
    Let $\admgts$ be perfect. Then $\leftlangof{\admgts}\separable\dycklangn{\dyckdims}$ if and only if $\sidesolutions{\leftside}(\admgts)\separable\sidesolutions{\rightside}(\admgts)$. 
\end{cor}
These insights allow us to decide regular separability.
\begin{proof}[Proof of Proposition~\ref{proposition:DyckSeparationDec}]
To ease the notation, we assume the subject VASS is actually a VAS, $\avas=((\set{\anode}, \dyckalphn{\dyckdims}, \leftside, \edges), (\anode, \aconf_1), (\anode, \aconf_2))$.
%
It is well-known that any VASS can be turned into a VAS with the same language 
by introducing auxiliary counters for the states. 
One can also adapt our procedure to directly work with VASS.  
%
    %
We first check $\langof{\avas}\cap\dycklangn{\dyckdims}=\emptyset$.
If the intersection is non-empty, our decision procedure returns inseparable. 
    %
If it is empty, we construct an initial DMGTS $\admgts$ with $\langof{\avas}=\leftlangof{\admgts}$. 
Checking $\langof{\avas}\separable\dycklangn{\dyckdims}$ then amounts to checking $\leftlangof{\admgts}\separable\dycklangn{\dyckdims}$. 

We define $\admgts=(\precovering, \mu)$ with $\charmod=1$. 
The precovering graph $\precovering$ uses the underlying VASS
$\avasp = (\set{\groot}, \dyckalphn{\dyckdims}, \leftside\uplus \rightside, \edges')$.
It has a single state and both sets of counters.  
The set of edges $\edges'$ contains a loop $(\groot, a, (\avar, \avarp), \groot)$ for every $(\anode, a, \avar, \anode)\in\edges$.
The vector~$\avarp$ modifies the counters in $\rightside$ as required by Dyck visibility. 
The precovering graph is 
$\precovering = (\avasp, (\groot, (\aconf_1, \zerovec)), (\groot, (\aconf_2, \zerovec)),  \bal)$.
The root state is decorated by $\omega$, expressed as $\bal=\nat^{\emptyset}$.
The initial and final valuations expect the counters in $\leftside$ to behave like in $\avas$, and the counters in $\rightside$ to go from zero to zero. 
For $\leftlangof{\admgts}=\langof{\avas}$, note that the modulo $\charmod=1$ constraints that $\leftlangof{\admgts}$ imposes on the Dyck-side do not mean a restriction.
%

To decide $\leftlangof{\admgts}\separable\dycklangn{\dyckdims}$, we invoke Lemma~\ref{Lemma:DecoupledDecomposition}. 
The required faithfulness of $\admgts$ is trivial: the extremal valuations are zero on $\rightside$, and since there are no intermediate precovering graphs, we have $\interacceptof{\ints}{\rightside}{\admgts}=\acceptof{\ints}{\rightside}{\admgts}$. 
The lemma yields finite sets of DMGTS $\opensystems$ and $\decidedsystems$ with $\leftlangof{\admgts} =\leftlangof{\opensystems}\cup\leftlangof{\decidedsystems}$.  
It moreover guarantees $\leftlangof{\decidedmgts}\separable\dycklangn{\dyckdims}$ for all~$\decidedmgts\in\decidedsystems$.
To decide $\leftlangof{\admgts}\separable\dycklangn{\dyckdims}$, it thus remains to check $\leftlangof{\openmgts}\separable\dycklangn{\dyckdims}$ for all $\openmgts\in\opensystems$.
If all checks succeed, our decision procedure returns separable, and if one check fails, it returns inseparable.
Since the DMGTS in $\opensystems$ are perfect, we can apply Corollary~\ref{Corollary:PerfectSeparationDec}. 
We compute~$\ints$-VASS for the languages $\sidesolutions{\aside}(\openmgts)$ 
using Lemma~\ref{Lemma:EasyInclusion}, and check $\sidesolutions{\leftside}(\openmgts)\separable\sidesolutions{\rightside}(\openmgts)$ with the algorithm from~\cite{IntVASS17} that is behind Theorem~\ref{Theorem:IntsVASSRegSep}.

The decomposition of the DMGTS takes resources $\fof{\sizeof{\leftside\cup\rightside}+4}$, followed by an elementary separability check.
By \cite[Lemma 4.6]{Schmitz16}, this yields an $\fof{\omega}$ upper bound. 
\end{proof}

%% file: separatingaut.tex
\newcommand{\autapprox}[1]{\annfa_{#1}}
\newcommand{\autapproxdef}{\autapprox{}}
\newcommand{\annfaph}{\annfap^{\sepsymbol}}
\newcommand{\annfah}{\annfa^{\sepsymbol}}
\section{Separability Transfer}\label{Section:SeparabilityResult}
We prove the separability transfer result in Lemma~\ref{Lemma:PerfectSeparation}.
\subsection{Regular Separator}\label{Subsection:TheSeparability}
For the direction from left to right, we show that every regular separator for the $\ints$-approximations $\leftsolutions(\admgts)$ and $\rightsolutions(\admgts)$ can be turned into a regular separator for $\leftlangof{\admgts}$ and $\dycklangn{\dyckdims}$. 
The idea is to rely on the common finite-state over-approximation of $\leftsolutions(\admgts)$ and $\rightsolutions(\admgts)$.
Faithfulness and modulo reasoning will thus play an important role. 
%
%

Let $\annfaph$ separate $\leftsolutions(\admgts)$ and $\rightsolutions(\admgts)$.
Recall that these languages are defined over $\analph_{n, \sepsymbol}$, and expect a letter of the form $(a, \sepsymbol)$ whenever an update is taken.
%
%
We write~$\annfap$ for the NFA that results from $\annfaph$ by replacing transition labels $(a, \sepsymbol)$ with~$a$.  
Our plan is to use $\annfap$ as a separator for $\leftlangof{\admgts}$ and $\dycklangn{\dyckdims}$. 
Note that this is not a trivial task.
The set $\rightsolutions(\admgts)$ (even without the $\#$ symbols) may not contain $\dycklangn{\dyckdims}$, since the former also imposes intermediate constraints.
The same reasoning also shows that $\dycklangn{\dyckdims}\cap\langof{\annfa}$ may also not be contained, where $\annfa$ is the control flow of $\admgts$.
For the separation to work, $\annfaph$ should be \emph{precise} as follows:
\begin{align}
\langof{\annfaph}\ \cap\ \dycklangn{\dyckdims}^{\sepsymbol}\ \subseteq\ \rightsolutions(\admgts)\ . \label{Equation:Preciseness}
\end{align}
Preciseness says that the language of $\annfaph$ is so small that it cannot intersect the Dyck language without already intersecting $\rightsolutions(\admgts)$.
The language $\dycklangn{\dyckdims}^{\sepsymbol}$ offers $a$ and $(a, \sepsymbol)$ whenever $\dycklangn{\dyckdims}$ has letter $a$. 

It is worth noting that the constructions in this subsection could have been done without the $\sepsymbol$ symbol, if we had also defined $\leftsolutions(\admgts)$ resp.\ $\rightsolutions(\admgts)$ without it.
However, the symbol plays a role for the reverse implication in Lemma~\ref{Lemma:PerfectSeparation} which we show in the next subsection.

The preciseness of $\annfaph$ is enough to realize our task.
\begin{lem}
If the NFA $\annfaph$ separates $\leftsolutions(\admgts)$ and $\rightsolutions(\admgts)$ and is precise, then $\annfap$ separates $\leftlangof{\admgts}$ and $\dycklangn{\dyckdims}$. 
\end{lem}
To see this, let $\annfaph$ be an NFA that is precise, separating $\leftsolutions(\admgts)$ from $\rightsolutions(\admgts)$.
We argue that $\annfap$ is a separator for $\leftlangof{\admgts}$ and $\dycklangn{\dyckdims}$.
The inclusion $\leftlangof{\admgts}\subseteq\langof{\annfap}$ follows, since $\leftsolutions(\admgts)\subseteq\langof{\annfap}$ relaxes the constraints defining $\leftlangof{\admgts}$, and marks some letters with $\sepsymbol$.
Now suppose that the disjointness $\langof{\annfap}$ and $\dycklangn{n}$ of does not hold, and let $\aword\in\langof{\annfap}\cap\dycklangn{n}$.
Since $\aword\in\langof{\annfap}$, we get $\aword_{\sepsymbol}\in\langof{\annfaph}$ for some $\aword_{\sepsymbol}$ that marks some letters with $\sepsymbol$.
Because $\dycklangn{n}^{\sepsymbol}$ allows arbitrary marking, $\aword_{\sepsymbol}\in\langof{\annfaph}\cap\dycklangn{n}^{\sepsymbol}$ as well.
Preciseness says $\langof{\annfaph}\cap\dycklangn{n}^{\sepsymbol}\subseteq\rightsolutions(\admgts)$, and yields $\aword_{\sepsymbol}\in\rightsolutions(\admgts)$.
But $\aword_{\sepsymbol}\in\leftlangof{\annfaph}$, and $\annfaph$ separates $\leftsolutions(\admgts)$ from $\rightsolutions(\admgts)$.
This is a contradiction. 
%


Our main finding is the following lemma, where the product captures language intersection, $\langof{\annfaph\times\annfah}=\langof{\annfaph}\cap\langof{\annfah}$.
%
\begin{lem}\label{Lemma:PrecisedSeparators}
Let $\admgts$ be faithful. 
Every  separator $\annfaph$ of $\leftsolutions(\admgts)$ and $\rightsolutions(\admgts)$  can be turned into a precise separator $\annfaph\times\annfah$.
The NFA $\annfah$ is independent of $\annfaph$. 
\end{lem}

We argue Lemma~\ref{Lemma:PrecisedSeparators}.
To explain our choices, we iteratively develop $\annfah$ by considering where preciseness may fail.
A first failure of preciseness may be due to the fact that $\annfaph$ accepts words that do not label a run through $\admgts$. 
To overcome the problem, we may understand $\admgts$ as an NFA $\annfa_{\admgts, \sepsymbol}$, and use this as $\annfah$ in the lemma. 
By this, we mean that $\annfa_{\admgts, \sepsymbol}$ follows the control flow of $\admgts$, and reads $\sepsymbol$-marked letters at every entry and exit.
%
%
%
If $\annfa_{\admgts, \sepsymbol}$ accepts a word from $\dycklangn{\dyckdims}^{\sepsymbol}$, then the word labels a run through $\admgts$ that takes the Dyck counters from zero to zero.
The latter holds, because the word is in the Dyck language and $\admgts$ is Dyck-visible.
Since $\admgts$ is faithful, the initial and final configurations are zero on~$\rightside$ by definition, and so the run belongs to $\acceptofshort{\rightside}{\admgts}$. 
Recall that this means the run is a $\rightside$-run (in $\N$ for $\rightside$ markings), and satisfies the reachability constraints at the entry and exit of $\admgts$.
Unfortunately, this does not suffice for preciseness.

The problem is that $\rightsolutions(\admgts)$ is not defined via  $\acceptofshort{\rightside}{\admgts}$, but via intermediate acceptance $\interacceptof{\ints}{\rightside}{\admgts}$. 
This means the run not only has to reach zero on the Dyck counters, but it also has to reach certain values at the entries and exits of the intermediate precovering graphs in~$\admgts$. 
This is where Inclusion~\eqref{Equation:Faithfulness} in the definition of faithfulness comes in.
It suggests we should define the NFA~$\annfah$ so that it (i) follows $\admgts$ like $\annfa_{\admgts, \sepsymbol}$ does, (ii) maintains the counters modulo the number $\charmod$ given by $\admgts$, and (iii) checks intermediate acceptance modulo $\charmod$. 
In other words, $\annfah$ should capture the the finite-state over-approximation of the $\leftside$ and $\rightside$ sides, as we discussed in \Cref{Section:Overview,Section:DMGTS}.
If $\annfah$ accepts a word from the Dyck language, we have a run in $\acceptofshort{\rightside}{\admgts}$ as before, but moreover we know that the run belongs to $\interacceptof{\rightside}{\restrictto{\omegalequiv{\charmod}}{\rightside}}{\admgts}$. %
%
Faithfulness now shows that the run is also in $\interacceptof{\ints}{\rightside}{\admgts}$. 

To be a separator,  $\langof{\annfaph\times\annfah}$  has to cover $\leftsolutions(\admgts)$. 
This holds, because $\leftsolutions(\admgts)$ is not only defined via ordinary acceptance on the counters in $\leftside$, but also via modulo~$\charmod$ acceptance on $\rightside$. 
The purpose of the constraint $\interacceptof{\ints}{\restrictto{\omegalequiv{\charmod}}{\rightside}}{\admgts}$ in the definition of~$\interacceptof{\ints}{\leftside}{\admgts}$ is thus to support the above restriction of a given separator. 
The disjointness $\langof{\annfaph\times\annfah}\cap\rightsolutions(\admgts)=\emptyset$ is by the fact that $\annfaph$ is a separator. 
This justifies the product construction $\annfaph\times\annfah$ in the sense that $\annfah$ alone may not be a separator. 

Assuming an NFA that satisfies (i) to (iii) above, these arguments yield Lemma~\ref{Lemma:PrecisedSeparators}.
It remains to define the NFA $\annfah$ with these properties:
\begin{align*}
\annfah = (\states\times [0,\charmod-1]^{\dims},\; \dyckalphn{\dyckdims, \sepsymbol},\;   
\transitions,\; &((\precovering_{\firstindex}, \genter), \remainderof{(\aconf_1, \zerovec)}{\charmod}), \\
&((\precovering_{\lastindex}, \gexit), \remainderof{(\aconf_2, \zerovec)}{\charmod}))\ .
\end{align*} 
The set $\states$ contains states $(\precovering, \genter)$ and $(\precovering, \gexit)$ for every precovering graph $\precovering$ in $\admgts$, and moreover all states in $\admgts$. 
For every transition $(\anode, a, \avarp, \anodep)$ in a precovering graph of $\admgts$ and for every counter valuation $\avar\in[0,\charmod-1]^{\dims}$, we have 
\begin{align*}
(\anode, \avar)\xrightarrow{a}(\anodep, \remainderof{\avar+\avarp}{\charmod})\in\transitions\ .
\end{align*}
We also have transitions that enter and exit a precovering graph $\precovering$, or move from $\precovering$ to $\precovering'$ via the update $\anupdate$:  
\begin{align*}
    ((\precovering, \genter), \avar)\xrightarrow{\emptyword}&(\precovering.\groot, \avar)\in\transitions,\hspace{0.85cm} \text{ if } x\omegalequiv{\charmod} \precovering.\incounters\\
    (\precovering.\groot, x)\xrightarrow{\emptyword}&((\precovering, \gexit), x)\in\transitions,\hspace{0.7cm}\text{ if } x\omegalequiv{\charmod} \precovering.\outcounters\\
    ((\precovering, \gexit), x)\xrightarrow{(\edgelabelof{\anupdate}, \sepsymbol)}&((\precovering', \genter), \remainderof{x + \actionbalof{\anupdate}}{\charmod})\in\transitions\ .
\end{align*}

We explain the construction.
The NFA $\annfah$ simulates the control flow, and tracks the counter valuations modulo $\charmod$.
If an update is taken, we expect it to be marked by $\sepsymbol$.
This realizes (i) and (ii).
To enter resp.\ exiting $\precovering$, we offer a special $\varepsilon$ transition.
To take this transition, the modulo $\mu$ values of the counter valuations must agree with the markings in $\admgts$.
This is ensured by the condition $x\omegalequiv{\charmod} \precovering.\incounters$ resp. $x\omegalequiv{\charmod}\precovering.\outcounters$.
Recall that the preorder $\omegalequiv{\charmod}$ requires modulo $\charmod$ equivalence at the counters where the right hand side has a non-$\omega$ value.
Thus $\annfah$ only accepts runs that reach the $\rightside$ markings modulo $\mu$.
This realizes (iii).

%% file: separatingpump.tex
\newcommand{\statecount}{\mathit{N}}
\newcommand{\dfaequiv}{\sim_{\adfa}}

\newcommand{\classof}[1]{[#1]_{\dfaequiv}}

\newcommand{\dfatrans}[1]{\mathop{\raisebox{-0.06cm}{$\xrightarrow{#1}$}}}

\subsection{Inseparability}\label{Subsection:TheInseparability}
We prove the missing direction of Lemma~\ref{Lemma:PerfectSeparation}, formulated as follows.
\begin{lem}\label{Lemma:Inseparable}
Consider a perfect DMGTS $\admgts$ that satisfies $\leftsolutions(\admgts)\inseparable\;\rightsolutions(\admgts)$. 
Then also $\leftlangof{\admgts}\inseparable\;\dycklangn{\dyckdims}$ holds.
\end{lem} 
The proof needs a classic definition~\cite{Buechi62}. 
A DFA $\adfa$ over $\analph$ induces the equivalence $\dfaequiv\ \subseteq\analph^*\times\analph^*$ that is defined by $\aword\dfaequiv\awordp$, if for all states $p, q$ in $\adfa$,  we have $p\dfatrans{\aword}q$ if and only if $p\dfatrans{\awordp}q$.
The equivalence says that the words lead to the same state changes in $\adfa$. 

To prove Lemma~\ref{Lemma:Inseparable}, we reason towards a contradiction, and assume there is a DFA~$\adfa$ that separates $\leftlangof{\admgts}$ from $\dycklangn{\dyckdims}$. 
We use the premise $\leftsolutions(\admgts)\inseparable\rightsolutions(\admgts)$ and Lambert's iteration lemma to construct words $o_{\leftside}\in\leftlangof{\admgts}$ and $o_{\rightside}\in\rightlangof{\admgts}$ with 
$o_{\leftside}\dfaequiv o_{\rightside}$. 
%
%
Then $\adfa$ must accept or reject both words.
%
%
If~$\adfa$ accepts~$o_{\rightside}$, we have a contradiction to $\langof{\adfa}\cap\dycklangn{\dyckdims}=\emptyset$ due to  Lemma~\ref{Lemma:EasyInclusion}.
If~$\adfa$ does not accept $o_{\leftside}$, we have a contradiction to $\leftlangof{\admgts}\subseteq\langof{\adfa}$.
This concludes the proof. 

To construct $o_{\leftside}$ and $o_{\rightside}$, we use the following lemma. Here, we need the $\sepsymbol$ symbols in the definition of $\sidesolutions{\aside}(\admgts)$. 
\begin{lem}\label{Lemma:IndistinguishablePairs}
Consider a DFA $\adfa$ such that for all pairs of words $\awordn{0}(a_1, \sepsymbol)\ldots\awordn{k}\in\leftsolutions(\admgts)$ and $\awordpn{0}(a_1, \sepsymbol)\ldots\awordpn{k}\in\rightsolutions(\admgts)$ there is $i\leq k$ with $\awordn{i}\not\dfaequiv\awordpn{i}$. Then $\leftsolutions(\admgts)\separable\rightsolutions(\admgts)$. 
\end{lem}
\begin{proof}
It is well-known that the equivalence $\dfaequiv$ has finite index and the equivalence classes $\classof{\aword}$ are regular languages~\cite{Buechi62}. Then the following is a finite union of regular languages:
    \begin{align*}
        \aseparator\quad=\quad\bigcup_{\awordn{0}(a_1, \sepsymbol) \ldots\awordn{k}\in\leftsolutions(\admgts)}\classof{\awordn{0}}(a_1, \sepsymbol) \ldots \classof{\awordn{k}}\ .
    \end{align*}
    
    We show that the regular language $\aseparator$ separates $\leftsolutions(\admgts)$ and $\rightsolutions(\admgts)$.  
    The inclusion $\leftsolutions(\admgts)\subseteq\aseparator$ is immediate. 
    Assume there is $\awordp\in\aseparator\cap\rightsolutions(\admgts)$.
    Then $\awordp=\awordpn{0}(a_1, \sepsymbol)\ldots\awordpn{k}\in\rightsolutions(\admgts)$ and there is $\awordn{0}(a_1, \sepsymbol)\ldots \awordn{k}\in\leftsolutions(\admgts)$ so that $\awordpn{i}\dfaequiv \awordn{i}$ for all $i\leq k$. 
    This is the conclusion that needs the $\sepsymbol$ symbols.
    Without them, the equivalent words may not align with the precovering graphs.
    The conclusion contradicts the lemma's premise, and $\awordp$ cannot exist. 
\end{proof}

\newcommand{\anysupsol}{\mathit{s}_{\aside}}
\newcommand{\sjsupsol}{\mathit{s}_{\leftside}}
\newcommand{\dysupsol}{\mathit{s}_{\rightside}}
\newcommand{\anyword}{\mathit{w}_{\aside}}
\newcommand{\sjword}{\mathit{w}_{\leftside}}
\newcommand{\dyword}{\mathit{w}_{\rightside}}
\newcommand{\anywordpar}[1]{\mathit{w}_{\aside, #1}}
\newcommand{\sjwordpar}[1]{\mathit{w}_{\leftside, #1}}
\newcommand{\dywordpar}[1]{\mathit{w}_{\rightside, #1}}
\newcommand{\anyscale}{\mathit{r}_{\aside}}
\newcommand{\sjscale}{\mathit{r}_{\leftside}}
\newcommand{\dyscale}{\mathit{r}_{\rightside}}
\renewcommand{\aplace}{\mathit{j}}
\newcommand{\diffword}{\mathsf{diff}}
\newcommand{\remword}{\mathsf{rem}}

We proceed with the definition of the words $o_{\leftside}$ and $o_{\rightside}$. 
%
%
We apply Lemma~\ref{Lemma:IndistinguishablePairs} in contraposition to $\leftsolutions(\admgts)\inseparable\rightsolutions(\admgts)$. 
This yields $\asolwordn{0}\ldots\asolwordn{k}\in\leftsolutions(\admgts)$ and $\asolwordpn{0}\ldots\asolwordpn{k}\in\rightsolutions(\admgts)$ so that  for all $i\leq k$ we have $\asolwordn{i}\dfaequiv\asolwordpn{i}$. 
%
%
The membership in $\leftsolutions(\admgts)$ resp.~$\rightsolutions(\admgts)$ gives us loops $\asolseqn{i}$ and~$\asolseqpn{i}$ in every precovering graph~$\graphn{i}$ of $\admgts$ that start in the root and are labeled by $\asolwordn{i}$ resp. $\asolwordpn{i}$. 
Moreover, since the languages are defined via intermediate acceptance, we know that 
$\sum_{i\leq k}\parikhof{\asolseq_{i}}$ and $\sum_{i\leq k}\parikhof{\asolseqp_{i}}$ solve $\leftcharof{\admgts}$ resp. $\rightcharof{\admgts}$. 
To sum up, the words given by Lemma~\ref{Lemma:IndistinguishablePairs} provide solutions to the characteristic equations as we need them to apply  Lambert's iteration lemma. 

The perfectness of $\admgts$ yields covering sequences  $\upcoveringn{i}'\in\coveringseqof{}{\precovering_{i}}$ and $\downcoveringn{i}'\in\downcoveringseqof{}{\precovering_{i}}$ for all $i\leq k$.     
We show how to construct new covering sequences $\upcovering_{i}$ and~$\downcovering_{i}$, as well as further rooted loops $\sjwordpar{i}$ and~$\dywordpar{i}$ in each precovering graph $\precovering_i$ so that the conditions on the homogeneous solutions formulated by Lambert's iteration lemma are met for both, $(\upcovering_{i}, \sjwordpar{i}, \downcovering_{i})_{i\leq k}$ and $(\upcovering_{i}, \dywordpar{i}, \downcovering_{i})_{i\leq k}$. 
In addition, we will guarantee that $\edgelabelof{\sjwordpar{i}}\dfaequiv\edgelabelof{\dywordpar{i}}$.
Applying Lemma~\ref{Lemma:ModifiedIteration} twice then yields a common $r\in\nat$ so that 
\begin{align*}
        o_{\leftside}=\edgelabelof{\upcovering_{0}^{r}\asolseqn{0} \sjwordpar{0}^{r} \downcovering_{0}^{r}\transn{0}\ldots\transn{k-1}\upcovering_{k}^{r}\asolseqn{k} \sjwordpar{k}^{r} \downcovering_{k}^{r}}&\in\leftlangof{\openmgts}\\ 
        o_{\rightside}=\edgelabelof{\upcovering_{0}^{r}\asolseqpn{0} \dywordpar{0}^{r} \downcovering_{0}^{r}\transn{0}\ldots\transn{k-1}\upcovering_{k}^{r}\asolseqpn{k} \dywordpar{k}^{r} \downcovering_{k}^{r}}&\in\rightlangof{\openmgts}\ . 
    \end{align*}
Note that $o_{\leftside}\in\leftlangof{\openmgts}$ also requires $\interacceptof{\ints}{\restrictto{\omegalequiv{\charmod}}{\rightside}}{\admgts}$. 
This is taken care of by the modulo constraints in $\leftcharof{\admgts}$. 

With this definition, the desired $o_{\leftside}\dfaequiv o_{\rightside}$ 
%
%
is a consequence of $\edgelabelof{\asolseqn{i}}\dfaequiv \edgelabelof{\asolseqpn{i}}$ and $\edgelabelof{\sjwordpar{i}}\dfaequiv \edgelabelof{\dywordpar{i}}$ for all $i\leq k$.      
%

We turn to the construction of $\upcovering_{i}$, $\downcovering_{i}$, $\sjwordpar{i}$, and~$\dywordpar{i}$. 
To ease the notation, we fix a precoving graph $\precovering$ and skip the index $i$.
So $\upcovering'$ and~$\downcovering'$ will be the pumping sequences for this precovering graph, $\edges$ will be the edges in this precovering graph, 
and $\upcovering$, $\downcovering$, $\sjword$, and~$\dyword$ are the sequences we want to construct.

With $\statecount$ being the number of states in $\adfa$, and $\diffword, \remword$ sequences of edges we define in a moment, we let
    \begin{align}
        \sjword\ =\ \diffword^{\statecount}.\remword\qquad\dyword\ =\ \diffword^{\statecount+c\cdot \statecount!}.\remword\ .\label{Equation:SJDY}
    \end{align}
The integer $c\geq 1$ will become clear when we define $\remword$.  
We first argue $\edgelabelof{\sjword}\dfaequiv\edgelabelof{\dyword}$ by using the form in \Cref{Equation:SJDY}.
Let $p$ and $q$ be states in $\adfa$. 
Since $\adfa$ is a DFA, there is a unique run from $p$ on $\lambda(\sjword)$. 
Consider the part of the run that reads $\lambda(\diffword^{\statecount})$.
By the pigeonhole principle, there are $0\leq i<j\leq \statecount$ where the state after reading $\lambda(\diffword^{i})$ and $\lambda(\diffword^{j})$ is the same.
This means we can repeat $\lambda(\diffword^{j-i})$ and still arrive at this state.
While we repeat $\diffword^{j-i}$ only once in $\sjword$, we repeat it an additional $c\cdot \statecount!/(j-i)$ times in $\dyword$. 
The sole purpose of the factorial $\statecount!$ is to guarantee that this division by $j-i$ results in an integer: $j-i\leq \statecount$ implies $\statecount!/(j-i)\in\nat$.
Since the states reached after $\lambda(\diffword^{\statecount})$ and $\lambda(\diffword^{\statecount+c\cdot \statecount!})$ coincide, we have that $\lambda(\sjword)$ leads from $p$ to $q$ if and only if this holds for $\lambda(\dyword)$.

It remains to construct $\upcovering$, $\downcovering$, $\diffword$, and $\remword$ for each precovering graph, along with a common $c\in\N$.
The key proof obligation is to realize the leftover solution in two parts, $\diffword$ and $\remword$.
We structure our argument by discussing the conditions that are needed for this.
The condition on $c$, which is needed to realize $\remword$, only becomes clear at the end of the discussion.
Therefore our arguments treat it as a parameter until we reach this step. 
%

By Lemma~\ref{Lemma:ModifiedIteration}, we can find maximal support solutions $\sjsupsol$ and~$\dysupsol$ that satisfy the Conditions~\eqref{Equation:Parikh} and \eqref{Equation:Enable} wrt. $\upcovering'$ and $\downcovering'$.
Later in the argument, we will need to realize $\dysupsol-\sjsupsol$ in $\diffword$.
For this reason, we find $\dysupsol$ such that $(\dysupsol-\sjsupsol)[\edges]\geq \onevec$ holds.
We can do this by repeating some maximal support solution $\dysupsol'$: we have 
$\dysupsol'\geq \onevec$ and $\ell\cdot\dysupsol'$ is a solution to the homogeneous system for all $\ell\in\N$.

We will not only construct cycles, but also new support solutions $\sjsupsol^*$ and $\dysupsol^*$. 
Our construction is guided by the following equations in Lemma~\ref{Lemma:ModifiedIteration}:
\begin{align}
\parikhof{\upcovering}+\parikhof{\downcovering}+\parikhof{\sjword}\ &=\ \restrictto{\sjsupsol^*}{\edges}\label{Equation:LambertSJ}\\
\parikhof{\upcovering}+\parikhof{\downcovering}+\parikhof{\dyword}\ &=\ \restrictto{\dysupsol^*}{\edges} \label{Equation:LambertDY}\ .
\end{align}
By inserting the shape of $\sjword$ and $\dyword$ required by Equation~\eqref{Equation:SJDY} and subtracting Equation~\eqref{Equation:LambertSJ} from ~\eqref{Equation:LambertDY}, we obtain 
\begin{align}
c\cdot \statecount!\cdot\parikhof{\diffword}\ =\ \coordacc{(\dysupsol^*-\sjsupsol^*)}{\edges}\ . \label{Equation:Diff} 
\end{align}
This leads us to define
\begin{align}
\anysupsol^*\ = \ c\cdot \statecount!\cdot\anysupsol\ .\label{Equation:SupSol} 
\end{align}
We can now divide Equation~\eqref{Equation:Diff} by $c\cdot \statecount!$ and obtain 
\begin{align*}
\diffword\ =\ \realizationof{\dysupsol-\sjsupsol}\ . 
\end{align*}    
Here, we use $\realizationof{v}$ to turn a Parikh vector $v\geq\onevec$ into a cycle.
Recall our assumption $\coordacc{(\dysupsol-\sjsupsol)}{\edges}\geq\onevec$, which ensures that $\realizationof{\dysupsol-\sjsupsol}$ is well-defined.

The new support solutions in Equation~\eqref{Equation:SupSol} suggest we should define the new covering sequences by repetition:
\begin{align} 
        \upcovering\ =\ (\upcovering')^{c\cdot \statecount!}
        \qquad\qquad
        \downcovering\ =\ (\downcovering')^{c\cdot \statecount!}\ .\label{Equation:Covering}
\end{align}

We insert the Equations~\eqref{Equation:SJDY}, \eqref{Equation:SupSol}, and~\eqref{Equation:Covering} into~\eqref{Equation:LambertSJ} and get  
\begin{align*}
c\cdot \statecount!\cdot(\parikhof{\upcovering'}+\parikhof{\downcovering'})+\statecount\cdot\parikhof{\diffword}+\parikhof{\remword}\ =\ c\cdot\statecount!\cdot\coordacc{\sjsupsol}{\edges}\ .
\end{align*}
This yields the missing
\begin{align*}
\remword\ =\ \realizationof{c\cdot \statecount!\cdot (\restrictto{\sjsupsol\;}{\edges}-\parikhof{\upcovering'}-\parikhof{\downcovering'})-\statecount\cdot \parikhof{\diffword}}\ .
\end{align*}
This is the moment we need the factor $c$: it has to be large enough so that 
$c\cdot \statecount!\cdot (\restrictto{\sjsupsol}{\edges}-\parikhof{\upcovering'}-\parikhof{\downcovering'})-\statecount\cdot \parikhof{\diffword}\geq \onevec$, and hence the vector can be realized as a cycle. 
Note that $c$ goes into the definition of the support solution, which is shared by all precovering graphs.
This means the choice of $c$ not only has to satisfy the inequality for~$\precovering$, but for all precovering graphs.
Since we only impose lower bounds on $c$, we can satisfy them by chosing a large enough $c$.
For this $c$, our construction ensures that $\upcovering$, $\downcovering$, $\diffword$, and $\remword$ fulfill the conditions imposed by Lemma~\ref{Lemma:ModifiedIteration} and \Cref{Equation:SJDY}.
This concludes the proof.

%% file: example.tex
\section{Decomposition on an Example}\label{Section:Example}

Before we move on to the details of the decomposition, we observe it on the example DMGTS~$\admgts_{ex}$, where $\leftlangof{\admgts_{ex}}\separable\dycklangn{n}$ holds.
We opt for a separable instance to keep the arguments short.
Our decomposition only adapts the KLMST decomposition, so the familiar reader will be able to follow the example.

\begin{figure}[ht]
    \begin{center}
        \includegraphics*[width=0.8\linewidth]{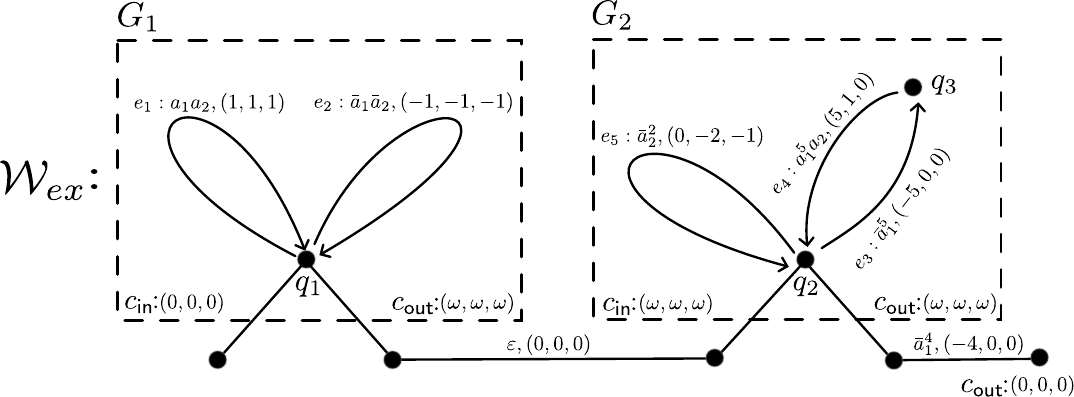}
    \end{center}
    \caption{Example DMGTS $\admgts_{ex}$}\label{Figure:DMGTSex}
\end{figure}

Consider the example DMGTS~$\admgts_{ex}$ in \Cref{Figure:DMGTSex}, with counters $\leftside\uplus\rightside$, where $\leftside=\set{3}$ resp.~$\rightside=\set{1, 2}$ and $\leftlangof{\admgts_{ex}}\subseteq\analph_{2}^{*}$.
We omit the consistent assignments $\bal$ in this discussion.
For the sake of compactness, we also summarize consecutive edges into one edge.
For example, the edge $\anedge_{1}$ represents three edges executed in a sequence, two reading $\incdyckn{1}$ and $\incdyckn{2}$, and one incrementing the $\leftside$ counter.
We also omit the trivial precovering graphs (such as the graph that would follow $\decdyckn{1}^{4}$) and only depict one marking.

We explain $\admgts_{ex}$.
First, note that the first two counters are updated Dyck-visibly.
For example, $\anedge_{1}$ reads $\incdyckn{1}\incdyckn{2}$, and increases the first and the second counters by $1$.
In this light, in the first precovering graph of $\admgts$, the $\leftside$ counter tracks the Dyck-counters, and ensures that only copies of $\incdyckn{1}\incdyckn{2}$ and $\decdyckn{1}\decdyckn{2}$ are read.
Then, the second precovering graph has a loop, $\anedge_{5}$, that reads $\decdyckn{2}^{2}$.
Thus, it decrements the second counter by $2$ when taken by the Dyck-side.
The loop has a different effect on the subject side.
Namely, it decrements the counter only by $1$.
The graph also has the cycle $\anedge_{3}.\anedge_{4}$ that the subject side may take freely, but requires $5$ tokens in the first counter for the Dyck-side.
The full cycle reads $\decdyckn{1}^{5}\incdyckn{1}^{5}\incdyckn{2}$.

Before we move on to the discussion, we recall the acceptance definitions with an example.
A run through $\admgts_{ex}$ is intermediate $\rightside$-accepting, if it is an $\N$-run, and projecting away the $\leftside$-counters, it is an intermediate accepting run in the underlying MGTS.
Note that since $\mu=1$, $\leftside$-acceptance is defined like $\rightside$ acceptance.
The run should be an $\N$-run, and intermediate accepting wrt. $\leftside$-counters. 
By our definition, for larger $\mu$, the acceptance will also weakly depend on the markings for $\rightside$-counters.
For example runs, consider the runs $\apath_{\leftside}$ and $\apath_{\rightside}$ below, which are intermediate $\leftside$- resp. $\rightside$- accepting in $\admgts_{ex}$.
It can be readily checked that $\apath_{\aside}$ does reach all $\aside$ markings.
To succinctly present the runs, we omit some configurations and only write the edges resp. labels of updates.
Note that we are ignoring (precise) reachability for a set of counters, so they may start from any valuation.
This makes positivity trivial for those counters.
{\small
\begin{align*}
    \apath_{\leftside}&=(20, 20, 0).\anedge_{1}^3.(23,23,3).\epsilon.(23, 23, 3).\anedge_{3}.(18, 23, 3).\anedge_{4}.(23,24,3).\anedge_{5}^{3}.(23,18,0).\decdyckn{1}^{4}.(19, 18, 0)\\
    \apath_{\rightside}&=(0, 0, 200).\anedge_{1}^{5}.(5, 5, 205).\anedge_{2}^{1}.(4, 4, 204).\epsilon.(4, 4, 204).\anedge_{5}^{2}.(4, 0, 202).\decdyckn{1}^{4}.(0, 0, 202).
\end{align*}
}%

\begin{figure}[ht]
    \begin{center}
        \includegraphics*[width=0.5\linewidth]{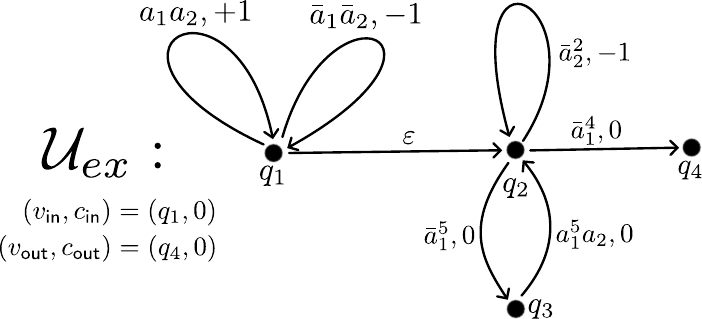}
    \end{center}
    \caption{VASS $\avas_{ex}$ accepting $\leftlangof{\admgts_{ex}}$}\label{Figure:VASSex}
\end{figure}

The example DMGTS $\admgts_{ex}$ encodes the problem $\langof{\avas_{ex}}\separable\dycklangn{n}$ for the initialized VASS~$\avas_{ex}$ given in \Cref{Figure:VASSex}.

We shortly explain the language $\leftlangof{\admgts_{ex}}=\langof{\avas_{ex}}$.
For the purposes of this section, define the $\precovering_{1}$-balance $\effectnof{\precovering_{1}}{\aword}$ of a word $\aword\in\analph_{2}^{*}$ as the number of $\decdyckn{1}\decdyckn{2}$ infixes substracted from the number of $\incdyckn{1}\incdyckn{2}$ infixes.
Further define $\dyckeffectof{\aword}\in\Z^{2}$ for a word $\aword\in\analph_{2}^{*}$ to be its effect on the Dyck VASS.
Let $A\subseteq\set{\incdyckn{1}\incdyckn{2}, \decdyckn{1}\decdyckn{2}}^{*}$ be the subset of words that have non-negative $\precovering_{1}$-balances in all prefixes.
Then, we have
\begin{align*}
    \langof{\avas_{ex}}&=\setcond{\aword.\awordp.\decdyckn{1}^{4}\in\analph_{2}^{*}}{\effectnof{\precovering_{1}}{\aword}=s\text{, }\aword\in A\text{, and }\awordp \in L^{s}}\text{ where }\\
    L&=(\decdyckn{1}^{5}.\incdyckn{1}^{5}.\incdyckn{2})^{*}.\decdyckn{2}^{2}.(\decdyckn{1}^{5}.\incdyckn{1}^{5}.\incdyckn{2})^{*}.
\end{align*}
We explain the reasoning.
Consider that in the first precovering graph $\precovering_{1}$ of $\avas_{ex}$, upon reading the prefix $\aword'\in\set{\incdyckn{1}\incdyckn{2}, \decdyckn{1}\decdyckn{2}}^{*}$, the counter records $\effectnof{\precovering_{1}}{\aword'}$. 
Let $\aword$ be the full prefix read in $\precovering_{1}$ with $\effectnof{\precovering_{1}}{\aword}=s$.
The second precovering graph $\precovering_{2}$ allows reading $\decdyckn{2}^{2}$ and decreasing the counter by $1$, or reading $\decdyckn{1}^{5}.\incdyckn{1}^{5}.\incdyckn{2}$ with no effect.
The final transition then reads $\decdyckn{1}^{4}$ and also preserves the counter value.
The run is accepted by $\avas$, if the initial and final counter valuations are $0$.
Thus, we must read $s$ copies of $\decdyckn{2}^{2}$ in $\precovering_{2}$.

We aim to deduce $\leftlangof{\admgts_{ex}}\separable\dycklangn{2}$, which holds if we can capture a disjointness argument using a regular language.
As preparation, we shortly argue $\leftlangof{\admgts_{ex}}\cap\dycklangn{2}=\emptyset$.
To this end, let $\awordpp=\aword.\awordp.\decdyckn{1}^{4}\in\langof{\avas_{ex}}$ with $\effectnof{\precovering_{1}}{\aword}=s$, $\aword\in A$, and $\awordp\in L^{s}$.
Now suppose $\awordpp\in\dycklangn{2}$.
Note that we may only see $\incdyckn{1}$ or $\decdyckn{1}$ in $\aword$, so $\dyckeffectof{\aword}[1]=\dyckeffectof{\aword.\awordp}[1]$.
Then, for $\dyckeffectof{u}=0$ to hold, the final $\decdyckn{1}^{4}$ must be negated by the prefix, and $\dyckeffectof{\aword}[1]=4$ must hold.
Upon entering $\precovering_{2}$, if we read $\decdyckn{1}^{5}$, then we are guaranteed to get negative Dyck-counters and leave~$\dycklangn{2}$.  
Then, only $\decdyckn{2}^{2}$ may be read, yielding $\awordp=\decdyckn{2}^{2s}$.
But, the $\incdyckn{1}$- and $\incdyckn{2}$-balances of $\aword$ must be equal, so we have $\dyckeffectof{\aword}[1]=\dyckeffectof{\aword}[2]=s=4$ as well.
Then, reading $\awordp=\decdyckn{2}^{2s}$ yields $\dyckeffectof{\aword.\awordp}[1]= -4$, meaning $\awordpp\not\in\dycklangn{2}$, a contradiction.

Now, we discuss how our algorithm discovers the disjointness argument, and concludes $\leftlangof{\admgts_{ex}}\separable\dycklangn{2}$.
Note that the initial DMGTS $\admgts_{ex}$ is faithful by default: it is zero reaching, and there are no intermediate markings.
As we discussed in \Cref{Section:Overview}, our algorithm follows the KLMST algorithm, but almost always considers only one side.
On $\admgts_{ex}$, it finds the following boundedness on the $\rightside$-side.
Similarly to our disjointness argument, in any word $\awordpp\in\rightlangof{\admgts}$, we have $\dyckeffectof{\aword}[1]=\dyckeffectof{\aword}[2]=4$ for the full prefix $\aword$ read in $\precovering_{1}$.
Therefore, both counters must have the valuation $4$ when exiting $\precovering_{1}$.
Given this information, our algorithm constructs $\admgts_{i, j}$ for all $i, j\leq 4$, where $\admgts_{i, j}$ is $\admgts$ with the markings of the $\rightside$ counters on the exit of $\precovering_{1}$ concretized to $(i,j)$, and $\mu$ updated to $5$.
The latter change preserves faithfulness in the general case. 
For presentational purposes, our example only allows the value $4$, so even $\mu=1$ would preserve the invariant.
However, when multiple values are present, the DMGTS $\admgts$ must be able to distinguish between them, so $\mu$ gets updated by $\mu\to \mu\cdot (k+1)$ where $k$ is the bound discovered by KLMST.
We also follow this step in the example.

Note that we do not only produce $\admgts_{4,4}$, even though $4$ is the only allowed value.
This is because constraining the $\rightside$ counters also constrains the $\leftside$-language.
Recall that the $\leftside$-language requires modulo $\mu$ acceptance on $\rightside$.
Thus, $\leftlangof{\admgts_{4,4}}$ misses words that exit $\precovering_{1}$ with a value not equivalent to $4$ modulo $\mu=5$. 
Only once we include all values, we get the desired $\leftlangof{\admgts}=\bigcup_{i, j\leq 4}\leftlangof{\admgts_{i,j}}$.
Here, our algorithm recognizes that $\admgts_{i,j}$ with $i<4$ or~$j<4$ do not need further decomposition.
To see why, consider that $\admgts_{i, j}$ guarantees that the value $4$ is not reached for at least one of the counters in $\rightside$.
Since only the value $4$ is allowed for $\Z$-reaching runs, $\rightlangof{\admgts_{i, j}}=\rightsolutions(\admgts_{i, j})=\emptyset$.
We conclude $\leftlangof{\admgts_{i, j}}\separable\dycklangn{2}$ by Lemma~\ref{Lemma:PerfectSeparation}.
Our algorithm returns such systems in $\decidedsystems$.

\begin{figure}[ht]
    \begin{center}
        \includegraphics*[width=0.8\linewidth]{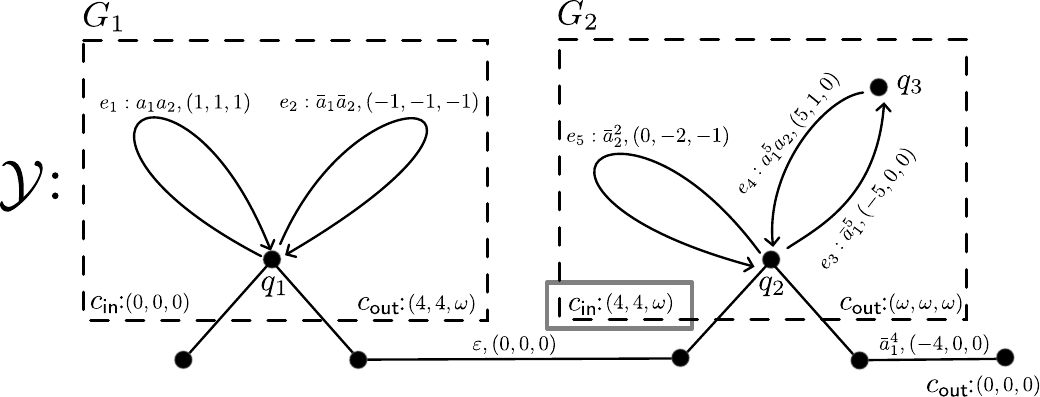}
    \end{center}
    \caption{The DMGTS $\admgtsp$ resulting from the decomposition.}\label{Figure:DMGTSex2}
\end{figure}

After the above argument, the only open fragment of $\leftlangof{\admgts_{ex}}\separable\dycklangn{2}$ is $\leftlangof{\admgtsp}\separable\dycklangn{n}$, with the DMGTS $\admgtsp$ obtained from $\admgts_{4, 4}$ as depicted in \Cref{Figure:DMGTSex2}.
It is easy to see that the equality~$\leftlangof{\admgts_{4, 4}}=\leftlangof{\admgtsp}$ holds and that $\admgtsp$ is faithful.

\begin{figure}[ht]
    \begin{center}
        \includegraphics*[width=0.8\linewidth]{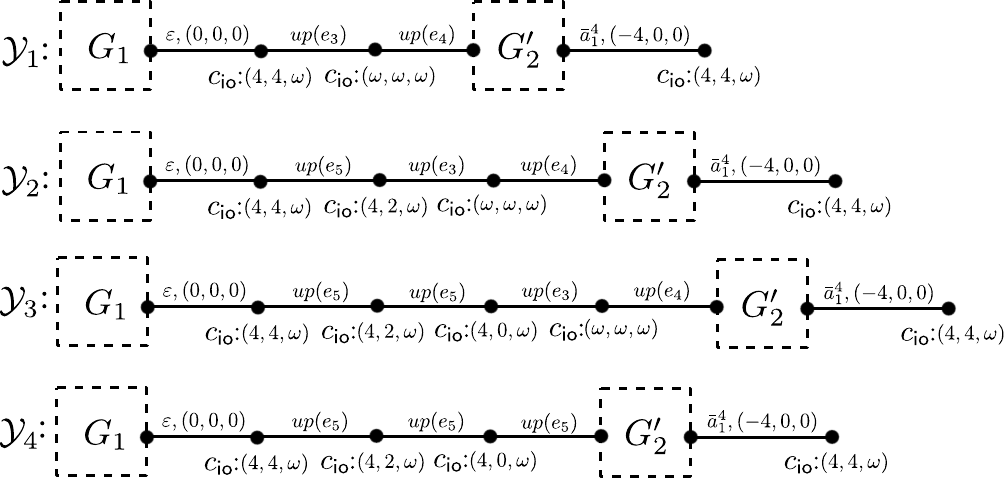}
    \end{center}
    \caption{The results $\admgtsp_{1}, \admgtsp_{2}, \admgtsp_{3}, \admgtsp_{4}$ obtained from unrolling $\admgtsp$.}\label{Figure:UnrolledExample}
\end{figure}

Now, the boundedness on the entry and exit markings of $\admgts_{ex}$ becomes visible to the separability algorithm via $\admgtsp$.
We apply the KLMST once more.
It is worth noting that we do not get a boundedness for counter $3$ on the exit of $\precovering_{1}$.
This is because when checking unboundedness wrt. markings, we follow perfectness, and consider each acceptance condition separately.
Thus the constraint for the $\leftside$ side is not to reach $4$ precisely, but to reach it modulo $\mu$.
For this reason, the set of reachable values $\set{4, 9, 14, \ldots}$ for the $\leftside$ counter is still unbounded.
Instead, we observe that $\coveringseqof{}{\precovering_{2}}=\emptyset$: the counters in $\precovering_{2}$ cannot be increased starting from $(4,4, \omega)$.
Note that this would be invisible to the decomposition, had we not concretized the markings of $\precovering_{1}$ beforehand.
In the formal construction, we implement a bounded counter for this case, similarly to \cite{Leroux19}.
For the sake of the example, we employ a construction that follows the coverability graph.
The result is the unrolling of $\precovering_{2}$, which yields the DMGTS~$\admgtsp_{1}, \admgtsp_{2}$, $\admgtsp_{3}$, and~$\admgtsp_{4}$, given in \Cref{Figure:UnrolledExample}.
Here, $up(\anedge_{i})$ represents the edge $\anedge_{i}$, turned into an update between precovering graphs.
The precovering graph $\precovering_{2}'$ is obtained by modifying both markings of $\precovering_{2}$ to $(\omega, \omega, \omega)$.
The remaining precovering graphs are trivial, and we depict $\inoutmarking=\inmarking=\outmarking$.

We explain the construction.
We start from the input marking, and construct the coverability graph for $\precovering_{2}$.
Since we are interested in separability, we also care about runs where $\rightside$ counters become negative.
Once this happens, we enter a sink where all transitions are allowed, and use $\omega$ markings for the $\rightside$ counters.
The words that enter this sink are guaranteed to not be in $\dycklangn{2}$, given that they start from the input marking.
After the graph construction, we represent the result as DMGTS by making the strongly connected components into precovering graphs.
We also lift the markings from the coverability graph.
In the example, the only non-trivial precovering graph is $\precovering_{2}'$, which is the sink visited after a $\rightside$ counter becomes negative.

Since the counters may only decrease, there is no pumping, and thus the unrolling stems from a finite set of runs (followed by $\precovering_{2}'$).
Note that $\leftlangof{\admgtsp}=\bigcup_{i\leq 4}\leftlangof{\admgtsp_{i}}$ even though there is no DMGTS that only takes $up(\anedge_{5})$ or $up(\anedge_{5}).up(\anedge_{5})$.
This is because for such DMGTS, the $\leftside$ language is empty.
We shortly argue this.
The modulo $\mu$ reachability constraint guarantees that the $\rightside$ counter is at least $4$ when leaving $\precovering_{1}$.
In $\precovering_{1}$, the value of the $\leftside$ counter is synchronized with the $\rightside$ counters, so it would also be at least $4$.
Then, only taking $up(\anedge_{5})$ or $up(\anedge_{5}).up(\anedge_{5})$ would not be enough to reach $0$ on the $\leftside$ counter.

It is easy to see that $\rightsolutions(\admgtsp_{i})=\emptyset$ for all $i\leq 4$: all DMGTS visit~$\precovering_{2}'$, and guarantee that a~$\rightside$ counter becomes negative.
Thus, assuming faithfulness, we can conclude as we did before, and get $\bigcup_{i\leq 4}\leftlangof{\admgtsp_{i}}\separable\dycklangn{2}$.
This closes the argument for $\leftlangof{\admgts_{ex}}\separable\dycklangn{2}$, as we have shown separability for all fragments.
We sketch out the faithfulness proof.
In \Cref{Section:OurDecomposition}, we generalize such arguments to handle all cases.
Intuitively, faithfulness is preserved, because the decomposition only extends the existing information locally.
For a precise argument, suppose that a run $\apath$ through $\admgtsp_{i}$ has a valuation in $\N^{2}$ at all entries and exits, reaches the intermediate $\rightside$ markings modulo $\mu$, and the extremal $\rightside$ markings precisely.
We derive a contradiction, which implies that no such run can exist.
Thus, faithfulness holds trivially. 
Let $i=1$ since the same argument applies to all cases.
Remark that the latter part of $\admgtsp_{1}$ is a result of unrolling $\precovering_{2}$, so we can find a run in $\admgtsp$ that visits the same valuations and corresponding states. 
Now, faithfulness of $\admgtsp$ implies that the markings on the exit of $\precovering_{1}$ and the entry of $\precovering_{2}$ are reached precisely.
This then carries over to $\admgtsp_{1}$, and implies that upon exiting $\precovering_{1}$, the value in counter $1$ is $4$.
Thus, taking $up(\anedge_{3})$ as prescribed by $\admgtsp_{1}$ must push the $\rightside$ valuations of $\arun$ outside of $\N^{2}$.
This contradicts that $\arun$ reaches an $\N^{2}$ valuation at all entries and exits.

%% file: decomposition.tex
\section{Decomposition}\label{Section:OurDecomposition}
We prove Lemma~\ref{Lemma:DecoupledDecomposition} by developing a decomposition algorithm that takes a faithful DMGTS and returns a finite set of perfect DMGTS and a finite set of DMGTS for which separability holds. 
At the heart of the algorithm is a single decomposition step, as described in Lemma~\ref{Lemma:Refinement} below. 
Given a DMGTS that is not perfect, the decomposition produces two sets of DMGTS, $X$ and $Y$, while preserving the $\leftside$-language.
The DMGTS in $X$ are guaranteed to decrease in the well-order, while those in $Y$ already have $\leftside$-languages that are separable from $\dycklangn{n}$. 
%
%
The separability stems from one of two reasons.
First, it may be that the $\rightside$-equations are infeasible, and therefore we get separability by Lemma~\ref{Lemma:PerfectSeparation}.
Second, the language may be separable by just counting modulo $\mu$.
To make this precise, we define the modulo counting languages $\modlangof{\mu, v}=\setcond{\aword\in\analph_{n}}{\effectof{\aword}\equiv_{\mu}v}$ for $v\in\Z^{n}$ and $\mu\in\N$.
\begin{lem}\label{Lemma:Refinement}
There is a function $\refineof{-}$, computable with elementary resources, that expects a faithful but imperfect DMGTS $\admgts$ with $\solutionsof{\leftcharof{\admgts}}\neq\emptyset$ and $\solutionsof{\rightcharof{\admgts}}\neq \emptyset$, and returns finite sets~$X, Y$  of DMGTS so that
\begin{itemize}
\item[(a)] for all $\admgts'\in X$ we have faithfulness and $\admgts'\rankless\admgts$, 
\item[(b)] for all $\admgts'\in Y$ we have $\leftlangof{\admgts'}\separable\dycklangn{n}$,
\item[(c)] $\leftlangof{\admgts}=\leftlangof{X\cup Y}$
\end{itemize}
\end{lem}

We proceed with the proof of Lemma~\ref{Lemma:DecoupledDecomposition}, which assumes Lemma~\ref{Lemma:Refinement}.
\begin{proof}[Proof of Lemma~\ref{Lemma:DecoupledDecomposition}]
    We formulate the overall decomposition algorithm and afterwards reason about its correctness.
    The input to the decomposition is a faithful DMGTS $\admgts$.
    If $\admgts$ is perfect, then we return $\opensystems=\set{\admgts}, \decidedsystems=\emptyset$.  
    If $\solutionsof{\leftcharof{\admgts}}=\emptyset$, then we return $\opensystems=\decidedsystems=\emptyset$. 
    If $\solutionsof{\rightcharof{\admgts}}=\emptyset$, then we return $\opensystems=\emptyset$ and $\decidedsystems=\set{\admgts}$. 
    If these conditions do not apply, we invoke $\refineof{-}$ from Lemma~\ref{Lemma:Refinement} to generate sets $X$ and $Y$ of DMGTS with the stated properties. 
    We recursively call our decomposition algorithm on each DMGTS $\openmgts\in X$, which returns $\opensystems_{\openmgts}$ and $\decidedsystems_{\openmgts}$.
    We include all DMGTS from $(\opensystems_{\openmgts})_{\openmgts\in X}$ in $\opensystems$, and all DMGTS from $(\decidedsystems_{\openmgts})_{\openmgts\in X}$ and from $Y$ in $\decidedsystems$.

    We reason about correctness with an induction on the height of the call tree. 
    The tree is finite as every recursive call decreases the well-founded order, each state has a finite outdegree, and hence König's lemma applies.
    For a perfect DMGTS, there is nothing to do.
    If $\solutionsof{\leftcharof{\admgts}}=\emptyset$, then $\leftlangof{\admgts}=\emptyset$ follows and we are done. 
    If $\solutionsof{\rightcharof{\admgts}}=\emptyset$, the set $Y=\set{\admgts}$ preserves the language.
    Since $\solutionsof{\rightcharof{\admgts}}=\emptyset$ implies $\rightsolutions(\admgts)=\emptyset$, the required separability holds by the forward direction of Lemma~\ref{Lemma:PerfectSeparation}.

    In the induction step, the induction hypothesis and Lemma~\ref{Lemma:Refinement} show that for the DMGTS in $\decidedsystems$ the required separability holds. 
    The induction hypothesis moreover shows that the DMGTS in $\opensystems$ are perfect.  
    We have $\leftlangof{\admgts}=\leftlangof{\opensystems\cup\decidedsystems}$, because $\leftlangof{\admgts}=\leftlangof{X\cup Y}$ by Lemma~\ref{Lemma:Refinement} and, by the induction hypothesis, our decomposition  yields $\leftlangof{\openmgts}=\leftlangof{\opensystems_{\openmgts}\cup\decidedsystems_{\openmgts}}$ for all $\openmgts\in X$.
    
    We sketch out an argument for the upper bound.
    The concrete proof would follow the same steps as the proof of \cite[Theorem 5.4]{Leroux19}, which makes a similar analysis.
    Perfectness and infeasibility can be checked with time and space elementary in the size of the input DMGTS \cite{Lambert92}. 
    By Lemma~\ref{Lemma:Refinement}, also $\refineof{\admgts}$ is computable with resources elementary in $\sizeof{\admgts}$. 
    The well-founded relation $\rankless$ is defined as a lexicographic order on $\nat^{\sizeof{\leftside\cup\rightside}+1}$.
    Thus, we have a computation that incurs an elementary sized blow-up at each step, while terminating thanks to a rank of the order $\nat^{\sizeof{\leftside\cup\rightside}+1}$.
    Elementary functions are contained in the third level of the hierarchy of fast growing functions.
    We may combine these ranks \cite{Schmitz16}, and get a complexity upper bound of $\fof{\sizeof{\leftside\cup\rightside}+1+3}=\fof{\sizeof{\leftside\cup\rightside}+4}$.
\end{proof}

Now, we prove Lemma~\ref{Lemma:Refinement}.
This lemma deals with faithful but imperfect DMGTS.
A faithful DMGTS $\admgts$ is not perfect if and only if it contains a precovering graph $\precovering$ for which one of the following conditions holds.

\begin{itemize}[leftmargin=2em]
\item[(i)] There are $\aside\in\set{\leftside, \rightside}$, a counter $\aplace\in\aside$, and $\inoutmarking\in\set{\incounters, \outcounters}$ so that $\coordacc{\inoutmarking}{\aplace}=\omega$ but $\coordacc{\avar}{\precovering, \inout, \aplace}\not\in\supportof{\sidecharof{\aside}{\admgts}}$. 
\item[(ii)] There are a side $\aside\in\set{\leftside, \rightside}$ and an edge $\anedge\in\edgesof{\precovering}$ so that $\coordacc{\avar}{\anedge}\not\in\supportof{\sidecharof{\aside}{\admgts}}$.
\item[(iii)] We have $\coveringseqof{}{\precovering}=\emptyset$ or $\downcoveringseqof{}{\precovering}=\emptyset$.
\end{itemize}

Case (iii) is part of the perfectness definition.
The Cases (i) and~(ii) follow from the fact that $\supportof{\sidecharof{\aside}{\admgts}}$ should capture the unboundedness of $\aside$ in $\admgts$.
We break down the proof of Lemma~\ref{Lemma:Refinement} into three arguments, one for each case.

On a high level, in cases (i) and (ii), the function $\refineof{-}$ applies one step of the KLMST decomposition by considering only the $\leftside$ or the $\rightside$ side.
In case (iii), the function $\refineof{-}$ applies KLMST by considering the full MGTS that tracks both counters.
The actual decomposition resp.\ the proof diverge from this summary on certain points.
The most important divergence is the new proof obligation, faithfulness.
Maintaining this property takes different approaches in (ii) and (iii) compared to (i).
In (ii) and (iii), we develop a general framework that allows us to extend faithfulness when the decomposition unrolls the structure locally.
In (i), we need to argue faithfulness by hand, as markings introduced here do not result from merely simulating existing markings along transitions.
There are also two slight divergences in the construction, which otherwise adapt the KLMST decomposition.
First, in the case (i) we deduce new concrete markings and need to increase our modulo base, $\charmod_{\mathsf{new}}=k\cdot\charmod_{\mathsf{old}}$.
This makes the language $\leftlangof{\admgts}$ smaller, since the modulo reachability constraints become tighter.
To ensure (c), we need to make sure that we also construct DMGTS in $Y$ that captures the words kicked out by the new restriction.
In all cases (i), (ii), and (iii), applying the KLMST decomposition as we outlined above would disregard some runs that are intermediate $\rightside$-rejecting, but intermediate $\leftside$-accepting.
This is too strict for (c), which only considers $\leftside$-acceptance.
Preserving such words in $Y$ requires slight modifications to the construction.
Note that in both situations, we rely on $Y$, which gives us the obligation of showing separability from $\dycklangn{n}$.

\subsection{Case (i)}\label{SubSection:Case1}
Consider a faithful but imperfect DMGTS $\admgts = (\avas, \charmod)$ that satisfies $\solutionsof{\leftcharof{\admgts}}\neq\emptyset\neq\solutionsof{\rightcharof{\admgts}}$. 
In Case~(i), there is a precovering graph $\precovering$, a side $\aside\in\set{\leftside, \rightside}$, a counter $\aplace\in\aside$, and a counter valuation $\inoutmarking\in\set{\inmarkingof{\precovering}, \outmarkingof{\precovering}}$ so that 
$\coordacc{\inoutmarking}{\aplace}=\omega$ but $\coordacc{\avar}{\precovering, \inout, \aplace}\notin\supportof{\sidecharof{\aside}{\admgts}}$.
If the variable is not in the support, the set of values $\anyA=\setcond{\coordacc{\asol}{\precovering, \inout, \aplace}}{\asol\in\solutionsof{\sidecharof{\aside}{\admgts}}}$ is finite.
We also know $\anyA\neq\emptyset$, because $\solutionsof{\sidecharof{\aside}{\admgts}}\neq\emptyset$.  
Finally, we have $\anyA\subseteq\nat$ by the shape of $\sidecharof{\aside}{\admgts}$. 
We show how to construct $(X, Y)=\refineof{\admgts}$ as required by Lemma~\ref{Lemma:Refinement}. 

\subsubsection*{\bfseries Case $\aside=\leftside$}
Let $\avas_{a}$ be the MGTS that results from $\avas$ by changing
 the entry or exit value of counter~$\aplace$ in $\precovering$  from $\omega$ to $a\in\nat$. 
We define 
\begin{align*}
X\ = \ \setcond{(\avas_{a}, \charmod)}{a\in \leftA}\qquad\text{and}\qquad Y\ =\ \emptyset\ .
\end{align*}

\begin{proof}
We begin with Property~(c) in Lemma~\ref{Lemma:Refinement}.
The difference between $\admgts$ and $\admgtsnew=(\avas_{a}, \charmod)$ is a single entry or exit value that changes from~$\omega$ to $a$. 
This makes intermediate acceptance stricter, 
$\interacceptofshort{\leftside}{\admgtsnew}\subseteq\interacceptofshort{\leftside}{\admgts}$, and so we have $\leftlangof{X\cup Y}\subseteq\leftlangof{\admgts}$. 
For the reverse inclusion, we use that every run $\apath\in\interacceptofshort{\leftside}{\admgts}$ induces a solution to $\leftcharof{\admgts}$. 
Then $\apath$ enters or exits the precovering graph $\precovering$ with a value $a\in \leftA$ on counter~$\aplace$. 
For Property~(b), there is nothing to show as $Y=\emptyset$. 
For Property~(a), note that we do not modify the edges, states, or $\rightside$-valuations when moving from~$\avas$ to $\avas_a$. 
Hence, the faithfulness of $\admgtsnew=(\avas_a, \charmod)$ follows from the faithfulness of $\admgts$.
We reduce the well-founded order, because we replace say an exit value $\omega$ by a concrete value, while keeping $\edgesof{\precovering}$, $\infinitiesof{\precovering}$, and $\inmarkingof{\precovering}$ unchanged.
The complexity follows from the fact that the set $\leftA$ can be constructed using resources elementary in the size of $\admgts$ \cite{Lambert92}.
\end{proof}

\subsubsection*{\bfseries Case $\aside=\rightside$}
The construction uses an equivalence $\openmgts_{1}\markingequiv_{\charmod}\openmgts_{2}$ on MGTS. 
It is the least equivalence that satisfies the following. 
For precovering graphs, we have $\precovering_{1}\markingequiv_{\charmod}\precovering_{2}$ if the states, the edges, and the root coincide, and moreover $\coordacc{\inoutmarkingof{\precovering_{1}}}{\leftside}=\coordacc{\inoutmarkingof{\precovering_{2}}}{\leftside}$ and $\coordacc{\inoutmarkingof{\precovering_{1}}}{\rightside}\equiv\coordacc{\inoutmarkingof{\precovering_{2}}}{\rightside}\modulo \charmod$, for both, $\inout=\mathsf{in}$ and $\inout=\mathsf{out}$. 
For composed MGTS, we use
\begin{align*}
\infer{
    \openmgts_1.\anupdate.\openmgts_2\markingequiv_{\charmod} \openmgts_1'.\anupdate.\openmgts_2'}{\openmgts_1\markingequiv_{\charmod} \openmgts_1'\qquad \openmgts_2\markingequiv_{\charmod} \openmgts_2'}\ .
\end{align*}
Equivalent MGTS may only differ in the entry and exit values of Dyck counters, and these values still have to coincide modulo~$\charmod$.
As a piece of notation, we use $0\leq \openmgts< c$ with $c\in\nat$ to mean $\inmarkingof{\precovering'}$ and~$\outmarkingof{\precovering'}$ only take values from $[0, c-1]\cup\set{\omega}$, for all $\precovering'$ in $\openmgts$.  

To define $X$ and $Y$, we choose the least value $l$ that is larger than the maximal value in $\rightA$ and moreover larger than any entry or exit value in a precovering graph of $\avas$.
We set $\charmodnew=l\cdot\charmod$ and 
\begin{align*}
    Z\ &=\ \setcond{(\openmgts, \charmodnew)}{\openmgts\markingequiv_{\charmod}\avas_{a},\;
    0\leq a, \openmgts<\charmodnew,\;
    \coordacc{\inmarkingof{\openmgts}}{\rightside}=\zerovec}\\
    X\ &=\ \setcond{(\openmgts, \charmodnew)\in Z}{\coordacc{\outmarkingof{\openmgts}}{\rightside}=\zerovec}\\
        Y\ &=\ Z\setminus X \ .
\end{align*} 
    The MGTS $\avas_{a}$ is defined as in the case of $\aside=\leftside$.

We discuss three important points before turning to the proof.
We deliberately not only replace $\omega$ by values from $\rightA$, 
but by all values $0\leq a< \charmodnew$. 
The reason is that $\leftlangof{\admgts}$ also checks intermediate acceptance for the Dyck counters, but only modulo~$\charmod$. 
The set $\rightA$ is constructed from runs where the Dyck counters reach intermediate values precisely. 
This means $\rightA$ may not offer enough values for  
$\leftlangof{\admgts}\subseteq \leftlangof{X\cup Y}$ to hold.

The definition of $\charmodnew$ addresses the main challenge in this case, namely the faithfulness of the DMGTS $\admgtsnew=(\openmgts, \charmodnew)\in X$. 
We have to show that we reach the value $0\leq a<\charmodnew$ that replaces $\omega$, whenever we reach it modulo~$\charmodnew$. 
The idea is this.
The left-hand side of Inclusion~\eqref{Equation:Faithfulness} will allow us to show that the run belongs to $\interacceptof{\ints}{\rightside}{\admgts}$. 
A consequence is that it reaches $b\in \rightA\subseteq\nat$ with $b<\charmodnew$. 
We thus reach $a$ if we can show $b=a$. 
We use the following property of the modulo equivalence.
\begin{fact}\label{Lemma:ModuloTrick2}
Consider $\charmodnew\in\nat$. For all $\avar, k\in\nat$, we have that $\avar, k<\charmodnew$ and $\avar\equiv k\modulo\charmodnew$ together imply $\avar=k$. 
\end{fact}
\noindent The missing $b\equiv a\modulo\charmodnew$ is by $\interacceptof{\ints}{\restrictto{\omegalequiv{\charmodnew}}{\rightside}}{\admgtsnew}$, which is a premise of faithfulness. 
To sum up, we let $\charmodnew$ exceed the counter values that any run (satisfying the premise of faithfulness) can take (in the moment we use $a$ for $\omega$), and so we do not lose information by only tracking counter values modulo $\charmodnew$.
%

The change from $\charmod$ to $\charmodnew$ brings $\markingequiv_{\charmod}$ to the definition of $X$ and~$Y$. 
The purpose of the equivalence is to modify the entry and exit valuations of the Dyck counters in all precovering graphs. 
To see the need for a modification, note that such a valuation is a constraint of the form $x \equiv k\modulo \charmod$. 
Imagine now we multiply $\charmod = 3$ by $l=4$ and get $\charmodnew=12$. 
Assume $k=2$. 
To obtain all solutions to  $\avar \equiv 2\modulo 3$, it is not sufficient to consider $\avar \equiv 2 \modulo 12$.
We need to join the solutions to $\avar \equiv i\modulo 12$ for all $i\in \set{2, 5, 8, 11}$.
The reason these values $i$ collect all solutions, is the following property.
\begin{fact} \label{Lemma:ModuloTrick1}
Let $\charmod$ divide $\charmodnew$.
For all $\avar, k\in\ints$ with $\avar\equiv k\modulo\charmod$, there is $0\leq i<\charmodnew$ with $\avar\equiv i\modulo\charmodnew$ and $i\equiv k\modulo\charmod$.
\end{fact}
\noindent The equivalence $\markingequiv_{\charmod}$ incorporates all and only these choices of $i$. 
%
%
    
\begin{proof}
We prove Property~(c) in Lemma~\ref{Lemma:Refinement} and begin with the inclusion $\leftlangof{X\cup Y}\subseteq\leftlangof{\admgts}$.
Let $\admgtsnew=(\openmgts, \charmodnew)\in X\cup Y$.
By definition, $\openmgts\markingequiv_{\charmod}\avas_{a}$ for some $0\leq a<\charmodnew$.
We argue that 
\begin{gather*}
\interacceptofshort{\leftside}{\admgtsnew}\subseteq \interacceptofshort{\leftside}{\openmgts, \charmod}\subseteq \interacceptofshort{\leftside}{\avas_{a}, \charmod} \subseteq \interacceptofshort{\leftside}{\admgts}\ . 
\end{gather*}
The first inclusion is by the fact that  $\charmod$ divides $\charmodnew$. 
The next uses the fact that $\markingequiv_{\charmod}$ preserves the valuations of the counters in $\leftside$  and the valuations of the counters in $\rightside$  modulo $\charmod$, and so $\interacceptofshort{\leftside}{-}$ is invariant under this equivalence. 
The last inclusion is by the fact that concrete values make intermediate acceptance stricter. 

For the inclusion $\leftlangof{\admgts}\subseteq\leftlangof{X\cup Y}$, consider 
 $\apath\in\interacceptofshort{\leftside}{\admgts}$.
As we do not change the entry and exit valuations for the counters in $\leftside$, we readily have $\apath\in
\interacceptof{\leftside}{\restrictto{\omegaleq}{\leftside}}{\admgtsnew}$ for all $\admgtsnew\in X\cup Y$. 
What remains is to argue that  
$\apath\in\interacceptof{\rightside}{\restrictto{\omegalequiv{\charmodnew}}{\rightside}}{\admgtsnew}$ for some $\admgtsnew=(\openmgts, \charmodnew)\in X\cup Y$.
Let $\apath$ enter or leave the precovering graph $\precovering$ of interest with counter valuation $c$.
Let $a\in[0, \charmodnew-1]$ be such that $\coordacc{c}{\aplace}\equiv a\modulo \charmodnew$, where $\aplace$ is the counter of interest. 
By Fact~\ref{Lemma:ModuloTrick1}, there is $(\openmgts, \charmodnew)\in X\cup Y$ with $\openmgts\markingequiv_{\charmod}\avas_{a}$ for which the run is intermediate accepting. 
%

For the properties stated in~(b), let $\admgtsnew=(\openmgts, \charmodnew)\in Y$.
The definition of $Y$ yields $\coordacc{\inmarkingof{\openmgts}}{\rightside}=\zerovec$. 
Moreover, we know that $0<\coordacc{\outmarkingof{\openmgts}}{\aplace}<\charmodnew$ for some counter $\aplace\in\rightside$.  
The final $\rightside$ marking of $\admgtsnew$ is concrete, because $\admgts$ is zero-reaching by faithfulness. 
It is 
bounded by $\charmodnew$ due to $0\leq \openmgts<\charmodnew$. 
It is 
different from zero by the definition of $Y$. 
The analysis of the initial and final values shows that every $\apath\in\interacceptofshort{\leftside}{\admgtsnew}\subseteq\interacceptof{\ints}{\restrictto{\omegalequiv{\charmodnew}}{\rightside}}{\admgtsnew}$ 
has an effect $\aconf\not\equiv \zerovec\modulo\charmodnew$ on the Dyck counter $\aplace$. 
This yields $\leftlangof{\admgtsnew}\subseteq\bigcup_{v\not\equiv_{\charmodnew} 0}\modlangof{\charmodnew, v}$.
%
%

For Property~(a), the argument that the well-founded relation decreases is the same as in the case $\aside=\leftside$.
For faithfulness, consider $\admgtsnew=(\openmgts, \charmodnew)\in X$.  
It is zero-reaching by definition. 
The challenge is to prove Inclusion~\eqref{Equation:Faithfulness}. 
We reason as follows:
\begin{align}
\hspace{-0.23cm}\acceptof{\ints}{\rightside}{\admgtsnew}\cap \interacceptof{\ints}{\restrictto{\omegalequiv{\charmodnew}}{\rightside}}{\admgtsnew} &\subseteq 
\interacceptof{\ints}{\rightside}{\admgts}\label{Equation:FaithfulnessEasy}\\
\hspace{-0.23cm}\interacceptof{\ints}{\rightside}{\admgts}\cap\interacceptof{\ints}{\restrictto{\omegalequiv{\charmodnew}}{\rightside}}{\admgtsnew} &\subseteq 
\interacceptof{\ints}{\rightside}{\admgtsnew}. \label{Equation:FaithfulnessTrick}
\end{align}

The first inclusion says that any $\Z$-run in $\admgtsnew$ that reaches the extremal $\rightside$ markings precisely, and remaining $\rightside$ markings modulo $\charmodnew$, then this is an intermediate $\rightside$-accepting $\Z$-run in $\admgts$.
The second inclusion intersects the righthand side of the former inclusion with $\interacceptof{\ints}{\restrictto{\omegalequiv{\charmodnew}}{\rightside}}{\admgtsnew}$, and says that this is in $\interacceptof{\ints}{\rightside}{\admgtsnew}$.
When combined, we get the desired $\acceptof{\ints}{\rightside}{\admgtsnew}\cap \interacceptof{\ints}{\restrictto{\omegalequiv{\charmodnew}}{\rightside}}{\admgtsnew}\subseteq \interacceptof{\ints}{\rightside}{\admgtsnew}$.
We show these inclusions.

Inclusion~\eqref{Equation:FaithfulnessEasy} is a consequence of the Inclusions~\eqref{Equation:InclusionAcc} and~\eqref{Equation:InclusionModulo}, which allow us to invoke the faithfulness of $\admgts$:
\begin{align}
\acceptof{\ints}{\rightside}{\admgtsnew}&\subseteq \acceptof{\ints}{\rightside}{\admgts}\label{Equation:InclusionAcc}\\
\interacceptof{\ints}{\restrictto{\omegalequiv{\charmodnew}}{\rightside}}{\admgtsnew}&\subseteq \interacceptof{\ints}{\restrictto{\omegalequiv{\charmod}}{\rightside}}{\admgts}\ \label{Equation:InclusionModulo}.
\end{align}
The implication follows from intersecting the respective sides of the inequalities.
We get 
\begin{align*}
    \acceptof{\ints}{\rightside}{\admgtsnew}\cap\interacceptof{\ints}{\restrictto{\omegalequiv{\charmodnew}}{\rightside}}{\admgtsnew}&\subseteq \acceptof{\ints}{\rightside}{\admgts}\cap\interacceptof{\ints}{\restrictto{\omegalequiv{\charmod}}{\rightside}}{\admgts}\\
    &=\interacceptof{\ints}{\rightside}{\admgts}
\end{align*}
where the last equality uses the faithfulness of $\admgts$.

Inclusion~\eqref{Equation:InclusionAcc} holds, as we only change an intermediate valuation from $\admgts$ to $\admgtsnew$, and acceptance does not take the intermediate valuations into account.
To see that we only change an intermediate valuation, note that also $\admgts$ is zero-reaching by faithfulness.
Inclusion~\eqref{Equation:InclusionModulo} holds with the same argument as Property~(c) above.

For Inclusion~\eqref{Equation:FaithfulnessTrick}, let $\apath\in \interacceptof{\ints}{\rightside}{\admgts}\cap\interacceptof{\ints}{\restrictto{\omegalequiv{\charmodnew}}{\rightside}}{\admgtsnew}$. 
Consider the moment the run enters or exits the precovering graph of interest, and the counter $j$ whose value changes from~$\omega$ in~$\admgts$ to $0\leq a<\charmodnew$ in~$\admgtsnew$. 
By the intermediate acceptance $\apath \in \interacceptof{\ints}{\rightside}{\admgts}$, the run induces a solution to $\rightcharof{\admgts}$.
The consequence is that, at this moment in the run, counter $j$ has a value $b\in \rightA\subseteq\nat$. 
Moreover $b <\charmodnew$, because we chose $l$ larger than all values in $\rightA$. 
With $\apath \in \interacceptof{\ints}{\restrictto{\omegalequiv{\charmodnew}}{\rightside}}{\admgtsnew}$, we additionally get $b\equiv a\modulo \charmodnew$ .
Fact~\ref{Lemma:ModuloTrick2} applies and shows $a=b$. 

For the remaining precovering graphs, and the precovering graph~$\precovering$ but a counter different from $j$, we show intermediate acceptance as follows. 
Let $\admgts$ carry value $b\in\nat$ at the moment of interest. 
Note that $b<\charmodnew$ by the choice of $l$, namely larger than all values in~$\admgts$.   
In $\admgtsnew=(\openmgts, \charmodnew)$, we find a value $b'$ at this moment.
As $0\leq \openmgts<\charmodnew$, we know $0\leq b'<\charmodnew$. 
We have  $\apath\in\interacceptof{\ints}{\rightside}{\admgts}\cap\interacceptof{\ints}{\restrictto{\omegalequiv{\charmodnew}}{\rightside}}{\admgtsnew}$.
We are thus sure to reach $b$ and reach $b'$ modulo $\charmodnew$.
This allows us to conclude $b\equiv b'\modulo\charmodnew$ with $0\leq b, b'<\charmodnew$.
Fact~\ref{Lemma:ModuloTrick2} applies and yields the desired $b=b'$. 

We analyze the complexity.
Every counter in every valuation may be replaced by $\charmodnew=l\cdot\charmod$ many values.
This limits the number of generated DMGTS to ${\charmodnew}^{\sizeof{\admgts}}$. 
%
The DMGTS have a maximal size of $\charmodnew\cdot\sizeof{\admgts}$. 
As we argued in the case $\aside=\leftside$, the value $l$ itself is of size elementary in $\sizeof{\admgts}$.
We conclude that the whole procedure takes elementary resources.
\end{proof}
\subsection{Reasoning Locally about Faithfulness}\label{SubSection:LocalReasoning}
In Case~(i), we modified the entry and exit valuations of every precovering graph in a DMGTS. 
In the remaining two cases, we will decompose a single precovering graph, in a way that is independent of the context. 
We now develop techniques that allow us to reason locally about the one precovering graph, and lift the results to the overall DMGTS. 
The decomposition steps will closely follow their reachability counterparts~\cite{Leroux19}. 
The novel property we need to argue for is faithfulness, and this will be the focus here.

To make our local argument precise, we define MGTS contexts.
An \emph{MGTS context} is an MGTS in which a distinguished variable~$\contextvar$ occurs precisely once:
\begin{align*}
    \acontext{\contextvar}\;\;::=\;\;\contextvar\;\;\mid\;\; \acontext{\contextvar}.\anupdate.\admgts \;\;\mid\;\; \admgts.\anupdate.\acontext{\contextvar}\ .
\end{align*} 
We write $\acontext{\admgts}$ for the MGTS that is obtained by replacing~$\contextvar$ with the MGTS $\admgts$. 
When $\admgts$ is a DMGTS $(\openmgts, \charmod)$, we also write $\acontext{\admgts}$ to mean $(\acontext{\openmgts}, \charmod)$. 
A first observation is that the well-founded relation is preserved when comparable MGTS and DMGTS are inserted into contexts, $\admgts_1\refineprecision\admgts_2$  implies $\acontext{\admgts_1}\refineprecision\acontext{\admgts_2}$. 

With contexts at hand, in the remaining cases we will start from $(\acontext{\precovering}, \charmod)$ and decompose $(\precovering, \charmod)$ into sets of DMGTS~$U$ and~$V$. 
The sets needed for Lemma~\ref{Lemma:Refinement} are then $X=\acontext{U}=\setcond{\acontext{\openmgts}}{\openmgts\in U}$ resp. $Y=\acontext{V}$.  
We also have $Y=\emptyset$ in one case. 
To show the faithfulness of these DMGTS, we use the following arguments. 

We define a relation called \emph{consistent specialization} between DMGTS.
The relation is defined by structural induction, where $(\acontext{\openmgts_{1}}, \charmod)$ is a consistent specialization of $(\acontext{\openmgts_{2}}, \charmod)$, if this relation holds between $(\openmgts_{1}, \charmod)$ and $(\openmgts_{2}, \charmod)$. 
In the base case, $(\openmgts, \charmod)$ is a consistent specialization of~$(\precovering, \charmod)$ if the following two conditions hold.
\begin{itemize}[leftmargin=2em]  
\item[(1)] We have $\inmarkingof{\openmgts}\omegaleq\inmarkingof{\precovering}$, $\outmarkingof{\openmgts}\omegaleq\outmarkingof{\precovering}$, and for all runs $\apath\in\pathsof{\ints}{\openmgts}$ there is $\apathp\in\pathsof{\ints}{\precovering}$ with $\apathp\pathequiv\apath$.
\item[(2)] For all $\apath\in\interacceptof{\ints}{\restrictto{\omegalequiv{\charmod}}{\rightside}}{\openmgts}$ with $\coordacc{\coordacc{\apath}{\firstindex}}{\rightside}\omegaleq\inmarkingof{\precovering}$ and $\coordacc{\coordacc{\apath}{\lastindex}}{\rightside}\omegaleq\outmarkingof{\precovering}$, we have $\apath\in\interacceptof{\ints}{\rightside}{\openmgts} $.
\end{itemize}
Note that $\charmod$ has to coincide for DMGTS that are related by consistent specialization. 
Also note that we expect $\apathp\pathequiv\apath$ instead of $\apathp=\apath$.
The former allows visiting different states as long as the labels and configurations are the same (as defined in \Cref{Subsection:VASS}), while the latter requires both runs to visit the same states.
Equality is too restrictive for the coming decompositions, since they introduce new states.
%
%
What we actually need is the equality of the visited counter values, and the labelings of the runs, which is precisely what $\apath\pathequiv\apathp$ expresses.
%

Condition (1) expects that every run through $\openmgts$ can be mimicked by $\precovering$. 
With this, consistent specializations have smaller languages.
Condition (2) requires all runs that are (i) modulo-$\mu$ intermediate-accepting in $\openmgts$ and (ii) agree with the enterance and exit markings of $\precovering$ to also be intermediate-accepting in $\openmgts$.
Observe that this is almost the definition of faithfulness for $\openmgts$.
The difference is that faithfulness assumes $\apath\in\acceptof{\ints}{\rightside}{\openmgts}$ instead of (ii) we state here.
The reason behind this choice becomes clear when lifting faithfulness to a context.
We sketch this out in the following.
The argument starts from a consistent specialization $\openmgts$ of $\precovering$, a faithful $\acontext{\precovering}$, and aims to show that $\acontext{\openmgts}$ is faithful.
The faithfulness of $\acontext{\openmgts}$ has the premise $\apath\in\interacceptof{\ints}{\restrictto{\omegalequiv{\charmod}}{\rightside}}{\acontext{\openmgts}}\cap\acceptof{\ints}{\rightside}{\acontext{\openmgts}}$, and we need to deduce $\apath\in\interacceptof{\ints}{\rightside}{\acontext{\openmgts}}$.
We partition the run $\apath=\apath_{0}.\apath_{1}.\apath_{2}$, with $\apath_{1}$ being the part in $\openmgts$.
The key point in the proof is showing $\apath_{1}\in\interacceptof{\ints}{\rightside}{\openmgts}$, since the rest of the run goes through the context $\acontext{\contextvar}$.
Thanks to $\apath\in\interacceptof{\ints}{\restrictto{\omegalequiv{\charmod}}{\rightside}}{\acontext{\openmgts}}$, we have $\apath_{1}\in\interacceptof{\ints}{\restrictto{\omegalequiv{\charmod}}{\rightside}}{\openmgts}$.
Using (1), we get an equivalent run $\apath_{1}'\in\pathsof{\ints}{\precovering}$, and the faithfulness of $\acontext{\precovering}$ yields $\apath\in\interacceptof{\ints}{\rightside}{\acontext{\precovering}}$.
Thus, we get $\apath_{1}'\in\interacceptof{\ints}{\rightside}{\precovering}$.
By run-equivalence, we get $\apath_{1}[\firstindex][\rightside]\omegaleq\inmarkingof{\precovering}$ and $\apath_{1}[\lastindex][\rightside]\omegaleq\outmarkingof{\precovering}$.
Here, (2) applies to show $\apath_{1}\in\interacceptof{\ints}{\rightside}{\openmgts}$.
In contrast, the faithfulness of $\openmgts$ would need the premise $\apath_{1}\in\acceptof{\ints}{\rightside}{\openmgts}$, which is too strong because of the additional intermediate constraints.
Therefore, Conditions (1) and (2) together show that consistent specializations preserve faithfulness.
\begin{lem}\label{Lemma:ConsistentSpec}
    Let $\admgts_{1}$ be a consistent specialization of $\admgts_{2}$. 
    Then $\leftlangof{\admgts_{1}}\subseteq\leftlangof{\admgts_{2}}$ holds.
    Moreover, if $\admgts_{2}$ is faithful, so is $\admgts_{1}$.
\end{lem}

We give an intuition as to why the decompositions for the Cases~(ii) and~(iii) will guarantee Condition~(2).  
%
%
%
%
%
The decompositions unroll the precovering graph $\precovering$ into DMGTS. 
The intermediate counter valuations of these DMGTS correspond to the consistent assignment in $\precovering$. 
This is the first point in our construction, where we handle this component of precovering graphs, which we defined in \Cref{Subsection:MGTS}.
The precovering graph only admits runs that respects the consistent assignment. 
As a consequence, every run through the new DMGTS will satisfy intermediate acceptance.

\subsection{Case~(ii)}\label{SubSection:Case2}
This is the case where an edge $\anedge$ belongs to a precovering graph $\precovering$, 
and hence can be taken in loops, but the characteristic equations for 
the DMGTS $(\acontext{\precovering}, \charmod)$ impose an upper bound $l_{\anedge}\in\nat$ on the number 
of times the edge can be taken. 
Let $\edges'$ be the set of all such edges, with the largest imposed upper bound $l\in\nat$. 
The decomposition unrolls $\precovering$ into 
MGTS where every copy of an edge in $\edges'$ leads to a new precovering graph. 
The MGTS thus count the number of times the edges in $\edges'$ are taken. 
We cover the details of the construction in Section~\ref{SubSection:Case2Deets}.

We give the specification of this case in Lemma~\ref{Lemma:EdgeDecomposition} below.
It is \cite[Proposition 3.3]{Lambert92} with information about faithfulness and the DMGTS in $V$ added. 
\begin{lem}\label{Lemma:EdgeDecomposition}
    Consider $(\acontext{\precovering}, \charmod)$ and $\aside\in\set{\leftside, \rightside}$. 
    Let $\edges'\neq\emptyset$ be the subset of edges $\anedge'$ in $\precovering$ with $\coordacc{\avar}{\anedge'}\not\in\supportof{\sidecharof{\aside}{\acontext{\precovering}, \charmod}}$.
    Using resources elementary in $\sizeof{(\acontext{\precovering},  \charmod)}$, we can compute sets $U$ and $V$ of consistent specializations of $(\precovering, \charmod)$, where 
    \begin{itemize}[leftmargin=2em]    
    \item for all $\openmgts\in U$, we have $\openmgts\rankless(\precovering, \charmod)$, 
    \item for all $\apath\in\interacceptofshort{\leftside}{\acontext{\precovering}, \charmod}$ there is $\apathp\in\interacceptofshort{\leftside}{\acontext{U\cup V}}$ with $\apath\pathequiv\apathp$,
    \item for all $\decidedmgts\in V$, we have that $\sidecharof{\rightside}{\acontext{\decidedmgts}}$ is infeasible.
    \end{itemize}
\end{lem}

The DMGTS in the set $U$ only admit runs where each edge in $\edges'$ is taken at 
most $l$ times. 
The precovering graphs in these DMGTS only contain edges that are in the support. Here we should have a references, as in~\cite{Leroux19}, the edges in the support span a vector space that has a smaller dimension, hence the well-founded preorder decreases. 

The DMGTS in $V$ count until an edge $\anedge'$ in $\edges'$ has been taken $l+1$ times, and then 
admit all edges while returning to the former root.
Any solution 
to their characteristic equations can be translated into a solution 
$\asol$ to the characteristic equations of $(\precovering, \mu)$ with $\coordacc{\asol}{\anedge'} > l$ for some $\anedge'\in E'$.
When 
the elements of $V$ are inserted into the context, this makes the 
characteristic equation infeasible as the edge count is too high.


We define the decomposition $(X, Y)=\refineof{\admgts}$ of the faithful DMGTS $\admgts=(\acontext{\precovering}, \charmod)$ whose precovering graph $\precovering$ contains the set of edges~$\edges'\neq \emptyset$ outside $\supportof{\sidecharof{\aside}{\admgts}}$. 
With Lemma~\ref{Lemma:EdgeDecomposition}, we compute the sets  $U$ and $V$. 
If $\aside=\leftside$, we set $X=\acontext{U}$ and $Y=\emptyset$.
If $\aside=\rightside$, we set $X=\acontext{U}$ and $Y=\acontext{V}$.
This should be read as follows.
If the subject VASS can only execute the edges in~$\edges'$ a bounded number of times, we use the usual decomposition and do not create elements in $Y$. 
If the Dyck-side can only execute the edges a bounded number of times, we split the runs of the subject VASS.
The set $X$ contains the runs where the edges are taken a bounded number of times. 
The set~$Y$ contains the runs where the edges may occur more often,  
and we have the guarantee to be separable from the Dyck language by Lemma~\ref{Lemma:PerfectSeparation}. 

\begin{proof}
We prove the properties promised by Lemma~\ref{Lemma:Refinement}. 
For~(a), we note that not only the DMGTS in $X$ but also the ones in $Y$ are faithful by 
Lemmas~\ref{Lemma:EdgeDecomposition} and ~\ref{Lemma:ConsistentSpec}. 
The well-founded relation decreases by Lemma~\ref{Lemma:EdgeDecomposition}.
It is stable under forming contexts as noted above.  
For (b), if $\aside=\leftside$ there is nothing to do, because $Y=\emptyset$. 
%
%
If $\aside=\rightside$, Lemma~\ref{Lemma:EdgeDecomposition} already yields the infeasibility of $\rightcharof{\acontext{\decidedmgts}}$ for all $\decidedmgts\in V$.
With Lemma~\ref{Lemma:PerfectSeparation}, this implies the desired $\leftlangof{\acontext{\decidedmgts}}\separable\dycklangn{n}$. 
%
%
%
   
For (c), we have $\leftlangof{X\cup Y}\subseteq\leftlangof{\admgts}$ by Lemmas~\ref{Lemma:EdgeDecomposition} and~\ref{Lemma:ConsistentSpec}. 
For reverse inclusion, consider a word $\edgelabelof{\apath}$ with $\apath\in\interacceptofshort{\leftside}{\admgts}$.
As $\admgts=(\acontext{\precovering}, \charmod)$, we have $\apath=\apath_{0}.\apath_{1}.\apath_{2}$, where $\apath_{1}$ is the part of the run through $\precovering$. 
Intermediate acceptance propagates down to the components of the DMGTS, which yields $\apath_{1}\in\interacceptofshort{\leftside}{\precovering, \charmod}$. 
By Lemma~\ref{Lemma:EdgeDecomposition}, there are $\openmgts\in U\cup V$ and $\apathp\in\interacceptofshort{\leftside}{\openmgts}$ with $\apath_{1}\pathequiv\apathp$.
The equivalence among runs guarantees that the labels and counter values coincide, only the visited states may differ.
Hence, we have $\apath_{0}.\apathp.\apath_{2}\in\interacceptofshort{\leftside}{\acontext{\openmgts}}$ and $\edgelabelof{\apath_{0}.\apathp.\apath_{2}}=\edgelabelof{\apath_{0}.\apath_{1}.\apath_{2}}=\edgelabelof{\apath}$. 
%
%
If $\aside=\rightside$, we are done.
If $\aside=\leftside$, we must additionally show $\openmgts\not\in V$, because the MGTS in $V$ are dropped by the construction. 
%
We reason with infeasibility, like we did for (b).
\end{proof}
\subsection{Preliminaries for Detailed DMGTS Constructions}
Before moving on to the detailed construction for Lemma~\ref{Lemma:EdgeDecomposition}, we formalize the notion of locally unrolling of (D)MGTS in the following with the definition of observers.

An \emph{observer} $\anobserver=(\states, \edgesof{\precovering}, \transitions)$ for $\precovering$ is a transition system that has the edges $\edgesof{\precovering}$ as its alphabet.
The language $\langof{\anobserver, \initstates, \finalstates}\subseteq\edgesof{\precovering}^{*}$ of an observer $\anobserver$ is defined relative to a set of final and initial states $\finalstates, \initstates\subseteq\anobserver.\states$.
We have $\aword\in\langof{\anobserver, \initstates, \finalstates}$ if $\aword$ labels a run that reaches $\finalstates$ from $\initstates$.
We underline that $\aword\in\precovering.\edges^{*}$.
The decompositions have the following form.
By constructing a product of an observer and a precovering graph, we get a larger observer that tracks the information we need. 
We then decompose this automaton into MGTS, in accordance with the valuations of $\precovering$.

We define the product $\precovering\times\anobserver$ between an observer $\anobserver=(\states, \edgesof{\precovering}, \transitions)$ and a precovering graph $\precovering$ to be the observer $(\nodesof{\precovering}\times\states, \edgesof{\precovering}, \transitions_{\times})$, where 
%
\begin{align*}
    \transitions_{\times}=\set{((\astate_0, \astate_1), &\anedge,(\astatep_0, \astatep_1))\mid \\ 
    &\anedge\in\edgesof{\precovering}, \anedge=(\astate_0, \aletter, x, \astatep_0), (\astate_1, \anedge, \astatep_1)\in\transitions}.
\end{align*}
Intuitively, $\precovering\times\anobserver$ simulates $\anobserver$ along the edges $\precovering$ can take during a run.
We construct MGTS along the runs in $\precovering\times\anobserver$.
Towards this construction, we define the precovering graphs below.
Namely, for each state $\astate$ of $\precovering\times\anobserver$, we define the precovering graph 
\begin{align*}
    \precovering_{\astate}^{\anobserver}&=(\avas_{\astate}^{\anobserver}, (\astate, \bal_{\precovering}(\astate)), (\astate, \bal_{\precovering}(\astate)), \bal_{\precovering})\\
    \avas_{\astate}^{\anobserver}&=(\sccof{\astate}, \dyckalphn{\dyckdims}, \leftside\cup\rightside, \edges_{\astate}^{\anobserver})
\end{align*}
where $\sccof{\astate}$ is the strongly connected component of $\astate$ in $\precovering\times\anobserver$, and $E_{\astate}^{\anobserver}$ is the set of edges $(\astatep, \aletter, x, \astatep')$ for which $\astatep, \astatep'\in\sccof{\astate}$, $(\astatep, \anedge, \astatep')\in\transitions_{\times}$ hold for some $\anedge=(\astatepp, \aletter, x, \astateppp)\in\edgesof{\precovering}$, and $\bal_{\precovering}(\astate, \astatep)=\precovering.\bal(\astate)$ for $(\astate, \astatep)\in\nodesof{\precovering}\times\states$.
We define the MGTS 
$$\observedmgts{\anobserver}{\apath, \mathsf{raw}}=\precovering_{\astate_{1}}^{\anobserver}.\anupdate_{1}\ldots\anupdate_{k-1}.\precovering_{\astate_{k}}^{\anobserver}$$
for each run $\apath=\astate_{1}.(\astate_{1}, \anedge_{1}, \astate_{1})\ldots (\astate_{k-1}, \anedge_{k-1}, \astate_{k}).\astate_{k}$ that starts from, and reaches root in $\precovering\times\anobserver$, where $\anupdate_{i}=(\aletter_{i}, x_{u})$ and $\anedge_{i}=(\astate_{i}, \aletter_{i}, x_{i}, \astate_{i+1})$ for all $i\leq k-1$.
To keep the valuations of $\precovering$ and $\observedmgts{\anobserver}{\apath}$ coherent, we define the MGTS $\observedmgts{\anobserver}{\apath}$, which is obtained from $\observedmgts{\anobserver}{\apath, \mathsf{raw}}$ by modifying the initial valuation, and the final valuation so that $\inmarkingof{\observedmgts{\anobserver}{\apath}}=\inmarkingof{\precovering}$ and $\outmarkingof{\observedmgts{\anobserver}{\apath}}=\outmarkingof{\precovering}$ hold.
Note that this modification is compatible with the consistent assignments of the first and the last precovering graphs of $\observedmgts{\anobserver}{\apath}$. 
This is a consequence of the precovering graph definition.

There are infinitely many runs in $\precovering\times\anobserver$.
Since we need a finite set for our decomposition, we define $\basicacceptof{\precovering\times\anobserver, \initstates, \finalstates}$ to be the set of accepting runs in $\precovering\times\anobserver$ that go from the states in $\initstates$ to the states in $\finalstates$ without repeating a state.
We are now ready to define the general form of our decomposition.
We write 
$$\decompalongof{\precovering, \charmod, \anobserver, \initstates, \finalstates}=\setcond{(\observedmgts{\anobserver}{\apath}, \charmod)}{\apath\in\basicacceptof{\precovering\times\anobserver, \initstates, \finalstates}}.$$

The form of the decomposition gives us guarantees that will be useful for proving Lemma~\ref{Lemma:EdgeDecomposition} and Lemma~\ref{Lemma:CoveringDecomposition}.
First consequence is that all DMGTS obtained this way are consistent specializations.
This is because we do not introduce any new counter constraints that were not implied by the already existing ones.
 
\begin{lem}\label{Lemma:ObserverConsistentSpec}
    Let $(\precovering, \charmod)$ be a precovering graph, $\anobserver$ an observer, $\initstates, \finalstates\subseteq\anobserver.\states$, and $\admgts\in\decompalongof{\precovering, \charmod, \anobserver, \initstates, \finalstates}$.
    Then, $\admgts$ is a consistent specialization of $\precovering$.
\end{lem}

Second consequence allows us to make assumptions about the runs accepted by $\openmgts\in\decompalongof{\precovering, \charmod, \anobserver, \initstates, \finalstates}$, and find a corresponding run in $\precovering$.

\begin{lem}\label{Lemma:ObserverInformation}
    Let $(\precovering, \charmod)$ be a precovering graph, $\anobserver$ an observer, and $\initstates, \finalstates\subseteq\anobserver.\states$.
    If $\apath\in\pathsof{\ints}{\precovering}$ takes the edges $\anedge_{1}\ldots\anedge_{l-1}\in\langof{\anobserver, \initstates, \finalstates}$, then there is a $\admgts\in\decompalongof{\precovering, \charmod, \anobserver, \initstates, \finalstates}$ with $\apathp\in\pathsof{\ints}{\admgts}$ and $\apathp\pathequiv\apath$.
\end{lem}

We lift the observation to intermediate $\leftside$-accepting runs.
The observation follows from Lemma~\ref{Lemma:ObserverInformation} and the construction of the valuations for DMGTS in $\decompalongof{\precovering, \charmod, \anobserver}$.
\begin{fact}\label{Remark:ObserverLanguage}
    Let $(\precovering, \charmod)$ be a precovering graph, $\anobserver$ an observer, and $\initstates, \finalstates\subseteq\anobserver.\states$.
    If $\apath\in\interacceptofshort{\leftside}{\precovering}$ takes the edges $\anedge_{1}\ldots\anedge_{l-1}\in\langof{\anobserver, \initstates, \finalstates}$, then there is a $\admgts\in\decompalongof{\precovering, \charmod, \anobserver, \initstates, \finalstates}$ with $\apathp\in\interacceptofshort{\leftside}{\admgts}$ and $\apathp\pathequiv\apath$.
\end{fact}

Lastly, we introduce a notational shorthand.
In both decompositions, we track a value as long as it is between $0$ and a bound $l\in\N$, and go into a sink state when this interval is left.
To express this succinctly, we define the abstracting function $\ctabstr{l}:(\Z_{\omega}\cup\bot)\to \set{\bot}\cup [l]_{\omega}$ for $l\in\N$, where 
$$[l]_{\omega}=[l]\cup\set{\omega}\qquad [l]=\set{0, 1, \ldots, l}\qquad\ctabstrof{l}{x}=\begin{cases} \bot, &x<0\\ \omega, &x>l\\ x,&\text{else.}\end{cases}$$
For all $x\in\Z_{\omega}\cup\set{\bot}$, we assume $\omega+x=x+\omega=\omega$ and $\bot+x=x+\bot=\bot$ if $x\neq\omega$.
We also extend $\ctabstr{l}$ to vectors component-wise.
Namely, we let $\ctabstrof{l}{x}\in ([l]_{\omega}\cup\set{\bot})^{d}$ with $\ctabstrof{l}{x}[i]=\ctabstrof{l}{x[i]}$ for all $x\in\Z^{d}$ and $i\leq d$.

%
%

\subsection{Case~(ii), Construction Details}\label{SubSection:Case2Deets}
We now prove Lemma~\ref{Lemma:EdgeDecomposition}.
The premise of Lemma~\ref{Lemma:EdgeDecomposition} gives us a precovering graph $\precovering$, $\mu\geq 1$, an MGTS context $\acontext{\contextvar}$, a set of counters $\aside\in\set{\leftside, \rightside}$, and an edge $\anedge\in\edgesof{\precovering}$, where $\anedge\not\in\supportof{\sidecharof{\aside}{\acontext{\precovering}, \charmod}}$.
We consider the set of all such edges $\edges\subseteq\edgesof{\precovering}$.
Because $\anedge\not\in\supportof{\sidecharof{\aside}{\acontext{\precovering}, \charmod}}$ for all $\anedge\in\edges'$, the set $A_{\aside}=\setcond{\coordacc{\asol}{\anedge}}{\anedge\in\edges',\;\asol\in\solutionsof{\sidecharof{\aside}{\acontext{\precovering}, \charmod}}}$ is finite, and computable with elementary resources.
In particular, we can compute an $l\in\nat$ with $A_{\aside}\leq l$.
In order to fulfill the conclusion of this lemma, we track how many times some $\anedge\in\edges'$ is taken with an observer.
If a run in $\admgts$ takes such an edge at least $l+1$ times in $\precovering$, then we know that it cannot belong to $\interacceptofshort{\aside}{\admgts}$.
This guarantees the infeasibility of $\sidecharof{\aside}{\admgts'}$ for $\admgts'=(\acontext{\openmgts}, \charmod)$, where the MGTS $\openmgts$ makes sure some $\anedge\in\edges'$ is taken at least $l+1$ times. 
We capture such $(\openmgts, \charmod)$ in $V$.
Note that $\rankless$ is not required to decrease in this case.
We also generate DMGTS that make sure that each $\anedge\in\edges'$ is taken $i_{\anedge}\leq l$ times.
Such DMGTS consist of precovering graphs $\precovering'$ whose edges are from $\edgesof{\precovering}\setminus\edges'$.
In this case, $\edgesof{\precovering}\setminus\edges'=\supportof{\sidecharof{\aside}{\admgts}}\cap\edgesof{\precovering}$ guarantees $\cyclespaceof{\precovering'}\subset\cyclespaceof{\precovering}$.
The argument, which can be found in the proof of \cite[Claim 4.7]{Leroux19}, is this.
If $\cyclespaceof{\precovering'}=\cyclespaceof{\precovering}$ held instead, we could realize some multiple of any edge effect $\effectof{\anedge}$ for $\anedge\in\edges'$  by adding or taking away applications of edges in $\edgesof{\precovering}\setminus\edges'$.
By this, we mean $\effectof{\precovering.\edges} \cdot v = k\cdot \effectof{\anedge}$ with $v[\edges']=0$ for a number of applications $v\in\Z^{\precovering.\edges}$ and $k\in\N_{\geq 1}$.
This allows us to insert $\anedge$ to any large enough support solution.
Let $\asol$ be such a solution with $\asol[\anedge']\geq 1 + \sum_{i\leq k}|v[i]|$ for all $\anedge'\in\edgesof{\precovering}$.
By adding $k\cdot\unitvecn{\anedge}-v$ to $\asol[\precovering.\edges]$ to get $\asol'$, we do not change the total effect, get $\asol'[\anedge]\geq 1$, and $\asol'[\precovering.\edges]\geq 0$. 
Then, $\asol'$ is also a support solution.
But this implies that all $\anedge\in\edges'$ are in support, which is a contradiction. 

We proceed with the formal construction of the observer.
$$\anobserver=([l]_{\omega}^{\edges'}, \edgesof{\precovering}, \transitions)$$ 
where
\begin{align*}
    \transitions=\setcond{(x, \anedge', x)}{x\in [l]_{\omega}^{\edges'} \anedge'\not\in\edges'}
    \cup\setcond{(x, \anedge, \ctabstrof{l}{x+\unitvecn{\anedge}})}{\anedge\in\edges'}.
\end{align*}
We define $\initstates=\set{\zerovec}$ and the initial state, and $\finalstates_{\ulabel}=[l]^{\edges'}$, $\finalstates_{\vlabel}=[l]_{\omega}^{\edges'}\setminus\finalstates_{\ulabel}$ as the final states we are interested in.
As we previously hinted at, $\anobserver$ counts how many times some $\anedge\in\edges'$ is taken.
Runs reaching $\finalstates_{\ulabel}$ respect the bound on the edges $\anedge\in\edges'$, and the remaining runs, those reaching $\finalstates_{\vlabel}$, do not respect the bound. 
We define 
$$U=\decompalongof{\precovering, \charmod, \anobserver, \initstates, \finalstates_{\ulabel}}\qquad V=
\begin{cases}
\decompalongof{\precovering, \charmod, \anobserver, \initstates, \finalstates_{\vlabel}},\;&\aside=\rightside\\
\emptyset,\; &\aside=\leftside
\end{cases}
$$
We proceed with the proof of the first property in Lemma~\ref{Lemma:EdgeDecomposition}.
First, we make an observation.
For any cycle of edges in $\precovering\times\anobserver$, which could be extended to reach $\finalstates_{\ulabel}$, we can find a cycle from the edges in $\edgesof{\precovering}\setminus\edges'$.
This is because taking an edge $\anedge\in\edges'$ at the state $(\astate, x)$ increases the counter irrevokably, unless $x=\omega$.
If $x\neq\omega$, then no cycle originating from $(\astate, x)$ in $\precovering\times\anobserver$ may take $\anedge$.
If $x=\omega$, then no path from $(\astate, x)$ can reach $\finalstates_{\ulabel}=\states\times[l]^{\edges'}$.
By construction, a cycle in a precovering graph $\precovering'$ that occurs in $U$, (i) has a corresponding cycle in $\precovering\times\anobserver$, and (ii) this cycle be extended to a run that reaches $\finalstates_{\ulabel}$.
As discussed, we refer to the arguments in the proof of \cite[Claim 4.7]{Leroux19} to obtain $\cyclespaceof{\precovering'}\subset\cyclespaceof{\precovering}$ for all $\openmgts\in U$ and $\precovering'$ in $\openmgts$.
The strict inclusion $\cyclespaceof{\precovering'}\subset\cyclespaceof{\precovering}$ implies that if $\cyclespaceof{\precovering}$ is $i$-dimensional, then $\cyclespaceof{\precovering'}$ is at most $i-1$ dimensional.
So $\coordacc{\rankingof{\precovering}}{d-i}>0$, while $\coordacc{\rankingof{\precovering'}}{d-i'}=0$ for all $i'\leq \sizeof{\leftside\cup\rightside}$ with $i\leq i'$, where $d=\sizeof{\leftside\cup\rightside}$.
Since $\openmgts$ consists of such $\precovering'$, we get $\coordacc{\rankingof{\precovering'}}{d-i'}=0$ for all $i'\leq \sizeof{\leftside\cup\rightside}$ with $i\leq i'$ as well.
We conclude $\rankingof{\openmgts}\rankless(\precovering,\charmod)$.

Now, we show the third property.
Let $(\openmgts, \charmod)\in V$.
Since $V=\emptyset$ if $\aside=\leftside$, we write $\aside=\rightside$.
Towards a contradiction, suppose $\asol\in\solutionsof{\sidecharof{\rightside}{\acontext{\openmgts, \charmod}}}$.
We sketch the construction of $\asol'\in\solutionsof{\sidecharof{\rightside}{\acontext{\precovering, \charmod}}}$ with $\coordacc{\asol'}{\anedge}\geq l+1$ for some $\anedge\in\edges'$, which contradicts the definition of $A_{\rightside}$.
Let $\openmgts=\precovering_{1}.\anupdate_{1}\ldots\anupdate_{r-1}.\precovering_{r}$ for some precovering graphs $\precovering_{1},\ldots, \precovering_{r-1}$ and updates $\anupdate_{1}\ldots\anupdate_{r-1}$ corresponding to edges $\anedge_{1}, \ldots, \anedge_{r-1}$.
As we argued before, any cycle in $\precovering\times\anobserver$ (and thus $\precovering_{i}$ for $i\leq r$) has a cycle in $\precovering$ with the same effect on the counters.
Furthermore, the definitions of $\precovering\times\anobserver$ and $\decompalongof{\precovering, \charmod, \anobserver}$ ensure that for $\anedge_{1}\ldots\anedge_{r}\in\basicacceptof{\precovering\times\anobserver}$, there is a cycle in $\precovering$ with the same effect that takes some $\anedge\in\edges'$ at least $l+1$ times.
Exchanging the values that stem from the edges of $\openmgts$ in $\asol$ by the sum of the Parikh images of these same effect cycles in $\precovering$, we get $\asol'\in\solutionsof{\sidecharof{\rightside}{\acontext{\precovering, \charmod}}}$ with $\coordacc{\asol'}{\anedge}\geq l+1$.
This contradicts $A_{\rightside}\leq l$.

Finally, we prove the second property in Lemma~\ref{Lemma:EdgeDecomposition}.
If $\aside=\rightside$, all runs are accounted for in $U\cup V$ by $\finalstates_{\ulabel}\cup\finalstates_{\vlabel}=\anobserver.\states$.
Let $\aside=\leftside$, and $\apath\in\interacceptofshort{\leftside}{\acontext{\precovering}, \mu}$.
We consider the factorization $\apath=\apath_{0}.\apath_{1}.\apath_{2}$, where $\apath_{0}.\contextvar.\apath_{2}$ is the part of the run in $\acontext{\contextvar}$, and $\apath_{1}$ the part of the run in $\precovering$.
We claim that the sequence of edges taken in $\apath_{1}$ are in $\langof{\anobserver, \initstates, \finalstates_{\ulabel}}$.
Towards a contradiction, suppose this is not the case. 
Then, the sequence must be in $\langof{\anobserver, \initstates, \finalstates_{\vlabel}}$.
Similarly to the proof of the third property, this would allow us to derive a solution $s'\in\solutionsof{\sidecharof{\rightside}{\acontext{\openmgts, \charmod}}}$.
This is a contradiction.
Then, the sequence is in $\langof{\anobserver, \initstates, \finalstates_{\ulabel}}$, which implies a run $\apath_{1}'\in\interacceptofshort{\leftside}{U}$ with $\apath_{1}\pathequiv\apath_{1}'$ by Remark~\ref{Remark:ObserverLanguage}.
We thus have $\apath_{0}.\apath_{1}.\apath_{2}\pathequiv\apath_{0}.\apath_{1}'.\apath_{2}\in\interacceptofshort{\leftside}{\acontext{U}}$.

\subsection{Case~(iii)}\label{SubSection:Case3}
This is the case where a precovering graph $\precovering$ does not have a covering sequence to arbitrarily increase or decrease the values of $\omega$-decorated counters. 
As $\downcoveringseqof{}{\precovering}$ is defined via $\coveringseqof{}{\precovering}$, we focus on $\coveringseqof{}{\precovering}=\emptyset$. 
We follow the construction by Leroux and Schmitz~\cite{Leroux19}, which employs a Rackoff argument~\cite{Rackoff78}, rather than the one by Lambert~\cite{Lambert92}, which works with coverability graphs~\cite{KM69}.

For each $\aplace\in\leftside\cup\rightside$ that is decorated by $\omega$ and has $\coordacc{\inmarkingof{\precovering}}{\aplace}\in\nat$, we unroll the precovering graph into a DMGTS that tracks $\aplace$ up to a bound $B$.
%
%
The bound $B$ is of size doubly exponential in $\sizeof{\precovering}$.
By the same Rackoff argument as in \cite{Leroux19}, we conclude that this captures all words in $\leftlangof{\precovering}$, because the opposite would imply $\coveringseqof{}{\precovering}\neq\emptyset$.
The graph has one peculiarity compared to \cite{Leroux19}.
Assume a Dyck counter has not yet exceeded $B$ and becomes negative.
Then we enter a sink state in which we enable all transitions.

The specification is given in Lemma~\ref{Lemma:CoveringDecomposition} below. 
It is based on \cite[Propositions 3.4 and 3.5]{Lambert92}.

\begin{lem}\label{Lemma:CoveringDecomposition}
    Let $(\precovering, \charmod)$ have $\coveringseqof{}{\precovering}=\emptyset$ or $\downcoveringseqof{}{\precovering}=\emptyset$.
    Using resources elementary in $\sizeof{(\precovering, \charmod)}$, we can compute sets $U$ and $V$ of consistent specializations of $(\precovering, \charmod)$, where 
\begin{itemize}[leftmargin=2em]  
    \item for all $\openmgts\in U$, we have $\openmgts\rankless(\precovering, \charmod)$, 
    \item for all $\apath\in\interacceptofshort{\leftside}{\precovering, \charmod}$ there is $\apathp\in \interacceptofshort{\leftside}{U\cup V}$ with $\apath\pathequiv\apathp$,
    \item for all $\decidedmgts\in V$, we have that $\rightcharof{\decidedmgts}$ is infeasible.
\end{itemize}
\end{lem}

The DMGTS in $U$ capture the runs that do not enter the sink and reach $\outmarkingof{\precovering}$.  
Since cycles cannot change counter $\aplace$, the dimension of the vector space decreases for each precovering graph.
This reduces the well-founded preorder.
We discuss the details of the construction in Section~\ref{SubSection:Case3Deets}.

The DMGTS in $V$ capture the runs that enter the sink and the runs that end in a valuation different from $\outmarkingof{\precovering}$.
In both cases, the characteristic equations for $\rightside$ become infeasible. 
Indeed, when a Dyck counter becomes negative, the construction forces us to leave a precovering graph towards the sink.  
But then we fail to satisfy the non-negativity requirement for entering the sink.

%
%

We explain the role of the DMGTS in $V$ that enter the sink as a Dyck counter becomes negative. 
How can they help with the second property, if the requirement there is that the Dyck counters stay non-negative? 
The observation is that intermediate acceptance modulo~$\charmod$ satisfies a monotonicity property: if $\apath\in \interacceptof{\ints}{\restrictto{\omegalequiv{\charmod}}{\rightside}}{\decidedmgts}$ then $\apath + k\cdot \charmod\in \interacceptof{\ints}{\restrictto{\omegalequiv{\charmod}}{\rightside}}{\decidedmgts}$. 
Here, we use $\apath + k\cdot \charmod$ to denote the run obtained from $\apath$ by raising the values of all Dyck counters in all configurations by $k\cdot \charmod$ with $k\in\nat$. 
%
%
A consequence of monotonicity is that even though a run starting from small  values has to enter the sink due to a negative counter, this negativity will disappear in properly scaled runs, and they will still take the same transitions. 
The set $V$ makes sure we do not miss the scaled runs.

We define $(X, Y)=\refineof{\admgts}$ for $\admgts=(\acontext{\precovering}, \charmod)$ faithful with 
$\coveringseqof{}{\precovering}=\emptyset$ or $\downcoveringseqof{}{\precovering}=\emptyset$. 
Lemma~\ref{Lemma:CoveringDecomposition} yields sets $U$ and~$V$ of consistent specializations. 
We set $X=\acontext{U}$ and~$Y=\acontext{V}$.
The requirements of Lemma~\ref{Lemma:Refinement} are derived like in Case~(ii).

%
%

\subsection{Case (iii), Construction Details}\label{SubSection:Case3Deets}

Before we move on to the decomposition, we need the observation formulated in Lemma~\ref{Lemma:Rackoffesque}.
This is an adaptation of Lemma A.1 from the appendix of \cite{Leroux19}.
For a set of counters $\countersp\subseteq\leftside\cup\rightside$, our version of the observation states the following.
If we have an integer run, where each counter (i) exceeds some doubly exponential bound in $C$, the number of counters $d$, and the largest effect of a transition $l$ along the run and (ii) remains positive until this happens, then there is an $\countersp$-run that reaches a value that covers $C$ in all counters $i\in\countersp$.
Our version of the observation differs from the one used in \cite[Lemma A.1]{Leroux19}, in that we allow counters to become negative after exceeding the bound.

\begin{lem}\label{Lemma:Rackoffesque}
    Let $\precovering$ be a precovering graph with the largest transition effect $l$, $C\geq 2$, $\countersp\subseteq\leftside\cup\rightside$, and $\apath\in\pathsof{\ints}{\precovering}$ 
    with $$\apath=(\astate_{1}, \aconf_{1})\anedge_{1}\ldots\anedge_{k-1}(\astate_{k}, \aconf_{k}).$$
    If for all $\aplace\in\countersp$, there is an $i\leq k$ with $\coordacc{\aconf_{i}}{\aplace}\geq(\sizeof{\nodesof{\precovering}}\cdot l\cdot C)^{\sizeof{\countersp}+1!}$, and $\coordacc{\aconf_{i'}}{\aplace}\geq 0$ for all $i'\leq i$, then there is a run $\apathp\in\pathsof{\countersp}{\precovering}$ of size at most $(\sizeof{\nodesof{\precovering}}\cdot l\cdot C)^{\sizeof{\countersp}+1!}$ where $\coordacc{\apathp}{\lastindex}=(\astate_{k}, \aconf)$, $\coordacc{\apathp}{\firstindex}=(\astate_{1}, \aconf_{1})$, and $\coordacc{\aconf}{\aplace}\geq C$ for all $\aplace\in\countersp$.
\end{lem}

The proofs is very similar to \cite[Lemma A.1]{Leroux19}, so we only sketch out the Rackoff inspired argument.
The proof is by induction on $|\countersp|$.
It relies on functions $f, g, r:\N\times\N\to\N$ to denote bounds, whose precise definitions are not important.
The statement we need to prove is this.
If there is a run $\apath$ from state $\astate$ to state $\astatep$, where for each counter $i\in\countersp$, it exceeds the bound $f(\sizeof{\countersp}, C)$ on this counter before becoming negative; then there is a run $\apath_{*}$ from $\astate$ to $\astatep$ of size at most $g(\sizeof{\countersp}, C)$ such that we reach a configuration where each counter $i\in\countersp$ exceeds $C$.
The base case is trivial. 
For the inductive case relies consider the suffix $\apath'$ after the first configuration where $f(\sizeof{\countersp}, C)$ is exceeded.
We may assume that this configuration appears in $r(\sizeof{\countersp}, C)$ steps by removing duplicate configurations.
Let $\apath'$ reach the state $\astate$.
Then, we apply the induction hypothesis, and construct a run $\apath''$ of size $g(\sizeof{\countersp}-1, C)$ that reaches a configuration with state $\astate$, where all counters $j\in\countersp\setminus\set{i}$ are larger than $C$.
But, $f(\sizeof{\countersp}, C)$ is chosen large enough to compensate for a run of this size, so replacing $\apath'$ with $\apath''$ yields the desired run $\apath_{*}$.
It has a size of at most $r(\sizeof{\countersp}, C)+g(\sizeof{\countersp}-1, C)$, which is defined to be less than $g(\sizeof{\countersp}, C)$.
By choosing fitting $f$, $g$, and $r$, we get the bounds given in the lemma.

We move on to the proof of Lemma~\ref{Lemma:CoveringDecomposition}.
The premise of Lemma~\ref{Lemma:CoveringDecomposition} yields a precovering graph $\precovering$, $\mu\geq 1$, with $\coveringseqof{}{\precovering, \mu}=\emptyset$ or $\downcoveringseqof{}{\precovering, \mu}=\emptyset$.
We only consider $\coveringseqof{}{\precovering, \mu}=\emptyset$, as the case of $\downcoveringseqof{}{\precovering, \mu}=\emptyset$ is symmetrical.
We will adapt the proof of \cite[Lemma 4.13]{Leroux19}.
However, to be able to apply the arguments in \cite[Lemma 4.13]{Leroux19}, we must deal with the fixed counters of $\precovering$.
In \cite{Leroux19}, they are handled in a cleaning step.
Fixed counters are counters where no cycle in $\precovering$ changes the valuation of this counter.
For $\precovering$, let the set of fixed counters be $\countersp$.
Formally, 
$$J=\setcond{\aplace\in\leftside\cup\rightside}{\text{for all cycles }\aloop\text{ in }\precovering, \coordacc{\effectof{\aloop}}{\aplace}= 0}.$$

Fixed counters allow us to derive consistent assignments.
For any fixed counter $\aplace\in\countersp$, there is a unique function $\fixedassig{\aplace}:\nodesof{\precovering}\to\ints$, where $\fixedassigof{\aplace}{\precovering.\groot}=\coordacc{\inmarkingof{\precovering}}{\aplace}$, and  $\fixedassigof{\aplace}{\astate}+\coordacc{\effectof{\anedge}}{\aplace}=\fixedassigof{\aplace}{\astatep}$ for all $\astate, \astatep\in\nodesof{\precovering}$ and $\anedge\in\edgesof{\precovering}$ between these states. 
This unique function can be constructed by setting $\fixedassigof{\aplace}{\precovering.\groot}=\coordacc{\inmarkingof{\precovering}}{\aplace}$, and then following the definition of $\fixedassig{\aplace}$ along the edges.
The construction yields a well-defined function.
The reasoning is as follows.
Thanks to strong connectivity, two distinct values for a counter $\aplace\in \countersp$ implies two cycles with distinct effects on counter $\aplace\in\countersp$.
One of them must have a non-zero effect, which contradicts the definition of $\countersp$.

With $\countersp$ and $\fixedassig{\aplace}$ at hand, we make a case distinction along their properties.
While handling a case, we assume that all the previous cases do not hold.
\begin{itemize}
    \item there is a $\aplace\in\countersp$ with $\coordacc{\inmarkingof{\precovering}}{\aplace}\neq\coordacc{\outmarkingof{\precovering}}{\aplace}\in\nat$
    \item there is a $\aplace\in \countersp$ with $\fixedassig{\aplace}:\nodesof{\precovering}\to\nat$
    \item there is $\aplace\in\countersp$ and $\anode\in\nodesof{\precovering}$ with $\fixedassigof{\aplace}{\anode}<0$
    \item  $\countersp\cap\infinitiesof{\precovering}=\emptyset$
\end{itemize}

We proceed with the first case.
Let $\aplace\in\countersp$ with $\coordacc{\inmarkingof{\precovering}}{\aplace}\neq\coordacc{\outmarkingof{\precovering}}{\aplace}$.
This contradicts the feasability of $\sidecharof{\aside}{\precovering, \charmod}$, where $\aplace\in\aside$.
The reason is that any solution to $\sidecharof{\aside}{\precovering, \charmod}$ implies a union of loops in $\precovering$ with the effect $\coordacc{\outmarkingof{\precovering}}{\aplace}-\coordacc{\inmarkingof{\precovering}}{\aplace}$ on $\aplace$, but loops have $0$ effect on $\aplace$ by the definition of $\countersp$.

For the next case, let there be $\aplace\in \countersp$ with $\fixedassig{\aplace}:\nodesof{\precovering}\to\nat$.
In this case, we enrich the consistent assignment using $\fixedassig{j}$ to get $\precovering'$, and we return $U=\set{(\precovering', \charmod)}$, $V=\emptyset$.
This also concretizes the the final valuation in $\precovering$.
Since we concretized counters without modifying the structure, the rank decreases: we have $\infinitiesof{\inmarkingof{\precovering'}}+\infinitiesof{\outmarkingof{\precovering'}}+\infinitiesof{\precovering'}<\infinitiesof{\inmarkingof{\precovering}}+\infinitiesof{\outmarkingof{\precovering}}+\infinitiesof{\precovering}$ and thus $\edgesof{\precovering}=\edgesof{\precovering'}$.
This yields $(\precovering', \charmod)\rankless(\precovering, \charmod)$.
Language equivalence, and the requirements for consistent specialization follow by simple arguments.

In the remaining two cases, the decomposition is more involved.
Let the largest transition effect in $\precovering$ be $l$.
Let $B=(\sizeof{\nodesof{\precovering}}\cdot l\cdot C)^{\sizeof{\countersp}+1!}$ and $C=\max(\setcond{\coordacc{\inmarkingof{\precovering}}{\aplacep}}{\aplacep\in\leftside\cup\rightside}\setminus\set{\omega})+1$.
We construct the observer 
$$\anobserver_{\aplace}=(\set{\bot}\cup [B]_{\omega}, \edgesof{\precovering}, \transitions_{\aplace})$$
for all $\aplace\in\leftside\cup\rightside$. 
%
%
The transitions
\begin{align*}
    \transitions_{\aplace}=\setcond{(i, \anedge, i')}{i, i'\in [B]_{\omega}\cup\set{\bot}, \anedge\in\edgesof{\precovering},\; i'=\ctabstrof{B}{i+\coordacc{\effectof{\anedge}}{\aplace}}}
\end{align*}
track the counter $\aplace$ from $\coordacc{\inmarkingof{\precovering}}{\aplace}$, as long as its value is positive and below $B$.
We have $\initstates_{\aplace}=\set{\inmarkingof{\precovering}[\aplace]}$ for the initial states, and 
\begin{align*}
    \finalstates_{\aplace,\ulabel}&=\setcond{i\in [B]}{i\omegaleq \coordacc{\outmarkingof{\precovering}}{\aplace}}\;&\text{for all }\aplace\in\leftside\cup\rightside\setminus\infinitiesof{\inmarkingof{\precovering}}\\
    \finalstates_{\aplace, \vlabel}&=[B]\cup\set{\bot}\setminus\finalstates_{\aplace, \ulabel}\qquad&\text{for all }\aplace\in\rightside\setminus\infinitiesof{\inmarkingof{\precovering}}\\
    \finalstates_{\aplace, \vlabel}&=\emptyset\qquad&\text{for all }\aplace\in\leftside\cup\infinitiesof{\inmarkingof{\precovering}}.
\end{align*}
for the final states we are interested in.
Intuitively, these states capture the runs from $\interacceptofshort{\leftside}{\precovering, \charmod}$ by counting up to bound.
The final states in $\finalstates_{\aplace, \ulabel}$ accept the runs whose edges would reach $\coordacc{\outmarkingof{\precovering}}{\aplace}$ from $\coordacc{\inmarkingof{\precovering}}{\aplace}$ on the counter $\aplace$ without exceeding the bound $B$.
With $\finalstates_{\aplace, \vlabel}$, we track the violation of the constraints in $\precovering$.
We only need to track the runs that violate constraints on $\rightside$ counters, as those that violate $\leftside$ counters were not in $\interacceptofshort{\leftside}{\precovering, \charmod}$ in the first place.
We do not track any runs if $\aplace$ is initially $\omega$.
We will argue that we do not miss any $\interacceptofshort{\leftside}{\precovering, \charmod}$ runs as follows.
If a $\interacceptofshort{\leftside}{\precovering, \charmod}$ run were to escape $\finalstates_{\aplace, \ulabel}$ for all $\aplace\in\leftside\cup\rightside\setminus\infinitiesof{\inmarkingof{\precovering}}$, and $\finalstates_{\aplace, \vlabel}$ for all $\aplace\in\rightside\setminus\infinitiesof{\inmarkingof{\precovering}}$, then its non-fixed counters must be pumpable by Lemma~\ref{Lemma:Rackoffesque}.
This contradicts $\coveringseqof{}{\precovering, \charmod}=\emptyset$.

For both of the cases, we define $V$ the same way,  
$$V=\bigcup_{\aplace\in\infinitiesof{\precovering}}\decompalongof{\precovering, \charmod, \anobserver_{j}, \initstates_{j}, \finalstates_{\aplace, \vlabel}}.$$
Since the last property of Lemma~\ref{Lemma:DecoupledDecomposition} only concerns $V$, we handle this property by the following common proof.

We show that for all $\admgts\in V$, $\rightcharof{\admgts}$ is infeasible.
Suppose that there is an $\admgts=(\openmgts, \charmod)\in V$, such that $\rightcharof{\admgts}$ is feasible.
Let $\aplace\in\infinitiesof{\precovering}$ with $\admgts\in\decompalongof{\precovering, \charmod, \anobserver_{\aplace}, \initstates_{\aplace}, \finalstates_{\aplace, \vlabel}}$.
Since $\finalstates_{\aplace, \vlabel}=\emptyset$ if $\aplace\in\leftside$, this implies $\aplace\in\rightside$.
Let $\admgts=\precovering_{1}.\anupdate_{1}\ldots\anupdate_{k}.\precovering_{k+1}$, and suppose $\asol\in\solutionsof{\rightcharof{\admgts}}$.
By the construction of $\decompalongof{\precovering, \charmod, \anobserver_{\aplace}, \initstates_{\aplace}, \finalstates_{\aplace, \vlabel}}$, there is a sequence of edges $\arunp\in\langof{\anobserver_{\aplace}, \initstates_{\aplace}, \finalstates_{\aplace, \vlabel}}$ that applies the same effects as $\anupdate_{1}\ldots\anupdate_{k}$.
Assume that $\arunp'$ is the largest prefix of this sequence, where the counter valuation of counter $\aplace$ remains in $\N$ when taken from $\inmarkingof{\precovering}[\aplace]$.
Let $\sizeof{\arunp'}=i$.
By a similar argument to Section~\ref{SubSection:Case2Deets}, no cycle in the first $i+1$ precovering graphs can change the valuation of the counter $\aplace$.
If $i+1=k+1$, then $\arunp$ reaches a non-$\bot$ state, meaning some $b\in [B]\setminus\finalstates_{\aplace, \ulabel}$.
We have $\finalstates_{\aplace, \ulabel}\neq [B]$ and thus $\finalstates_{\aplace,\vlabel}\neq\emptyset$ if and only if $\outmarkingof{\precovering}[j]\neq\omega$.
Then, $\finalstates_{\aplace, \ulabel}=\set{\outmarkingof{\precovering}[j]}$ must hold. 
Since counter $\aplace$ was concretely tracked in $\anobserver_{\aplace}$ along the run of $\arunp$, we know that $\asol[\precovering_{\lastindex}, \mathsf{out}, \aplace]\in \finalstates_{\aplace, \vlabel}=[B]\setminus\set{\outmarkingof{\precovering}}$, where $\precovering_{\lastindex}$ is the last precovering graph in $\admgts$.
This is a contradiction.
Now, let there be at least $i+2$ precovering graphs in $\admgts$, and let $\anedge$ be the edge corresponding to the $i+1$-th update.
This means that $\arunp'.\anedge$ reaches the state $\bot$ or that $\arunp'$ reaches $\omega$ in $\anobserver$.
Since $\arunp'$ is a prefix of $\arunp$ and $\arunp$ reaches a non-$\omega$ state, only the former case can hold.
In this case, we have $\asol[\precovering_{i+2}, \mathsf{in}, \aplace]<0$, which contradicts $\asol\in\solutionsof{\rightcharof{\admgts}}$.

We handle the construction of $U$ and the remaining properties individually for each case.
We proceed with the case of a negative fixed assignment.
Let $\aplace\in\countersp$ and $\anode\in\nodesof{\precovering}$ with $\fixedassigof{\aplace}{\anode}<0$.
We construct $U$.
Because $\precovering$ is strongly connected, there is an ingoing edge $\anedge=(\anode', \aletter, x, \anode)\in\edgesof{\precovering}$ for some $\anode'$, $\aletter$, and $x$.
We remove $\anedge$ from $\precovering$ as follows.
We construct $\precovering'$ from $\precovering$ by removing $\anedge$, along with all $\anodep\in\nodesof{\precovering}$ that are not in the same strongly connected component as $\precovering.\groot$ after the removal of $\anedge$.
The result is strongly connected, and we let $U=\set{\precovering'}$.

Towards a proof of the second property of Lemma~\ref{Lemma:EdgeDecomposition}, we show that $V$ captures all $\leftside$-accepting runs in $\precovering$ that take $\anedge$.
If $\aplace\in\leftside$, then any run that is $\leftside$-accepting cannot take $\anedge$: since an $\leftside$-accepted a run starts from $\fixedassigof{\aplace}{\groot}=\coordacc{\inmarkingof{\precovering}}{\aplace}$, it must reach $\anode$ with $\fixedassigof{\aplace}{\anode}<0$ upon taking $\anedge$, which contradicts the positivity requirement of $\interacceptofshort{\leftside}{\precovering, \charmod}$.
Now let $\aplace\in\rightside$, let $\apath\in\interacceptofshort{\leftside}{\precovering, \charmod}$ take $\anedge$.
Then, some prefix $\apath'$ of $\apath$ must have an effect $u$ on counter $j$, such that $\coordacc{\outmarkingof{\precovering}}{\aplace}+u<0$.
This is required by $\fixedassig{\aplace}$.
We know that $\apath'$ cannot have a prefix that has an effect $u'$ on $\aplace$ with $\coordacc{\outmarkingof{\precovering}}{\aplace}+u'\geq B$.
We know this, because $\fixedassig{\aplace}(\anode)$ can be at most $l\cdot \sizeof{\nodesof{\precovering}}$ larger than $\coordacc{\inmarkingof{\precovering}}{\aplace}$ and $B\geq  \sizeof{\nodesof{\precovering}}\cdot l\cdot \coordacc{\inmarkingof{\precovering}}{\aplace}$ is guaranteed by the construction.
Thanks to this boundedness, the edges of $\apath$ have a run in $\anobserver_{\aplace}$ that reaches $\bot$ from $\initstates$.
By Remark~\ref{Remark:ObserverLanguage} there is $\apathp\in\interacceptofshort{\leftside}{\openmgts}$ for some $\openmgts\in\decompalongof{\precovering, \charmod, \anobserver_{\aplace}, \initstates_{\aplace}, \finalstates_{\aplace, \vlabel}}$ with $\apathp\pathequiv\apath$.
Since $V$ captures all $\leftside$-accepting runs that use $\anedge$, all remaining runs must avoid $\anedge$.
Furthermore, because $\leftside$-accepting, uncaptured runs in $\precovering$ are cycles on $\precovering.\groot$, they cannot visit the removed states.
So, for all $\apath\in\interacceptofshort{\leftside}{\precovering, \charmod}$, there is a $\apathp\in\interacceptofshort{\leftside}{U\cup V}$ with $\apath\pathequiv\apathp$.
Now we show the first property of Lemma~\ref{Lemma:EdgeDecomposition}, $\rankingof{\precovering'}<_{\mathsf{lex}}\rankingof{\precovering}$.
 It suffices to observe $\edgesof{\precovering'}\subset\edgesof{\precovering}$, since this also implies $\cyclespaceof{\precovering'}\subseteq\cyclespaceof{\precovering}$.
The argument for consistent specialization is the same as the one we gave in the case with $\aplace\in\countersp$ and $\fixedassig{\aplace}:\nodesof{\precovering}\to\nat$.

Finally, let the last case $\infinitiesof{\precovering}\cap\countersp=\emptyset$ hold.
We let 
$$U=\bigcup_{\aplace\in\infinitiesof{\precovering}}\decompalongof{\precovering, \charmod, \anobserver_{\aplace}, \initstates_{\aplace}, \finalstates_{\aplace, \ulabel}}.$$
We proceed with the first property of Lemma~\ref{Lemma:CoveringDecomposition}, and show $\openmgts\rankless\precovering$ for all $(\openmgts, \charmod)\in U$.
%
%
Let $(\openmgts, \charmod)\in\decompalongof{\precovering, \charmod, \anobserver_{\aplace}, \initstates_{\aplace}, \finalstates_{\aplace, \ulabel}}$ for some $\aplace\in\infinitiesof{\precovering}$.
%
%
We argue that for all precovering graphs $\precovering'$ in $\openmgts$, $\cyclespaceof{\precovering'}\subset\cyclespaceof{\precovering}$ holds.
This shows $(\openmgts, \charmod)\rankless(\precovering, \charmod)$, similarly to the proof of Lemma~\ref{Lemma:EdgeDecomposition}.
Let $\precovering'$ be a precovering graph in $\openmgts$.
Recall that each precovering graph in $(\openmgts, \mu)$ corresponds to a strongly connected component in $\precovering\times\anobserver_{\aplace}$.
Since $\openmgts$ is constructed using a basic path in $\precovering\times\anobserver_{\aplace}$ that reaches $\finalstates_{\aplace, \ulabel}\subseteq [B]$, said basic path cannot visit the sinks $\omega$ and $\bot$.
Thus, the strongly connected component that $\precovering'$ corresponds to, must also remain outside of the sinks, i.e. in $[B]$, since we can construct a run from these states to reach $[B]$.
Since all cycles in $\precovering'$ can be imitated in $\precovering$,
we only need to show $\cyclespaceof{\precovering'}\neq\cyclespaceof{\precovering}$ to show $\cyclespaceof{\precovering'}\subset\cyclespaceof{\precovering}$.
We know that $\aplace\in\infinitiesof{\precovering}$ is not fixed by $\infinitiesof{\precovering}\cap \countersp=\emptyset$, so there is a cycle $\aloop$ in $\precovering$ with $\effectof{\aloop}[\aplace]\neq 0$.
However, since the strongly connected component that corresponds to $\precovering'$ remains in $[B]$, this counter is concretely tracked.
This means that any cycle $\aloop'$ in $\precovering'$ has $\effectof{\aloop'}[\aplace]= 0$.
This means that no linear combination of cycles can have the same effect as $\aloop$, so $\effectof{\aloop}\not\in\cyclespaceof{\precovering}$, and thus $\cyclespaceof{\precovering}\neq\cyclespaceof{\precovering'}$.

Finally, we show the second property of Lemma~\ref{Lemma:CoveringDecomposition}.
Let $\apath=(\astate_{1}, \aconf_{1}).\anedge_{1}\ldots\anedge_{r-1}.(\astate_{r}, \aconf_{r})\in\interacceptofshort{\leftside}{\precovering, \charmod}$.
We must show the existence of $\apathp\in\interacceptofshort{\leftside}{U\cup V}$ with $\apathp\pathequiv\apath$.
By Fact~\ref{Remark:ObserverLanguage}, this is implied by $\anedge_{1}\ldots\anedge_{r-1}\in\langof{\anobserver_{\aplace}, \initstates_{\aplace}, \finalstates_{\aplace, \ulabel}\cup\finalstates_{\aplace, \vlabel}}$ for some $\aplace\in\infinitiesof{\precovering}$.
For the moment, assume that there is a $\aplace\in\infinitiesof{\precovering}\setminus\infinitiesof{\inmarkingof{\precovering}}$ such that $\anedge_{1}\ldots\anedge_{r-1}$ never visits $\omega$.
We postpone the proof of this claim.
Let $\aplace\in\leftside$.
We know that $\anedge_{1}\ldots\anedge_{r-1}$ does not reach $\bot$ in $\anobserver_{\aplace}$, since this would imply that the valuation of $\aplace\in\leftside$ becomes negative along the run.
We also know that $\anedge_{1}\ldots\anedge_{r-1}$ must reach $\omega$ or some $x\omegaleq\outmarkingof{\precovering}[\aplace]$, since $\arun\in\interacceptofshort{\leftside}{\precovering}$ and $\aplace$ is concretely tracked in $\anobserver_{\aplace}$ up to $B$.
Since $\anedge_{1}\ldots\anedge_{r-1}$ never visits $\omega$ in $\anobserver_{\aplace}$, we know that it reaches some $x\omegaleq\outmarkingof{\precovering}[\aplace]$.
We have $\anedge_{1}\ldots\anedge_{r-1}\in\langof{\anobserver_{\aplace}, \initstates_{\aplace}, \finalstates_{\aplace, \ulabel}}$,  since $\finalstates_{\aplace, \ulabel}=\setcond{x\in [B]}{x\omegaleq\outmarkingof{\precovering}}$.
Now let $\aplace\in\rightside$.
First, assume that $\anedge_{1}\ldots\anedge_{r-1}$ reaches $\bot$ from $\initstates_{\aplace}$ in $\anobserver_{\aplace}$.
We have $\bot\in\finalstates_{\aplace, \vlabel}$, and thus $\anedge_{1}\ldots\anedge_{r-1}\in\langof{\anobserver_{\aplace}, \initstates_{\aplace}, \finalstates_{\aplace, \vlabel}}$.
Now assume that $\anedge_{1}\ldots\anedge_{r-1}$ reaches some non-$\bot$ state in $\anobserver_{\aplace}$.
Since it cannot visit $\omega$ by our assumption, it must reach $[B]$.
Since $[B]\subseteq\finalstates_{\aplace, \ulabel}\cup\finalstates_{\aplace, \vlabel}$ in this case, we have $\anedge_{1}\ldots\anedge_{r-1}\in\langof{\anobserver_{\aplace}, \initstates_{\aplace}, \finalstates_{\aplace, \ulabel}\cup\finalstates_{\aplace, \vlabel}}$.

Now we show our assumption.
We reason with a contradiction.
Suppose for each $\aplace\in\infinitiesof{\precovering}\setminus\infinitiesof{\inmarkingof{\precovering}}$, the path $\anedge_{1}\ldots\anedge_{r-1}$ reaches $\omega$ in $\anobserver_{\aplace}$.
This means that, for each $\aplace$, there is an $m\leq r$ with $\coordacc{(\inmarkingof{\precovering}+\effectof{\anedge_{1}\ldots\anedge_{m}})}{\aplace}> B$, and $\coordacc{(\inmarkingof{\precovering}+\effectof{\anedge_{1}\ldots\anedge_{m'}})}{\aplace}\geq 0$ for all $m'\leq m$.
We execute $\anedge_{1}\ldots\anedge_{r-1}$ from $(\rootof{\precovering}, c_{1})$ for some $c_{1}\omegaleq\inmarkingof{\precovering}$ to get $\arun'$.
We apply Lemma~\ref{Lemma:Rackoffesque} to $\arun'$ and obtain a $\apathp\in\pathsof{\infinitiesof{\precovering}}{\precovering}$ with $\coordacc{\apathp}{\firstindex}=(\precovering.\groot, \aconf_{1})$, $\aconf_{1}\omegaleq\inmarkingof{\precovering}$, and $\coordacc{\apathp}{\lastindex}=(\precovering.\groot, x')$ with $\coordacc{x'}{\aplace}\geq C >\max(\setcond{\coordacc{\inmarkingof{\precovering}}{\aplacep}}{\aplacep\in\leftside\cup\rightside}\setminus\set{\omega})$.
We argue that $\apathp\in\coveringseqof{}{\precovering}$, which is a contradiction to the premise of case (iii).
Membership to $\coveringseqof{}{\precovering}$ does not impose any restrictions on counters $\aplace\in\infinitiesof{\inmarkingof{\precovering}}$.
The argument for the counters $\aplace\in\infinitiesof{\precovering}\setminus\infinitiesof{\inmarkingof{\precovering}}$ is clear from the properties delivered by Lemma~\ref{Lemma:Rackoffesque}.
The counters $\aplace\not\in\infinitiesof{\precovering}$, are already tracked by $\precovering.\varphi$.
Then, these counters can neither become negative nor have $\coordacc{x'}{\aplace}\neq\coordacc{\inmarkingof{\precovering}}{\aplace}$.
This concludes the proof.

%% file: conclusion.tex
\section{Conclusion and Future Work}

%
We have settled the decidability of the regular separability problem of VASS-reachability languages.
We gave a tight upper-bound of $\fof{\omega}$ for this problem.
Our development first used a transducer to fix one language to be $\dycklangn{n}$, and then employed a decomposition to reduce $\N$-separability to $\Z$-separability.
This relied on the key notion of faithfulness, which is based on modulo reachability constraints.

For future work, one might look in two directions.
The first direction is practical.
Recall that the VASS-reachability problem is $\fof{\omega}$-hard, which means that we have no hope of solving the worst-case instances.
However, this does not rule out an algorithm that runs in feasible time bounds on ``better-case'' instances.
Development of such an algorithm will need deeper insights into the reachability problem.
It would be interesting to see whether the insights we develop here can contribute to this.

The remaining direction is theoretical.
Namely, it would be interesting to explore other approaches to the problem.
Our approach fixes one language to $\dycklangn{n}$, and only decomposes the other langugage.
It would be interesting to explore an approach that treats the input languages symetrically.
It would also be interesting to see, how an algorithm using basic separators \cite{CZ20,kocher2023} would look.
Future work may also explore whether the existence of semilinear invariants \cite{Leroux10} can be used towards simpler proof of the decidability result.